# Labor Market Policies in High- and Low-Interest Rate Environments: Evidence from the Euro Area[*]

Povilas Lastauskas[†]
Homerton College, University of Cambridge, and Vilnius University

Julius Stakėnas[‡]
Vilnius University

**Abstract**

Do labor market policies initiated in periods of loose monetary policy yield different outcomes from those introduced when monetary tightening prevails? Using data from 11 euro-area members up to 2010–and extending to 17 countries up to 2020–we analyze three labor market policies: replacement rates, spending on active labor market policies (ALMPs), and employment protection. We find that these policies deliver different macroeconomic outcomes in low- and high-interest rate environments. In particular, ALMPs reduce unemployment if implemented under a loose monetary policy but not otherwise, whereas higher employment protection delivers expansionary effects under a tight monetary policy. These findings highlight that the effectiveness of labor market policies is significantly influenced by the monetary policy environment, emphasizing the need for coordinated policy design. Methodologically, we contribute by proposing to average local projections using Mallow's $C_p$ criterion, allowing for inferences that are robust to mis-specification and accommodate non-linearities.

[*]We are grateful, without implication, to anonymous referees, the participants of the ECB MPC task-force on structural reforms, particularly Klaus Masuch. Moreover, thanks to participants at the CESifo Venice Summer Institute Workshop "The Future of Europe" at the CESifo Venice Summer Institute Workshop. Many thanks to Romain Duval, Balázs Égert, Yuemei Ji, and Sushanta K. Mallick. Needless to say, all errors remain our own. This paper is a substantially updated and revised version of and a replacement for the CESifo Working Paper 7844, circulated under the title "Does It Matter When Labor Market Reforms Are Implemented? The Role of the Monetary Policy Environment". Lastauskas is grateful for the support from Iceland, Liechtenstein, and Norway through the EEA grants (Project No S-BMT-21-8 (LT08-2-LMT-K-01-073)) under a grant agreement with the Research Council of Lithuania.

[†]Homerton College, University of Cambridge, United Kingdom; Faculty of Economics and Business Administration, Vilnius University, Saulėtekio av. 9, 2nd building, Vilnius 10222, Lithuania. *Email*: P.Lastauskas@cantab.net. *Web*: www.lastauskas.com.

[‡]Faculty of Economics and Business Administration, Vilnius University, Saulėtekio av. 9, 2nd building, Vilnius 10222, Lithuania. *Email*: stakenas@gmail.com.

# 1 Introduction

There is large empirical literature dealing with the evaluation of labor market policies. However, it mainly focuses on the microeconomic evidence, whereas the macroeconomic evaluation of intervention into the labor market is quite limited. And when macroeconomic impacts are evaluated, they are often confined to the low frequency (e.g. annual) panel data studies (e.g., Bassanini and Duval, 2006; Blanchard and Wolfers, 2000; Scarpetta, 1996). Instead, our paper focuses on the dynamic paths of key macroeconomic variables in response to changes in labor market policies at quarterly data frequency, which allows us explore short term effects of policy changes as well as interactions with the monetary policy. Our target is the set of advanced and highly integrated economies within the euro area. To answer the question about the impact of the monetary policy environment on the macroeconomic impact of labor market policies, we have to tackle three empirical challenges. First, the precise interactions between labor and monetary policies are unknown. Second, those interactions can be non-linear. And, third, policy inference may be affected by agents' expectations about economic evolution and aggregate demand fluctuations over the business cycle.

We build a single empirical model that addresses all these issues – non-linearities, model uncertainty, and anticipation effects – as well as developing a statistical test to evaluate the importance of the monetary policy environment on the dynamic responses of macroeconomic variables to labor market policies for any horizon and the entire impulse response function. Our proposed methodology can be applied in other contexts where dynamic policy impacts are of interest.[1] We thus contribute by showing how to use Mallow's weighted local projections, recovering the significant role of monetary policy on the macroeconomic effects of labor policies, and helping to resolve the empirical active labor market policies (ALMP) puzzle, showing how the monetary policy stance can impact ALMP's macroeconomic effects.

In our baseline specification, we analyze data on 11 euro-area countries in the period 1985-2010. The end date helps us use the interest rate as the key monetary policy measure before new monetary policy tools take precedence.[2] Therefore, we do not need to take unorthodox monetary policy instruments into account; rather, our monetary policy definition includes decreasing (or low) interest rates and increasing (or high) interest rates over time periods. For robustness purposes, however, we expand the sample to include six new euro-area members with varying data availability, reaching the year 2020. Even with the expanded sample, we found that reducing employment protection and increasing ALMP spending are more effective under looser monetary policy, as they reduce unemployment more and can boost international competitiveness. Conversely, more labor protection is warranted under tighter monetary policy.

We consider a drop in the short-term interest rate on an annual basis as well as a lagged measure on a quarterly basis. When it comes to labor market policies, we analyze employment protection legislation, spending on active labor market policies, and the unemployment insurance replacement rate (the share of a salary one can expect to receive when unemployed). These policy measures received most of the literature attention, particularly in the European context (refer to Boeri (2011) and Eichhorst et al. (2017) who cover major institutional reforms in European labor markets, focusing on employment protection, unemployment benefits and activation schemes). Our interest lies in interactions between policies themselves, and *not* in identified monetary and labor market policy shocks. In the identified shock case, the question is about the interaction between surprise (unexpected) monetary policy (or interest rate) shocks and labor policies, whereas our focus is on the monetary policy environment within which labor market policies *happened to be*

---

[1] In the Appendix we also show how to accommodate a horizon-specific error factor structure. This feature is particularly useful for a set of interdependent economies affected by time-varying common shocks.

[2] There is policy tightening in 2011, which is retrospectively considered a mistake in the middle of a supply shock and debt crisis, leading to the expansion of the ECB balance sheets (largest increases followed refinancing operations with a maturity of thirty-six months in December 2011 and March 2012 and a series of targeted longer-term refinancing operations in 2014 and in later years), forward guidance ("whatever it takes") in 2012 and a negative deposit facility rate in 2014.

*initiated.*[3]

Theoretical work has shed light on expected labor market outcomes under different economic conditions. For instance, as shown theoretically in Cacciatore et al. (2016a), more flexible labor markets (lower firing costs) have steep adverse effects on employment and output in the short run (in particular, if implemented in recession), whereas a drop in unemployment benefits boosts the economy more in recession than in normal times. Yet empirically these findings are less clear-cut and often opposite to the theoretical predictions. For example, Duval and Furceri (2018) find that weaker employment protection and lower unemployment benefit have expansionary effects in good times, but can become contractionary in periods of slack. Similarly, Duval et al. (2020); Thommen (2022) find that job protection deregulation works during a positive state of the business cycle but it is costly during bad times. Finally, Lastauskas and Stakenas (2020a) show theoretically and empirically that larger unemployment benefits are contractionary and unemployment-increasing, unless implemented in a loosening monetary policy environment. Even though a more flexible labor market brings about a considerable increase in real GDP and a drop in unemployment in normal times, a more rigid labor market is preferred over the longer run during a crisis period and, in particular, if interest rates cannot be lowered.[4] These studies, however, abstract from model uncertainty, including how different states of an economy (e.g., different monetary policy stances) impact labor market policies' effect on the macroeconomy.[5]

Even though international organizations make structural labor market policies as key ingredients to revive productivity, improve economic performance, and enhance resilience to shocks (see, for example, OECD 2011; ECB 2017; Banerji et al. 2017), there is a lot of uncertainty about the required pre-conditions for the policy changes to deliver desired results, in particular in the short to medium term. The lack of theoretical and empirical consensus (see summaries in Crépon and van den Berg 2016; McKenzie 2017; Ulku and Georgieva 2022, which demonstrate difficulties in establishing robust impacts of labor policies) and uncertainty about the model call for a robust empirical strategy to infer the dynamic path of policy effects. Some policy reforms (like activation schemes) remain heavily suggested by international bodies as ways to deal with adverse shocks, though solid empirical evidence of their macro effect is still missing.[6]

We tackle this gap by first analyzing the effects that labor market policies have on the macroeconomy, depending on whether the policy is implemented in a low- or high-interest rate environment. Second, we explore the interactions of realized labor and monetary policies, allowing for different definitions as well as regimes (independent monetary policy versus common policy in the monetary union) and non-linearities. Third, we merge Jordà and Hansen into a methodological framework which does not require knowledge of the exact functional form, is robust to mis-specification, admits non-linearities, and addresses uncertainty regarding interactions between labor policies and the macroeconomy. It is a frequentist alternative to Bayesian model averaging. It uses key features of a popular local projections methodology, namely a least squares estimator for each spanning horizon, enabling the uncovering of horizon-specific optimal weights for an average dynamic response. Equipped with these tools, we document monetary policy's role in shaping macro responses to labor policies.

The paper is organized as follows. In Section 2, we briefly describe the economics of labor market policies, data, econometric methodology, a baseline model and a few extensions. Section

---

[3]Suppose that agents expect 1 p.p. decrease in the policy rate and it actually drops by 1 p.p. Though the shock is zero, agents' behavior may nevertheless be different in a higher and in a lower interest rate environment.

[4]In addition to monetary policy, fiscal stance might also matter. For instance, Burlon et al. (2021) studies fiscal devaluation's impact on the euro area's labor markets and finds important interactions with the monetary policy stance. We leave it for future research to explore how fiscal and monetary policy interactions affect the macroeconomic impacts of labor policies.

[5]An opposite perspective, examining the impact of labor reforms on monetary policy, particularly inflation persistence, is explored in Geronikolaou et al. (2016).

[6]For instance, active labor market policies (ALMPs) are said to be "key for a well-functioning labor market and for recovery of jobs and incomes" by a joint ILO, ISSA and OECD report.

[3](#) describes the empirical results of local projections, by splitting the entire sample into pre- and post-euro periods (to capture monetary policy differences) as well as covering the overall sample. A formal statistical test is developed and carried out to evaluate the importance of the monetary policy environment on the dynamic responses of macroeconomic variables to labor market policies. Section [4](#) sets out a unified framework wherein model uncertainty is taken into account. It describes the choice of weights for each time horizon, each response variable and each labor market policy. After constructing and testing an average model, we extend our framework to the case where anticipation effects and aggregate demand fluctuations are taken into account in Section [5](#). We also expand our sample until 2020 and include new euro-area members. Section [6](#) concludes and specifies policy implications and directions for further research. The Online Appendix provides all the supporting material.[7]

## 2 Framework

We briefly review different statistical issues that need to be addressed in order for the empirical results to be valid and robust. We thus set out by describing the empirical literature of labor policies macroeconomic evaluation and then proceed to the econometric framework, a baseline model and some extensions.

### 2.1 Economics of Labor Market Policies

Our focus on three types of policies makes it possible to shed light on important dimensions of the labor market.[8] The first variable, employment protection legislation, measures the difficulty in hiring and firing workers; the second one – spending on active labor market policies – accounts for spending on programs that help the unemployed find work; whereas the last one – the replacement rate – accounts for the expected unemployment benefits as a share of salary before losing a job. The interaction of labor market institutions with macroeconomic fluctuations has recently been investigated by Gnocchi et al. (2015). The authors empirically demonstrate that a higher replacement rate and employment protection are associated with the volatility of unemployment. We thus consider aggregate variables from a standard small open economy macroeconomic model (refer, for instance, to the canonical model in Gali and Monacelli (2005) and its extensions with empirical considerations in Pesaran and Smith (2006) or Dees et al., 2007). It is standard to cover real output, inflation, exchange rate and interest rate with some additions for the small open economy context (to account for the trade network wide effects, we deal with the real effective exchange rate). We complement this setup with unemployment, as it is of genuine interest to the policymakers who devise and implement labor market policies.

The efficacy of labor market policies remains an open question in the literature, so our work also contributes to that stream. More generous unemployment benefits generally have an increasing effect on unemployment (similar to a tax wedge, which also tends to make it rise), both due to a change in job search motive and a sharp contraction in vacancy creation (Hagedorn et al., 2015). However, that is not necessarily true in the case of incomplete markets (Kekre, 2022); and things are even less transparent for active labor market policies (ALMP) and depend on the specific ALMP categories (see Bassanini and Duval, 2006, Nickell et al., 2005, and Orlandi, 2012). It constitutes an empirical puzzle since there is no clear-cut evidence on the efficacy of ALMP. However, it remains one of the most used and recommended labor policies, extending beyond traditional shock absorption, but also recently suggested by the IMF (refer to Brollo et al., 2024)

---

[7]The accompanying Supplementary Material file, codes and data sets will be posted on the corresponding author's website.

[8]Our approach differs from the recent literature on large (major) reforms only (e.g., Aumond et al. 2022; Duval and Furceri 2018; Duval et al. 2021; Rünstler 2021) as we focus on all (conditional) changes rather than just a few instances considered to be large enough to qualify as a "large reform." In other words, the interpretation of our results is different: we are interested in learning the average response to *all* historical movements in labor policy variables.

to cushion disruptions associated with the enhanced use of generative artificial intelligence.

Though policy relevant (for instance, ECB (2017) emphasizes that business leaders consistently rank labor market reforms as the most pressing area for further work), there are challenges in robustly identifying the impacts of reforms. In a macro model with a spatial dimension and cross-country inter-dependencies, Felbermayr et al. (2013) find that labor market institutions that are prone to increase unemployment in a home economy spill over to foreign countries and make unemployment rise there.[9] We omit a number of important dimensions that the literature has explored (income and substitution, labor supply, and other channels) but instead offer new light on the importance of non-linearities, model uncertainty and interactions between labor market policies and business-cycle movements.

As our identification stems from both temporal and cross-country variation, we abstract from country-specific experiences and concentrate on a monetary union instead.[10] We hypothesize that the low- or high-interest rate environment as well as common and independent monetary policy may lead to different effects of labor policies.[11] There is a growing literature that sheds light on the potential effects of labor market policies and their changes, yet relatively little is known about interactions between labor market policies and stances of monetary policy.[12] Bassanini and Duval (2006) find evidence that high replacement rates and tax wedges are associated with higher unemployment, as is anti-competitive product market regulation.

Boeri (2011) reviews a large strand of literature on reforms of employment protection, unemployment benefits, active labor market policies and employment subsidies. He reports that effects are very sensitive to the nature of the reform, its magnitude, as well as the phasing-in and phasing-out stages, and stresses the need to account for severe asymmetries that many real-world reforms create as they rarely affect the entire population. Yet, little is known if monetary policy, affecting firms' and households' decisions, makes labor policies deliver different outcomes. Adjustment mechanisms for labor policies (in fact, to a substantially lesser extent for product market changes, which we abstract from) are found to be different in normal times compared to recessions in Cacciatore et al. (2016a). Intuitively, when aggregate productivity is below trend, job creation and destruction react differently to policy changes due to their effects on outside options and wages. Since monetary policy targets a cyclical component, it is of interest to learn whether job creation and destruction frictions deliver different outcomes over the business cycle frequency.[13]

---

[9] Emphasis on cross-country spillovers is another emerging aspect that has recently been examined in a number of contributions. Dao (2008) finds that German labor market reforms create positive spillovers for the rest of the euro-area countries. Felbermayr et al. (2013) analyze 20 OECD countries and find that more rigid labor market increase unemployment at home and abroad, and the spillover magnitude depends on the relative size of countries and the trade costs. The theoretical accounts of Felbermayr et al. (2015) conclude that labor market reforms benefit trading partners, whereas Lastauskas and Stakenas (2020b) stress that spillover effects depend on the particular labor policy reform.

[10] We thus abstract from country-specific experiences such as the flexicurity reforms in Denmark and the Nordic countries and the so-called 'Kurzarbeit' reduced-working-hours program and Hartz reforms in Germany (see Moeller (2010) on the buffer capacity within firms, Faia et al. (2012) on large fiscal multipliers from 'Kurzarbeit' policies and Burda and Hunt (2011) for the effects of German labor market regulation during the financial crisis).

[11] By combining synthetic control ideas with the impulse response functions, Lastauskas and Stakėnas (2022) demonstrate that monetary policy shocks yield different outcomes for countries within a monetary union compared to those with independent monetary policies, suggesting that labor policies may also produce varying results and interact differently with monetary policy depending on a country's membership in a monetary union.

[12] There is, however, a rather rich literature on monetary policy and fiscal policy over business cycles. While these are usually analyzed separately, there is an emerging strand on the interactions of these factors. However, labor market policies, despite being one of the most heavily used policies in practice, Boeri (2011), do not receive the same attention. For more general definitions regarding the fiscal side, see, among many others, Auerbach and Gorodnichenko (2013) who, using a panel of OECD countries, establish the importance of cross-country spillovers and a larger impact of fiscal shocks when the affected country is in recession. Owyang et al. (2013), for their part, find that government spending multipliers are no greater during periods of high unemployment in the United States but there is some evidence otherwise for Canada. On the monetary policy side, Cloyne and Hürtgen (2016) demonstrate that an increase in monetary policy rate reduces output and inflation. Belinga and Ngouana (2015) employ quarterly U.S. data and show that the federal government spending multiplier is substantially higher under accommodative than non-accommodative monetary policy. On the interactions, Rossi and Zubairy (2011) analyze monetary and fiscal policy shocks simultaneously and conclude that fiscal shocks are more important in explaining medium-cycle fluctuations, whereas monetary policy shocks matter more for business cycle fluctuations. Tenreyro and Thwaites (2016) report that monetary policy is less powerful in recessions, and gather some evidence that fiscal policy has counteracted monetary policy in recessions but may have reinforced it in booms.

As is clear from the above discussion, monetary policy interactions with labor market policies are far from settled. Economic theory, however, helps to justify shocking labor market variables and conditioning on the macroeconomic aggregates. The theoretical literature, for example, Helpman et al. (2010) derive results when the labor market can be separated from the goods market in the sense that first labor market conditions (such as search costs and labor market tightness) are computed and then the goods market is solved for (average or aggregate productivity), given labor market conditions. Felbermayr et al. (2011), in a different environment, also find that the average productivity is independent of labor market outcomes, thus also yielding a recursive structure. Following the theoretical trade literature, we thus condition on all macroeconomic variables, and consider first changing labor market policies and exploring how aggregate variables evolve. Economic theory, however, falls short of guiding medium-run (business-cycle) adjustments and interactions with loosening and tightening monetary policy, in particular for economies with independent monetary policy and for monetary union members. We therefore have to address model uncertainty along with the different, possibly non-linear, channels that might describe interactions between labor and monetary policies. Instead of trying to pin down the true data generating process, we instead work out a combined (weighted) model.[14]

## 2.2 Empirical Macroeconomic Evaluation of Labor Policies

Since monetary policy targets a cyclical component, it is of interest to learn whether job creation and destruction frictions deliver different labor market policy outcomes over the business cycle frequency. As has already been summarized, the literature is not settled on macroeconomic effects of labor market institutions. Given the uncertainty regarding the way labor market policy affects macroeconomic outcomes, the timing of labor policies and the time span over which they yield results as well as uncertain responses of cross-country interactions, we propose to estimate a dynamic path of macroeconomic outcomes in a flexible way using cross-country variation to average out country-specific as well as spillover effects. It is clear that econometric methodology should allow for flexible functional form to ensure that monetary policy and labor market policies are adequately captured. What is more, abstracting from a particular economic framework, we should ensure that model uncertainty is part of the modeling strategy.

Since our main goal is the dynamic response of macroeconomic variables to changes in the labor market policies under different monetary policy stances, we focus on the impulse response function. However, instead of the use of vector autoregressions (VARs) in the context of labor market policy evaluation (see, among others, Abbritti and Weber, 2018; Boeck et al., 2022; Lastauskas and Stakenas, 2018), we will make use of local projections (Jordà, 2005). Plagborg-Moeller and Wolf (2021) prove that under the unrestricted lag structure, local projections and VARs estimate the same impulse responses. However, local projection is more robust to lag length mis-specification (Jordà, 2005) and is not subject to propagation of mis-specification across equations as is the case in the multivariate model as VAR. What is more, Olea and Plagborg-Moeller (2021) show that local projections are robust to highly persistent data and the inference of response at long horizons.

---

[13]There are also a number of possible channels which are relevant for households such as the role of financial frictions (borrowing constraints), planning horizon, asset types – everything that affects labor supply and income is relevant for unemployment and the macroeconomy (recent literature on the heterogeneous agent New Keynesian models point to the importance of labor market adjustments; see Kaplan et al. (2018) and references therein). Our intention, however, is limited to identifying the existence of interactions, not the channels that give rise to them.

[14]Instead of testing different models against each other and hoping that the true one is part of the considered set, model averaging turned out to be particularly powerful in prediction problems (as is our main focus, the impulse response function). For macroeconomic predictions refer to Stock and Watson (2006), whereas for the machine learning approach that applies ensemble methods, see Athey et al. (2019) and references therein.

For the above ends, we combine Jordà, 's local projections with the Mallow's $C_p$ criterion (Hansen, 2007) to arrive at an inference that does not require knowledge of the exact functional form, is robust to mis-specification, admits non-linearities, and addresses uncertainty regarding interactions between labor policies and the macroeconomy. The dynamic effects of policy changes are conveniently captured by the impulse response function. It has a connection to the causal inference and treatment effect: an impulse response can be considered as an average treatment effect provided conditional independence holds. As labor market policies do not follow established rules (as, for instance, in the monetary policy case), we resort to the regression control strategy and do not expect that, after controlling, specific labor market policies (which are fiscally rather negligible from a macroeconomic perspective) are correlated with an unobserved variable which is correlated with macroeconomy-wide outcomes.[15] What is more, we model independent monetary regimes and the monetary union, where the latter provides us with the more causal interpretation about policy outcomes (independent monetary policy at the Euro area level is legally and institutionally separated from the individual country's labor market changes). Suspecting anticipation effects and common shocks across economies, as a robustness check, we further control for the forecasts of GDP, output gaps and time-varying unobserved dynamic factors with country-specific loadings.

Note, however, that conditions for model consistency are weak by construction; Jordà (2005) shows that impulse responses can be calculated by a sequence of projections of the endogenous variables shifted forward in time onto its lags. The residuals from the local projection are a moving average of the forecast errors from time $t$ to $t+k$ and as such are uncorrelated with the regressors, which are dated $t-1$ to $t-p$ under weak requirements of no announcement effects (which are very likely to hold in our application specific labor market policies, observed over quarterly frequency).[16] Since labor market policies are usually set having long-term developments in mind (structural changes), our business-cycle frequency in outcome variables alleviates endogeneity problem of the labor policy variable (even having in mind a political cycle argument, large policy changes are not implemented having a few quarters ahead in mind). What is more, projections are local to each forecast horizon and therefore more robust to mis-specification of the unknown data generating process. We also do not face a problem of unobserved shock or response variables, so impulse responses can be simply uncovered using a standard regression method. We will make use of this property whilst dealing with the model uncertainty.

In fact, we employ a method to average a model by selecting the weights that minimize a Mallow's $C_p$, which is an estimate of the average squared error from the model average fit and is an unbiased estimate of the expected squared error (Hansen, 2007). The method thus targets least squares regressions, and our local projections happen to be them. Uncovered weights are optimal in the sense of asymptotic minimization of conditional squared error (with no restrictions on the largest model).[17]

Some caveats of our joint approach, which uncovers dynamic paths of policies and model uncertainty, are as follows. First, Mallows' criterion depends on unknown variance, $\sigma^2$, which needs to be estimated and thus a bias may be introduced. The Mallow's criterion assumes homoskedasticity. Liu et al. (2016) consider generalized least squares in the presence of heteroskedasticity, as it is well known that they have smaller variances. The authors use Mallow's $C_p$ weights of various GLS estimators where each model uses different set of regressors.[18] Inference about uncertainty using model average estimates is difficult, however. There are no useful standard errors for the model average estimators. The estimates have non-standard distributions and are non-pivotal. It is also unclear how to form confidence intervals from the estimates. Despite these limitations, however, our approach provides gains on robustness, in particular when it comes to mis-specification errors and empirically important non-linear effects.

---

[15]This condition is also known as the "selection on observables" assumption. The literature has entertained a number of potential solutions, such as instrumental variables, but they do not naturally exist in our context. Some alternatives include inverse probability estimators and matching, as well as a combination of these methods. They, however, require the existence of a policy model, which is used to predict a policy and is later used to weigh treated and non-treated variables. As our "treatment" is continuous and not discrete, applies to all countries, and is observable (versus latent), we instead rely on main macro variables and unobserved heterogeneity (fixed effects) as sufficient controls. Instead of addressing all sources of endogeneity, we rather focus on non-linear effects and model uncertainty. We leave it for future research to extend our approach.

[16]This guess has been confirmed in our robustness checks, where we expanded our baseline model to address announcement effects.

[17]Wan et al. (2010) show that there is an alternative proof compared to the one proposed by Hansen (2007) in which the optimality of the Mallow's criterion is preserved for continuous model weights and under a non-nested model environment that allows any linear combination of regressors in the approximating models that make up the model average estimator. This property is particularly useful in our environment where the impact of labor policies may depend on different measures of interest rate changes and their interactions and makes the methodology widely applicable in other circumstances.

## 2.3 Data Description

Before delving into the results, we shall briefly review the data. We run a baseline panel model that includes the following euro-area members: Austria, Belgium, Finland, France, Germany, Greece, Ireland, Italy, the Netherlands, Portugal and Spain. Basic summary statistics for each economy are reported in Table 2.1. The expanded sample includes six more euro-area members (Estonia, Latvia, Lithuania, Slovakia, Slovenia, and Luxembourg). Note that we deal with the quarterly changes in log variables, namely quarterly growth rates. We analyze the following three labor market policies. Public expenditure on active labor market policies is defined as expenditure per number of unemployed (we take the number of unemployed in the previous year) divided by GDP per capita. Employment protection legislation index accounts for procedures and costs involved in dismissing and hiring of individuals or groups of workers. And unemployment replacement rate is defined as the proportion of net in-work income that is maintained when unemployed.

In our baseline sample, the expenditure on active labor market policies was growing for all economies but Finland and Germany. Unemployment benefits were mainly on the rise except for Belgium, Germany and the Netherlands. Finally, employment protection was declining – labor markets were more liberalized – everywhere except for Belgium. The key macroeconomic variables, that is, gross domestic product (GDP), inflation, unemployment, real effective exchange rate (REER), and short interest rate are summarized in Table 2.1 and elaborated further in Appendix A.2.

## 2.4 Baseline Model and Extensions

Since local projection methodology (Jordà, 2005) permits simple non-linear modeling, we use interest rate variable as an interaction term, both as a level variable and as a dummy, to explore differences in macroeconomic responses to labor market policies. Which channel is more important to impact the effect of labor policies on the macroeconomy is an empirical question. The economic agent might respond to the fact that the policy rate changed or its size; it could be that a longer rate is of importance since it drives the mortgage market and may impact businesses making investment decisions, or it could be that accommodative monetary policy and the size of a drop in the policy rate matter. We shall consider one baseline model (absent interactions with the monetary policy environment) and its five extensions (A-E) which include non-linear interactions between labor market and monetary policies:[19]

---

[18]Liu and Okui (2013) extend the method of Hansen (2007) to models with heteroskedastic errors. Hansen and Racine (2012) propose jackknife model averaging that is also robust to heteroskedasticity. Jackknife model averaging is extended to the case of dependent data by Zhang et al. (2013).

[19]Note important differences compared to the current literature on labor market reforms such as Duval and Furceri (2018). In our model, unlike theirs, there is no contemporaneous effects from explanatory variables; i.e., they are *predetermined* from the outcome variable's perspective. Therefore, such concerns as simultaneity or contamporenous determination of variables are not relevant in our case. The potential anticipation effects are dealt with in Section 5.1.

Table 2.1: Basic summary statistics of model variables

| Country | GDP | | Inflation (CPI) | | Unemployment | | REER | | Short interest rate | | ALMP | | UB | | EPL | |
|---|---|---|---|---|---|---|---|---|---|---|---|---|---|---|---|---|
| | mean | sd | mean | sd | mean | sd | mean | sd | mean | sd | mean | sd | mean | sd | mean | sd |
| Austria | 0.56 | 1.00 | -0.00 | 0.34 | 0.09 | 5.19 | 0.00 | 1.23 | -0.06 | 0.50 | 0.71 | 6.32 | 0.08 | 1.15 | -0.14 | 0.89 |
| Belgium | 0.50 | 0.76 | -0.01 | 0.45 | -0.28 | 4.13 | 0.22 | 1.30 | -0.09 | 0.57 | 0.68 | 2.95 | -0.04 | 0.51 | 0.05 | 0.76 |
| Finland | 0.53 | 1.47 | 0.00 | 0.42 | 0.46 | 6.13 | -0.23 | 3.18 | -0.12 | 0.92 | -0.07 | 5.83 | 0.00 | 0.84 | -0.20 | 0.89 |
| France | 0.45 | 0.54 | -0.01 | 0.33 | -0.12 | 2.34 | 0.05 | 1.39 | -0.09 | 0.64 | 0.48 | 3.13 | 0.15 | 0.86 | -0.04 | 0.66 |
| Germany | 0.46 | 0.93 | -0.00 | 0.42 | -0.19 | 2.95 | -0.07 | 1.95 | -0.05 | 0.46 | 0.03 | 5.83 | -0.22 | 0.89 | -0.16 | 0.56 |
| Greece | 0.47 | 2.27 | -0.04 | 0.74 | 0.54 | 3.38 | 0.19 | 2.29 | -0.15 | 3.23 | 0.15 | 11.52 | 0.86 | 2.60 | -0.16 | 0.91 |
| Ireland | 1.12 | 1.74 | -0.01 | 0.48 | -0.10 | 6.07 | 0.11 | 2.61 | -0.13 | 1.43 | -0.28 | 4.97 | 0.41 | 1.40 | -0.10 | 0.75 |
| Italy | 0.34 | 0.72 | -0.02 | 0.28 | 0.04 | 2.78 | 0.05 | 2.75 | -0.15 | 0.82 | 0.44 | 3.97 | 4.44 | 13.92 | -0.10 | 0.29 |
| Netherlands | 0.55 | 0.65 | -0.00 | 0.37 | -0.54 | 4.45 | 0.10 | 1.94 | -0.05 | 0.47 | 0.87 | 5.78 | -0.45 | 1.52 | -0.10 | 0.40 |
| Portugal | 0.62 | 1.03 | -0.05 | 0.55 | 0.30 | 5.12 | 0.43 | 1.52 | -0.20 | 1.00 | 0.76 | 7.42 | 0.78 | 1.83 | -0.23 | 0.60 |
| Spain | 0.64 | 0.59 | -0.01 | 0.53 | -0.00 | 4.21 | 0.38 | 2.04 | -0.11 | 0.95 | 0.90 | 7.03 | 0.00 | 0.63 | -0.36 | 2.03 |

Note: the variables are quarterly, spanning from 1985Q1 to 2010Q4. The summary statistics are calculated for 100*log(quarterly changes), except for the interest rate which is simply a quarterly change.

$$
\begin{aligned}
\triangle y_{i,t+k} = \alpha_i + \boldsymbol{\beta}' \boldsymbol{X}_{it} + \boldsymbol{\gamma}' \boldsymbol{X}_{i,t-1} + \delta_1 \triangle \ln LMP_{it} + \delta_2 \triangle \ln LMP_{i,t-1} + \quad & k = 1, \ldots, 12, \\
+ \delta_3 \triangle \ln LMP_{it} \times \mathbb{I}_{\triangle i_{it} < 0} + u_{i,t+k} \quad & \text{Model A}, \\
+ \delta_3 \triangle \ln LMP_{it} \times \mathbb{I}_{\triangle^a i_{it} < 0} + u_{i,t+k} \quad & \text{Model B}, \\
+ \delta_3 \triangle \ln LMP_{it} \times \triangle i_{it} + u_{i,t+k} \quad & \text{Model C}, \quad (2.1)\\
+ \delta_3 \triangle \ln LMP_{it} \times \triangle^a i_{it} + u_{i,t+k} \quad & \text{Model D}, \\
+ \delta_3 \triangle \ln LMP_{it} \times \mathbb{I}_{\triangle i_{it} < 0} + \delta_4 \triangle \ln LMP_{it} \times \triangle i_{it} + & \\
\delta_5 \triangle \ln LMP_{it} \times \triangle i_{it} \times \mathbb{I}_{\triangle i_{it} < 0} + u_{i,t+k} \quad & \text{Model E},
\end{aligned}
$$

where $\triangle y_{it}$ stands for a change in either log consumer price index or CPI (resulting in log inflation),[20] log real GDP, log real effective exchange rate or log unemployment, a set of controls $\boldsymbol{X}_{it}$ include two lags of a change in log CPI (two lags of log inflation), log real GDP, log real effective exchange rate, log unemployment and short nominal interest rate, $\triangle \ln LMP_{it}$ refers to a change in log labor market policies (expenditure on active labor market policies, replacement rate and employment protection legislation index)[21] and $u_{i,t+k|t}$ stands for the forecast error at time $t$ for forecasting $k$ periods ahead.

Our model, as defined in (2.1), explores how monetary policy influences the effects of labor policies on the macroeconomy through various scenarios. A quarterly change in the interest rate is denoted by $\triangle i_{it}$, whereas an annual change is written as $\triangle^a i_{it}$. Model A suggests that the impact may differ during periods of negative quarterly interest rate changes ($\mathbb{I}_{\triangle i_{it} < 0}$, where $\mathbb{I}$ is an indicator function that splits into periods when the condition in the subscript is satisfied and when it is not), while Model B focuses on the significance of negative annual rate changes ($\mathbb{I}_{\triangle^a i_{it} < 0}$). Model C examines the interaction between quarterly interest rate changes $\triangle i_{it}$ (specifically, not the timing of these changes) and labor market policies. In contrast, Model D considers the interaction with annual rate changes ($\triangle^a i_{it}$). Model E identifies periods of negative quarterly interest rate changes ($\mathbb{I}_{\triangle i_{it} < 0}$) and examines both the interaction of interest rate changes ($\triangle i_{it}$) with labor policies and a triple interaction involving labor policy changes, interest rate changes ($\triangle i_{it}$), and the timing of negative interest rate changes ($\mathbb{I}_{\triangle i_{it} < 0}$). Collectively, these models highlight the uncertainty of how the interest rate environment and labor policies together influence the macroeconomy.

A baseline model, for instance, for the real GDP looks like

$$
\triangle \ln RGDP_{i,t+k} = \alpha_i + \boldsymbol{\beta}' \boldsymbol{X}_{it} + \\
+ \boldsymbol{\gamma}' \boldsymbol{X}_{i,t-1} + \delta_1 \triangle \ln LMP_{it} + \delta_2 \triangle \ln LMP_{i,t-1} + u_{i,t+k}.
$$

The power of the approach lies in the fact that an impulse response can be thought of as an average treatment effect, often the main object of interest.[22] Furthermore, note that we are dealing with a case when both outcome and policy (shock) variables are observed unlike, for instance, structural VAR literature. In such a case, the average (both over time and across countries) dynamic response of variable $y_{it}$ to an initial shock to labor market policies, $LMP_{it}$, coined as impulse response function, is nothing else but a vector of 12 elements for each horizon that we consider: $\tilde{\boldsymbol{\delta}} = \left( \tilde{\delta}_1, \ldots, \tilde{\delta}_{12} \right)$. Notice that in the linear world $\tilde{\delta}_k$ represents a single parameter, whereas in the non-linear applications $\tilde{\delta}_k$ includes a number of other elements that capture interaction and other terms ($\partial \triangle y_{i,t+k} / \partial \triangle \ln LMP_{it} = \delta_1$ for the baseline model, but this is no longer true for the extensions A-E).

---

[20]The approximation works best for small values of inflation since $(\text{CPI}_t / \text{CPI}_{t-1} - 1) \approx \ln(\text{CPI}_t / \text{CPI}_{t-1})$.

[21]For the setting where identified shocks are used, refer to Duval and Furceri (2018); Lastauskas and Stakenas (2020a), among others. In this paper we are interested in all labor market changes, *conditional* on macroeconomic history (inflation, real GDP, real effective exchange rate, unemployment and short nominal interest rates), and their impacts on the macroeconomy (similarly to Hur (2019); Goulas and Zervoyianni (2018); Lastauskas and Stakenas (2018), among others).

[22]There is no accepted equivalent of the Taylor rule for labor market policies, we thus resort to the regression control strategy.

Once (2.1) is run for each horizon, we obtain Jordà's local projection IRF, which is a consistent estimate of the impulse response. Single response is still prone to the functional form uncertainty. In fact, uncertainty regarding multi-step forecasts has led to a large literature on forecast combination, initiated by Bates and Granger (1969) and Granger and Ramanathan (1984). Our motivation revolves around model uncertainty for policy evaluation, but the theoretical basis is alike: impulse responses are functions of multi-step forecasts, and, as such, all the issues surrounding forecast combinations are applicable to our context, too.

## 3 Local Projections

Though standard economic theory suggests that employment should increase with subsidies and decrease with benefits, we lack knowledge of *a priori* effects on other macro variables (refer to the recent literature on market incompleteness, which impact how labor market policies affect not only the macroeconomy, but also labor market variables, often in opposite than expected ways, see, for instance, Den Haan et al. 2018; Esteban-Pretel and Kitao 2021; McKay and Reis 2016). We cover all five models in Figures 3.1-3.2 and focus on those instances when responses are visually different for different monetary policy environments. The focus at this stage is on visual differences in macroeconomic reactions, with the formal test being introduced later; we relegate graphs with confidence intervals to Online Appendix D.1.

### 3.1 Before and After The Euro

As covered in the literature section, recent theoretical contributions predict that labor policies interact with the business cycle conditions, we conjecture that independent monetary policy may have yielded different outcomes compared to the monetary union member. This is because one policy tool for all member-states is likely to fail to address individual country needs, absent perfect business cycle synchronization. In Figures 3.1-3.2, monetary policy is modeled by introducing an indicator function that changes a slope parameter for labor market policies, depending on the positive or negative change in the short interest rate, for the two periods before (left graph) and after (right graph) the introduction of the euro (1985-1998 and 1999-2010, respectively). Firstly, we focus on Model A results. As for the replacement rate, after an introduction of euro, reactions become strikingly similar, independent of the monetary policy stance. Before 1999, changes in the replacement rate triggered quite diverse responses. Notably, however, replacement rate results in substantially larger effect (note scale differences in graphs) for the period of the monetary union. This difference is not so clear for other policy variables. Therefore, unemployment benefits are more effective and do not depend as much on the cyclical component.

In a more recent period, changes in the generosity of unemployment benefits are more deflationary, at least on impact and over the long run (they are inflationary in the medium run). This, at first surprising impact, aligns well with the recent evidence due to Krusell et al. (2010). They accommodate incomplete asset markets and endogenous labour-market frictions to show that unemployment benefits, even though increased wages and enhanced insurance, lead to lower firm entry, fewer jobs, and lower capital demand. The welfare might suffer due to higher benefits and the unemployed consumer, while benefiting in the short run, might suffer in the future from having a lower wage when employed. Exploring the importance of incomplete markets, Den Haan et al. (2018) further show that individuals facing fears of future unemployment tend to adopt precautionary saving behaviours in economic downturns. Their result highlights that even a small rise in money demand, fueled by the desire to hold money as a form of precautionary savings, can exert deflationary pressure on the economy.

In line with the incomplete market findings, we find that unemployment benefits tend to reduce real GDP (a recessionary, welfare-reducing effect), increase REER on impact with a dying-out effect and increase unemployment. The long-run effect on unemployment is less pronounced for loose monetary policy. It is likely that the credit channel is at work: an increase in unemployment benefits makes the employees' outside option more attractive, which places pressure on prices. In a cheap lending environment, unemployment rate increases to a lesser extent than in a costly credit environment.

Expenditure on active labor market policies (ALMP) is more deflationary in the euro area than before euro, with real GDP reducing and unemployment increasing patterns, in contrast to pre-euro times. This can, at least to a certain extent, contribute to the literature that fails to find positive effects of ALMP, especially at the aggregate level. This result can be partially attributed to the monetary policy stance, something yet to be confirmed by testing. Changes in the real effective exchange rate (REER) are also remarkably different for the two periods. The recent literature (e.g., Abbritti and Weber, 2018; Benda et al., 2020; Gehrke and Weber, 2018; Ulku and Georgieva, 2022) has moved to emphasizing the importance of institutional setup by considering other policies in place and development level for the impact of ALMP (for the meta analysis refer to Vooren et al. (2019) showing that ALMP might have different impacts in the short and long run; it is often negative for the former and positive for the latter).

Finally, an increase in the employment protection legislation index (EPL) – leading to more rigid labor market – yields substantially more inflationary pressures after 1999, especially if implemented during times of tightening monetary policy. Interestingly, higher rigidity happened to be associated with higher unemployment (over the longer run) before the euro introduction, but it can be unemployment-reducing if implemented at the same time as the tightening monetary policy, after the euro has been introduced. Hence, faced with higher borrowing costs and unable to adjust their workforce easily, firms may pass these increased costs onto consumers in the form of higher prices, leading to inflationary pressures, but no unemployment in the short run.

We summarize our key findings in the following takeaways:

**Takeaway 1. Unemployment Benefits** Post-euro, unemployment benefits have a more deflationary effect initially but can be inflationary in the medium term. Enhanced unemployment benefits can lead to lower firm entry, fewer jobs, and lower capital demand due to incomplete asset markets and labor-market frictions, among other factors.

**Takeaway 2. Active Labor Market Policies (ALMP)** ALMP expenditure has become more deflationary in the euro area, contrasting with pre-euro times, also demonstrating a shift in the impacts of ALMP on real GDP and unemployment.

**Takeaway 3. Employment Protection Legislation (EPL)** Post-1999, increased EPL, especially during monetary tightening, leads to more inflationary pressures. While EPL was associated with higher unemployment before the euro introduction, it can potentially keep unemployment intact or even lower it when implemented in conjunction with tightening monetary policy after the euro's introduction.

Other variants regarding monetary policy include an average (annual) change in the short interest rate, entering as an indicator function (Model B), a quarterly and four-quarter changes, entering multiplicatively (Models C and D), and the model that combines indicator function and multiplicative effects (Model E). We report results for the cases when changes in short interest rates have corresponded to the first and third historical quartiles (in particular, they refer to -0.40 and 0.23 p.p. and -1.35 and 0.66 p.p. for quarterly and annual frequencies, respectively) for Models C, D and E.[23] Though dynamic responses are quite comparable, there are notable differences, too. An increase in EPL during the pre-euro time yields considerably different reactions in Model B compared to the model A. For instance, EPL is deflationary and makes REER appreciate in Model B, unlike Model A. Model D, for example, delivers inflationary effects of replacement rates for both loosening and tightening monetary policies as well as longer-run reductions in unemployment before the euro. Model E provides evidence for expansionary and contractionary effects of ALMP, depending on the monetary policy stance, before 1999. These differences illustrate uncertainty about the true effects of labor market policies on the macroeconomy, and thus necessitate a framework which takes all of that into account. Before addressing these concerns, we shall first summarize the key takeaway and, next, discuss results for the combined sample.

---

[23]The interest rate below which monetary policy is accommodative or tight is unobserved. Our approach places no parametric, long-run value or filtering restrictions, except for the choice of a particular historical quantile. For

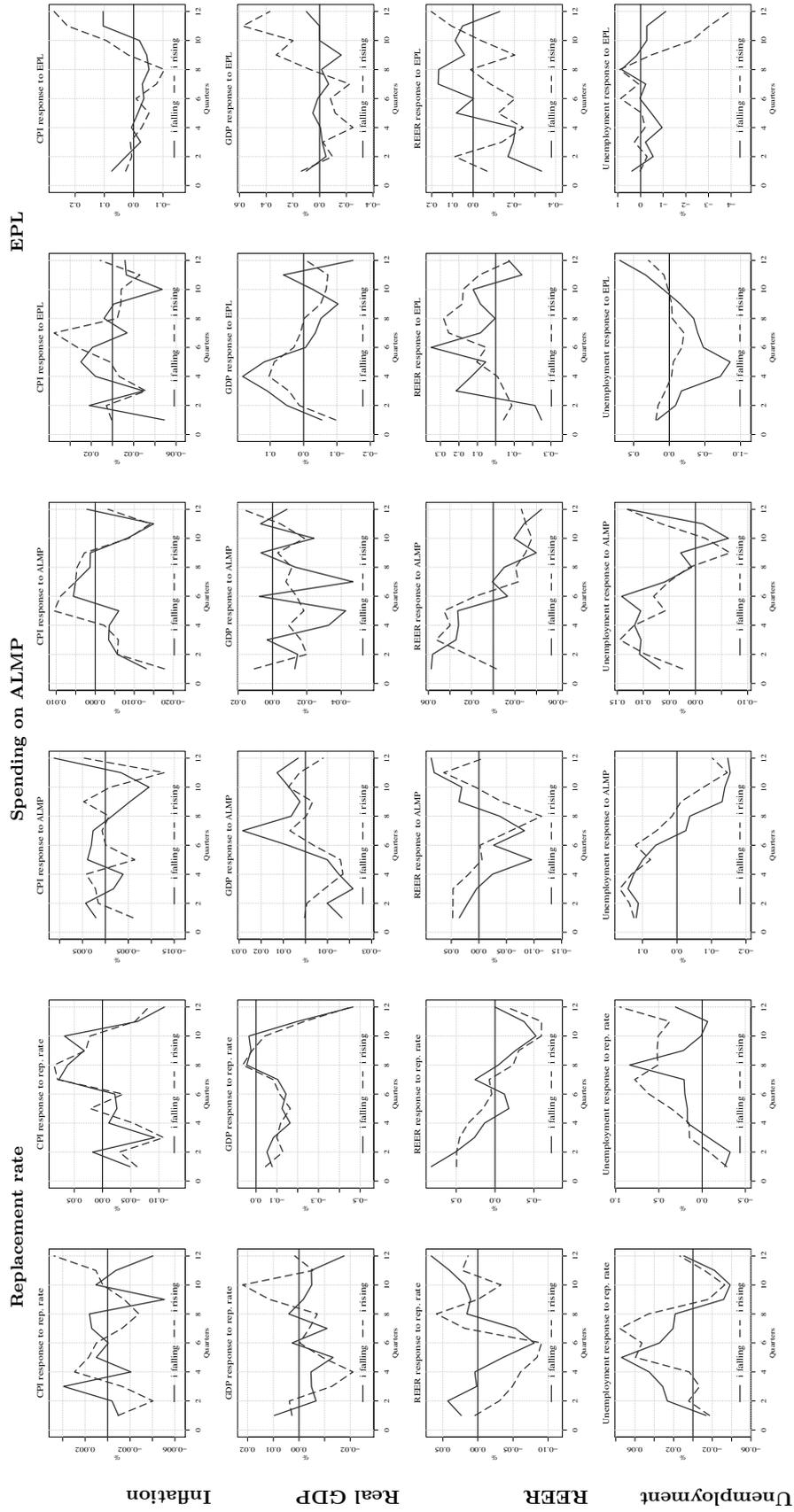

Figure 3.1: Changes in macroeconomic variables due to the 1% increase in labor market policies (replacement rate, ALMP and EPL), 1985-1998 (left) and 1999-2010 (right), model A

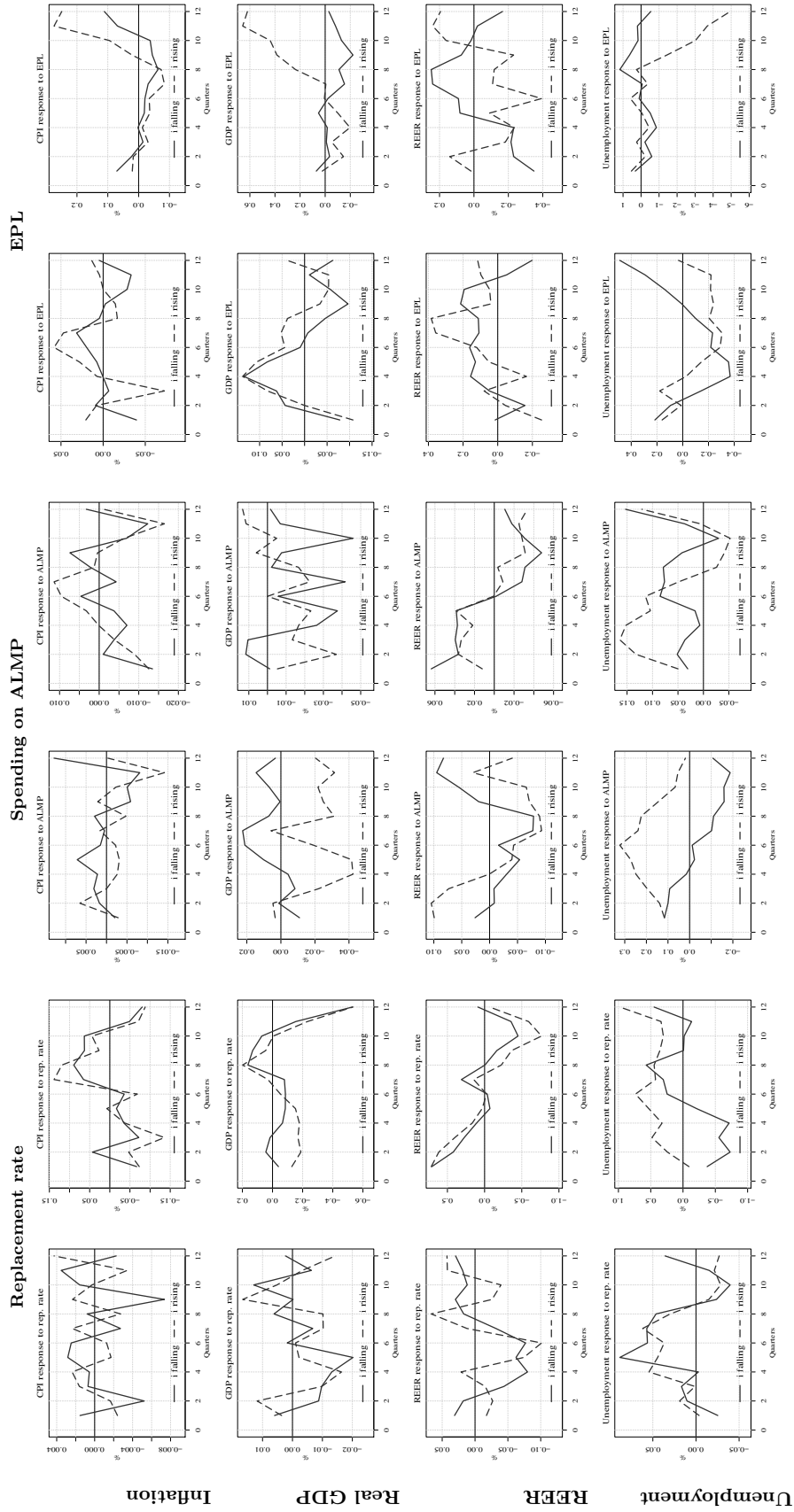

Figure 3.2: Changes in macroeconomic variables due to the 1% increase in labor market policies (replacement rate, ALMP and EPL), 1985-1998 (left) and 1999-2010 (right), model B

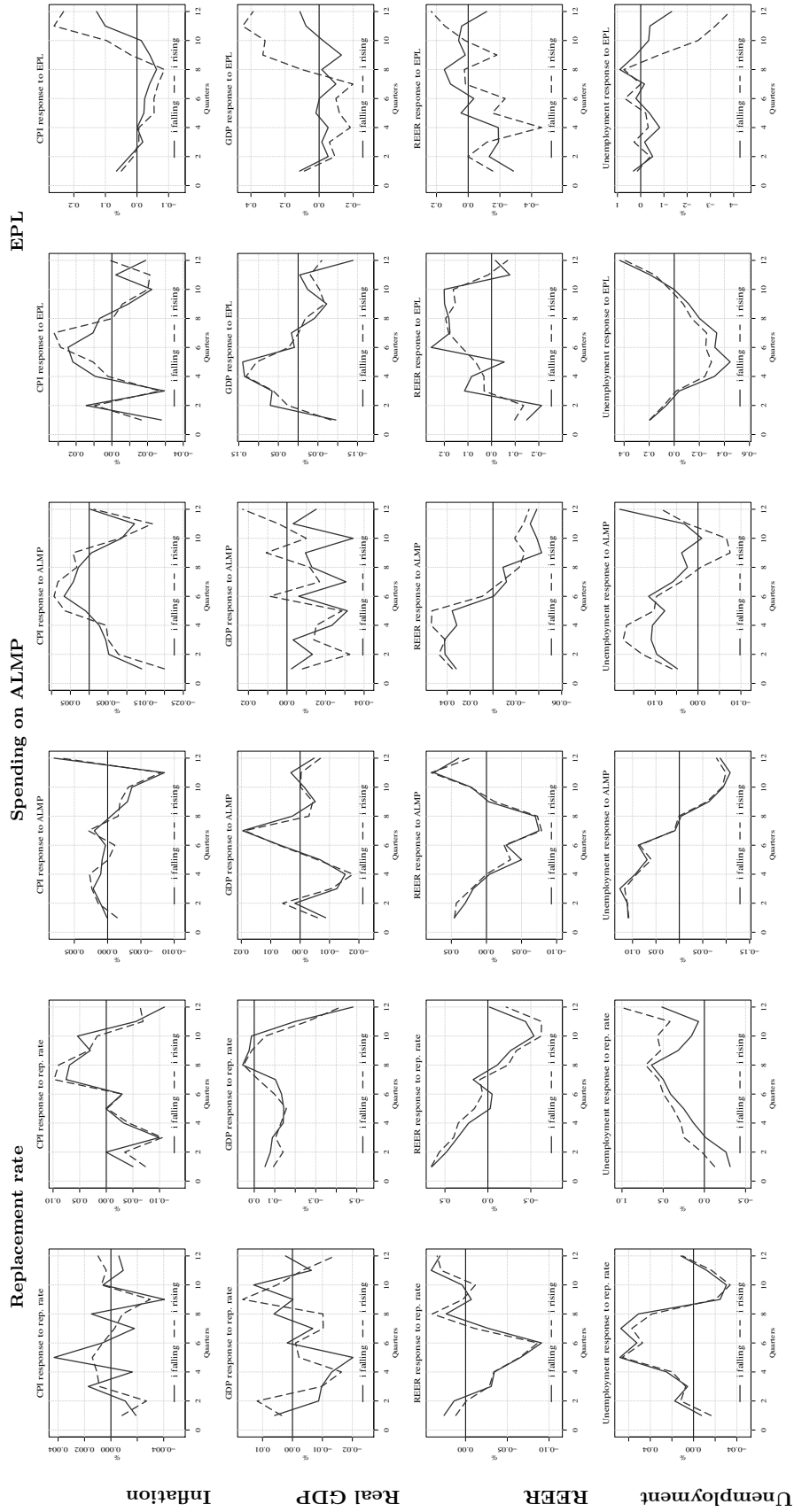

Figure 3.3: Changes in macroeconomic variables due to the 1% increase in labor market policies (replacement rate, ALMP and EPL), 1985-1998 (left) and 1999-2010 (right), model C

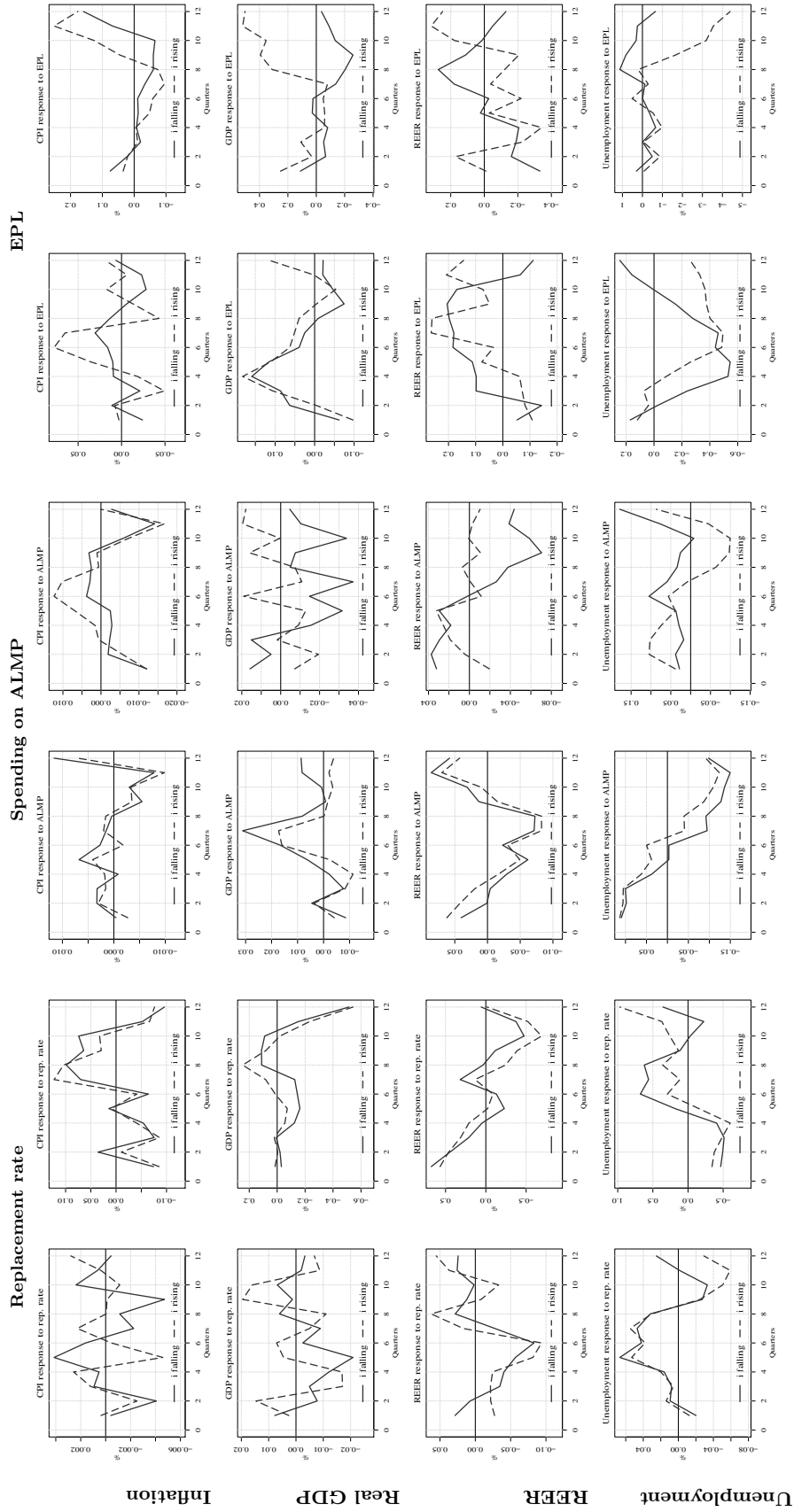

Figure 3.4: Changes in macroeconomic variables due to the 1% increase in labor market policies (replacement rate, ALMP and EPL), 1985-1998 (left) and 1999-2010 (right), model D

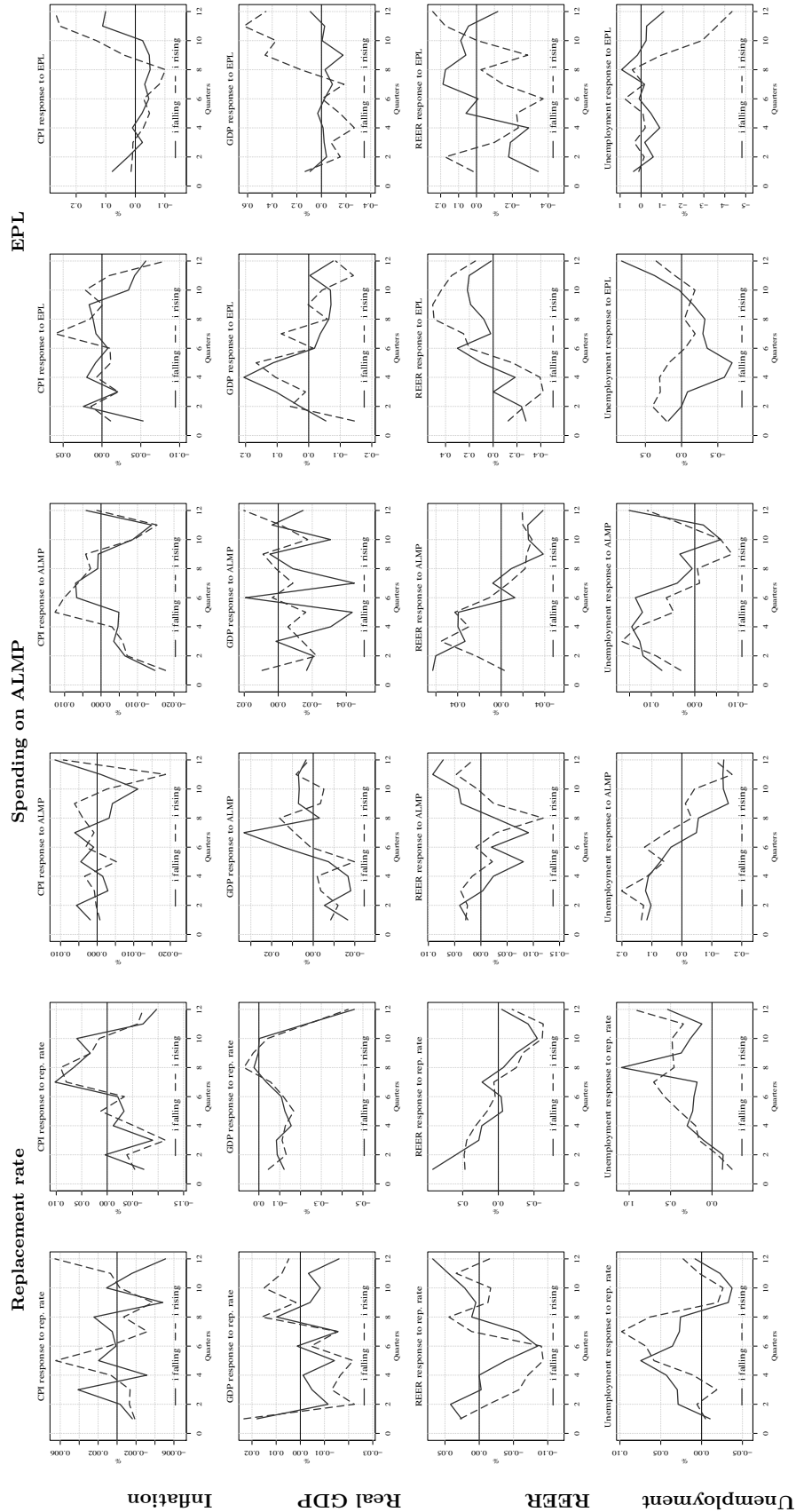

Figure 3.5: Changes in macroeconomic variables due to the 1% increase in labor market policies (replacement rate, ALMP and EPL), 1985-1998 (left) and 1999-2010 (right), model E

**Takeaway 4. Model Uncertainty** Each model (A-E) yields distinct insights into the macroeconomic impact of labor market policies under different monetary conditions. To draw robust inference about monetary policy's role in shaping labor market impacts on the macroeconomy, we require a robust average model.

## 3.2 Full Sample

Figures 3.6-3.7 report responses for the entire sample period, 1985-2010. Model A tells that the replacement rate is deflationary on impact, tends to be contractionary, is associated with the appreciated REER and a rise in unemployment (unless implemented under loosening monetary policy). ALMP is also deflationary on impact, and surprisingly contractionary with reductions in unemployment being realized over the medium run only (again, implementation during loosening monetary policy times receives more support to fight unemployment). The latter observation can be rationalized by the demand-side arguments: lowering interest rates usually coincides with insufficient demand and poor expectations, then expenditure shock (e.g. in terms of wage subsidies) would be more efficient than in the environment where the economy may be overheated. This result is important in explaining an observed failure in the literature to find a robust effect of expenditure on activation schemes. Business cycle conditions and monetary policy stance seem to be causing heterogeneity in outcome, at least over the medium run. The falling interest rate environment does not only attenuate an unemployment-increasing effect of the replacement rate but is also conducive to unemployment-reducing effect of ALMP (though ALMP is not modeled directly, these results can be viewed from the lenses of a simulation of fiscal interventions and accommodative monetary policy in Coenen et al., 2012).

EPL (an increase in rigidity), on the contrary, is more inflationary, and also contractionary (unless implemented during a tightening monetary policy), with a competitiveness gain on impact and an increase in international prices (REER) being realized over time, and unemployment first plummeting and then starting to increase. This may capture the fact that firing becomes more difficult but the adverse effect on the intensity of hiring kicks in with a lag only.

As before, though responses are quite comparable across models, there are also non-negligible differences, such as EPL being contractionary for loosening and expansionary for tightening monetary policies after the euro in Models A, B and E, but merely contractionary or merely expansionary in models C and D, respectively. Model C seems to deliver small differences for different monetary policy regimes, unlike other models. This hints that, economically if not statistically, quarterly changes are not as important as a sign of a change (Model A and B), which captures a monetary policy trend. Model D is also interacted with the annual change, thus providing some support to the monetary policy trend interpretation. Of course, the combination of two time intervals helps with some uncertainties by making use of more degrees of freedom (at a cost of homogeneous responses), but it does not resolve uncertainty in macroeconomic reactions fully, an undertaking that we pursue next.

**Takeaway 5. Labor Market Policies and the Macroeconomy** Models A-E demonstrate that replacement rates exhibit deflationary effects and ALMPs are initially deflationary but can reduce unemployment over the medium term, especially under loose monetary policies. EPL increases inflation and unemployment, though the effect depends on the model choice and the stance of

---

alternatives and references to compute a natural interest rate refer to Holston et al. (2017).

monetary policy, both influencing the macroeconomic impact of labor market policies.

## 4 Least-Squares Model Averaging

As shown above, different modeling strategies yielded a number of non-negligible differences, especially for policy purposes, and the monetary policy environment throughout which labor policies get implemented happened to be an important factor in picking up effects on the macroeconomy. Unfortunately, economic theory can hardly narrow down potential effects of the monetary policy environment on the efficacy of labor market policies. We thus develop an empirical framework to uncover a dynamic path of policy effects which admits a number of different channels for the labor and monetary policies to interact.

Let us rewrite the model in (C.1) as

$$y_{t+k}(m) = z_t(m)' a(m) + u_{t+k}(m), \tag{4.1}$$

where all variables are combined into

$$z_t(m) = (1, X_t, X_{t-1}, \triangle \ln LMP_t, \triangle \ln LMP_{t-1}, g(\triangle \ln LMP_t(m)), \ldots)', \tag{4.2}$$

and $m$ denotes one of the models (A-E) that are covered in specifications (2.1). To pin down the optimal averaging, we apply the Mallow's criterion for the weight of each model, given by

$$C_T(w) = \frac{1}{T} \sum_{t=1}^{T} \left( \sum_{m=1}^{5} w(m) \hat{u}_t(m) \right)^2 + \frac{2\hat{\sigma}_T^2}{T} \sum_{m=1}^{5} w(m) \dim(z_t(m)), \tag{4.3}$$

such that $\hat{w} = \arg \min C_T(w)$, $0 \leq w(m) \leq 1$, $\sum_m w(m) = 1$, with the average local projection given by $\hat{y}_{t+k}(\hat{w})$ for each horizon (with $T$ depending on the exact horizon considered). It is clear that the criterion requires one to estimate $\hat{\sigma}_T^2$, residual variance estimate, which can be done, following Hansen (2008), by $\hat{\sigma}_T^2 = (T - \dim(z_t(M)))^{-1} \sum \hat{u}_t^2(M)$, where $M$ stands for the largest fitted model. Hansen (2008) provides a simple proof to demonstrate that the expectation of the Mallow's criterion is asymptotically unbiased for stationary dependent data (Hansen (2007) established this result for IID observations). The main difference between our approach and the methodology in Cheng and Hansen is that we apply Mallow's criterion for each horizon, thus making it possible for a dynamic response to be determined flexibly at each point in time, with different weights assigned to different models. In the robustness checks, we will also admit a factor structure *unique* to each time horizon. Unlike Cheng and Hansen (2015), therefore, our approach does not suffer from the fixed (unknown) number of factors issue.[24]

**Proposition 1.** *Impulse response function, measured by local projection and weighted by Mallow's weights $C_T(w)$, is asymptotically unbiased in the sense that such averaging delivers asymptotically unbiased estimator of the mean squared (forecast) error in the presence of stationary, dependent data, and even a multi-factor error structure.*

*Proof.* The result follows by combining Theorem 1 in Hansen (2008) and Theorem 1 in Cheng and Hansen (2015) (along with the assumptions R and F, for the case without and with multi-factor error structure, respectively) with the definition of the local projection as proposed by Jordà (equation 2, 2005). Local projections are conducted at each horizon separately, and the proof follows from the theory developed in Hansen to one-step forecasts applied separately for each horizon.[25] As for the factor, given $N, T \to \infty$, the estimated factor (estimated principal component) issue is negligible and the Mallow's weights minimize mean squared (forecast) error. For a full discussion and implementation, refer to Online Appendix. □

---

[24]Cross-validation techniques have also been developed for further robustness issues. Following Hansen and Racine (2012), one can introduce a leave-1-out cross-validation criterion $CV_1(w) = \frac{1}{T} \sum_{t=1}^{T} \left( \sum_{m=1}^{5} w(m) \hat{u}_{t,k}(m) \right)^2$, where $\hat{u}_{t,k}$ is the residual obtained by the least-squares omitting one observation $T$ times, or extending this approach to the leave-$k$-out cross-validation criterion. As demonstrated analytically by Hansen (2010), this is a robust criterion once heteroskedasticity and serial correlation are severe. As our approach is direct for each horizon, and we focus on local effects with a large number of considered angles (split in time, horizon-specific error factor structure, results for

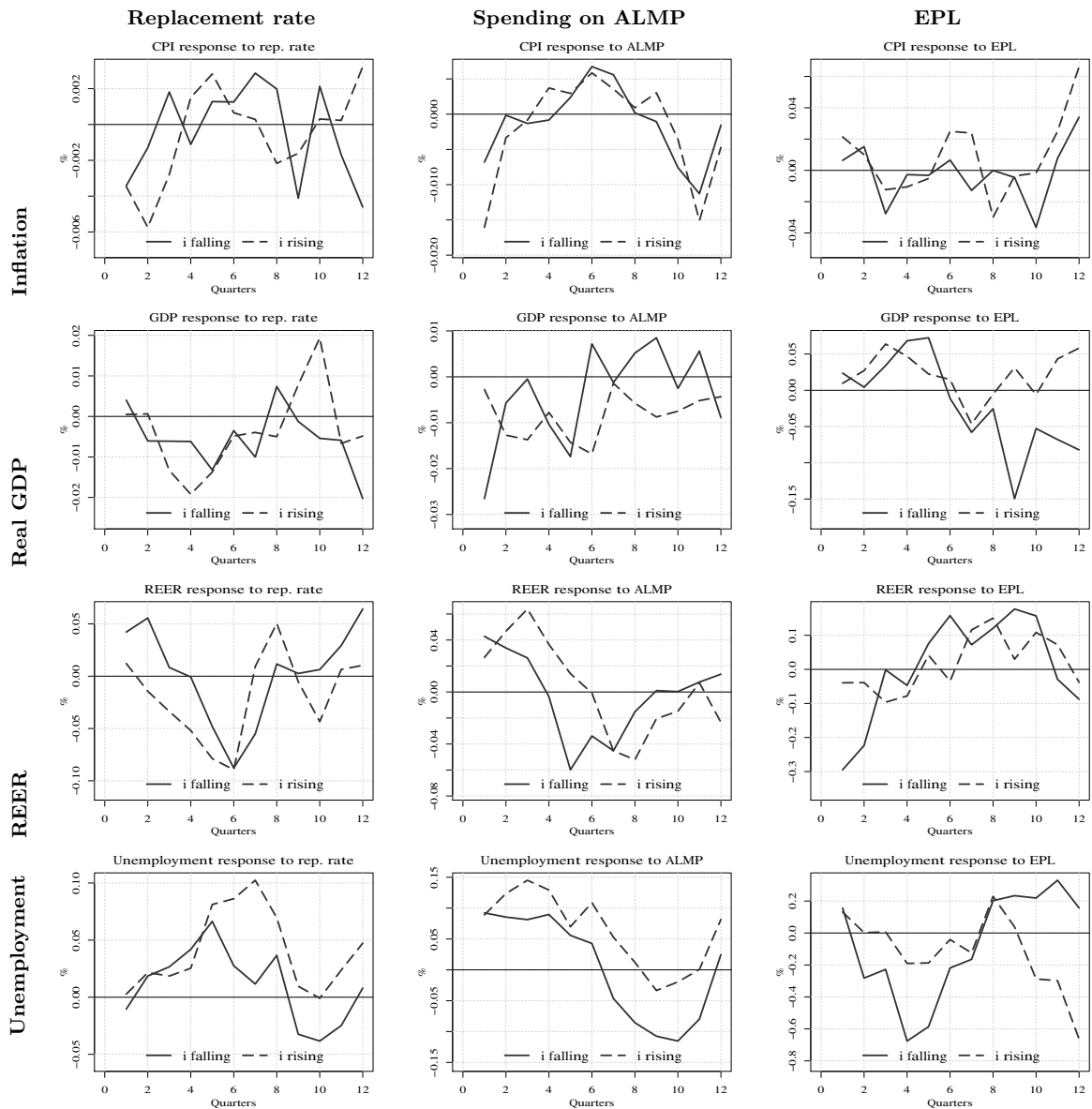

Figure 3.6: Changes in macroeconomic variables due to the 1% increase in labor market policies (replacement rate, ALMP and EPL), 1985-2010, model A

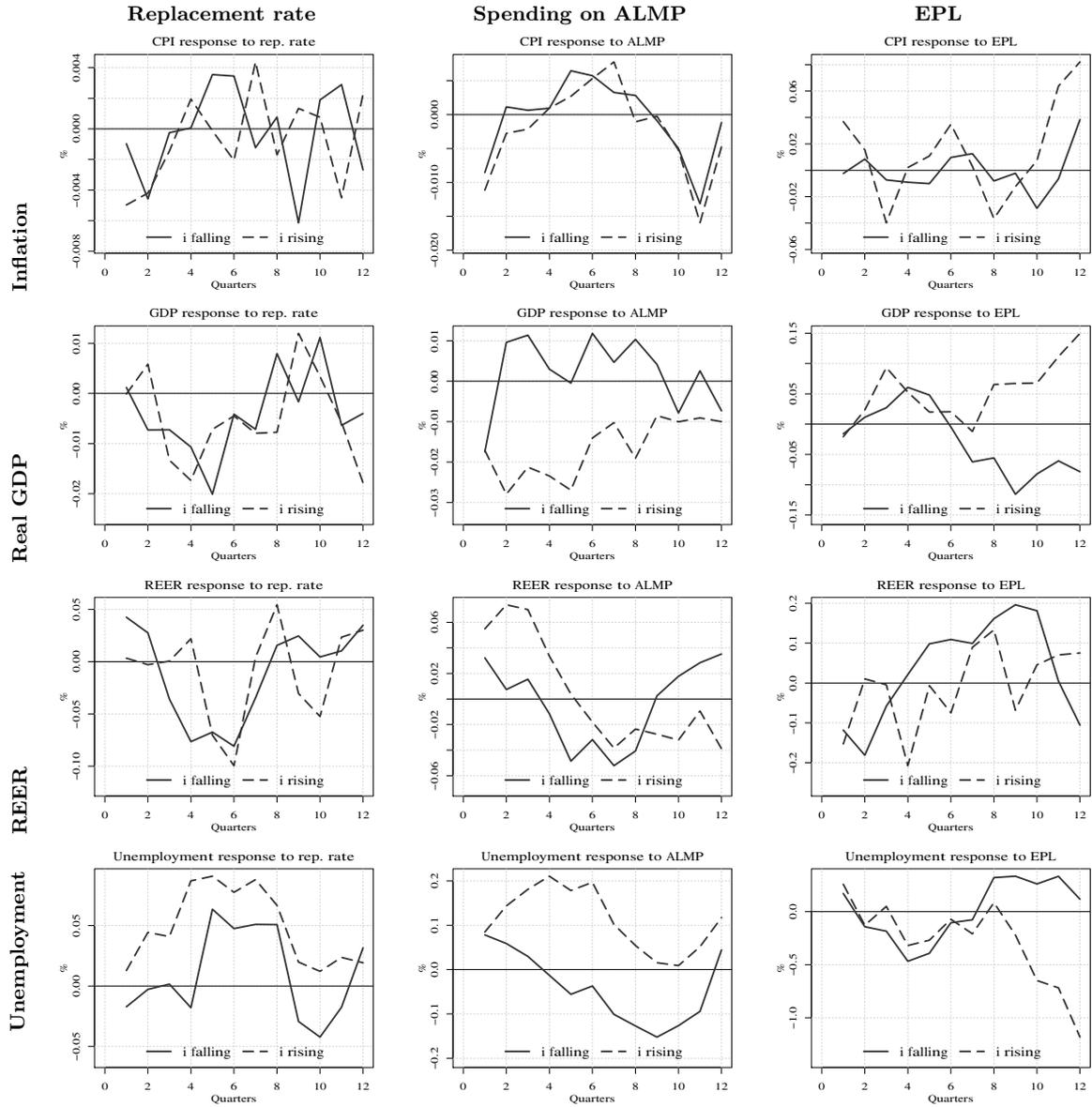

Figure 3.7: Changes in macroeconomic variables due to the 1% increase in labor market policies (replacement rate, ALMP and EPL), 1985-2010, model B

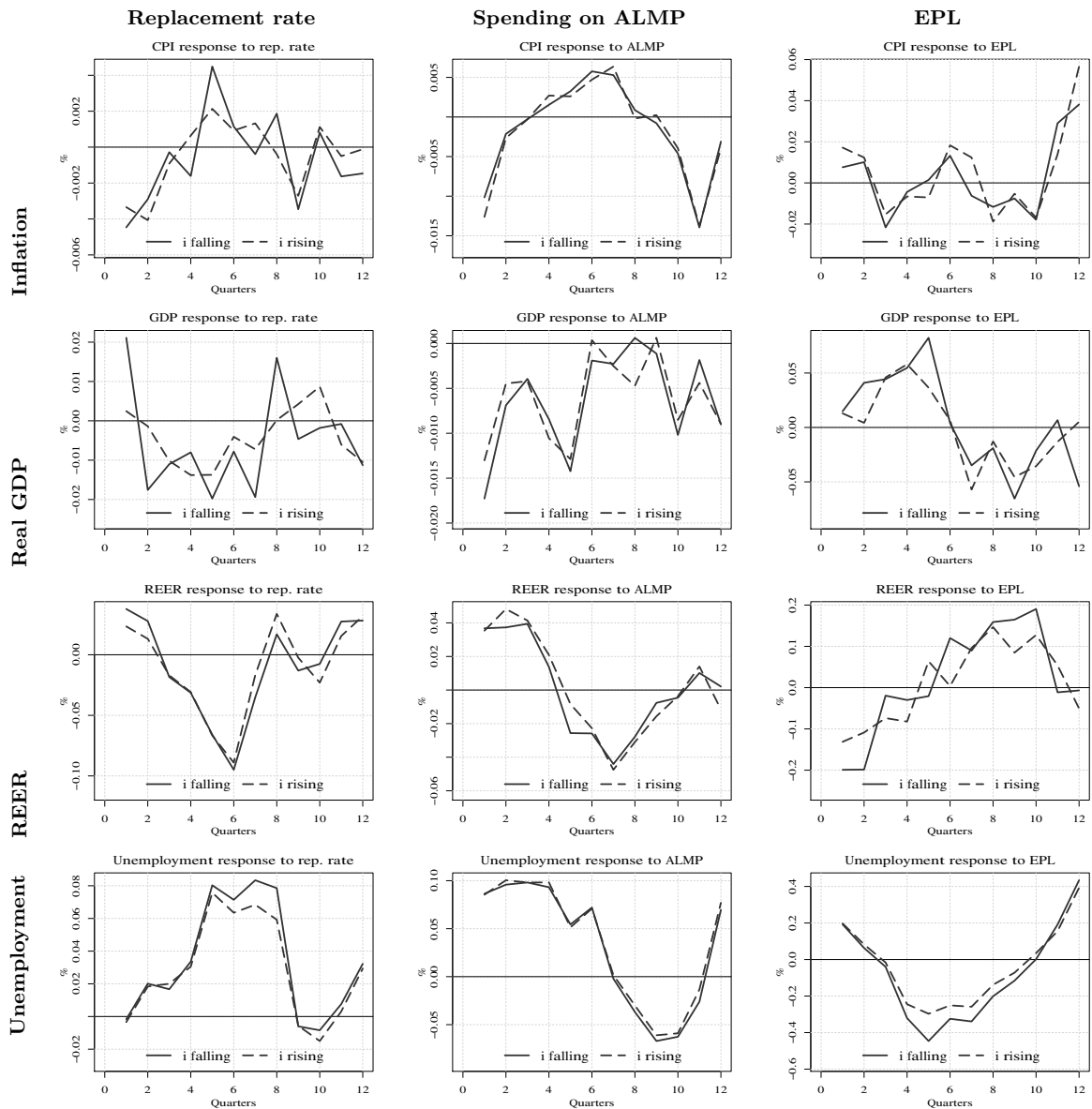

Figure 3.8: Changes in macroeconomic variables due to the 1% increase in labor market policies (replacement rate, ALMP and EPL), 1985-2010, model C

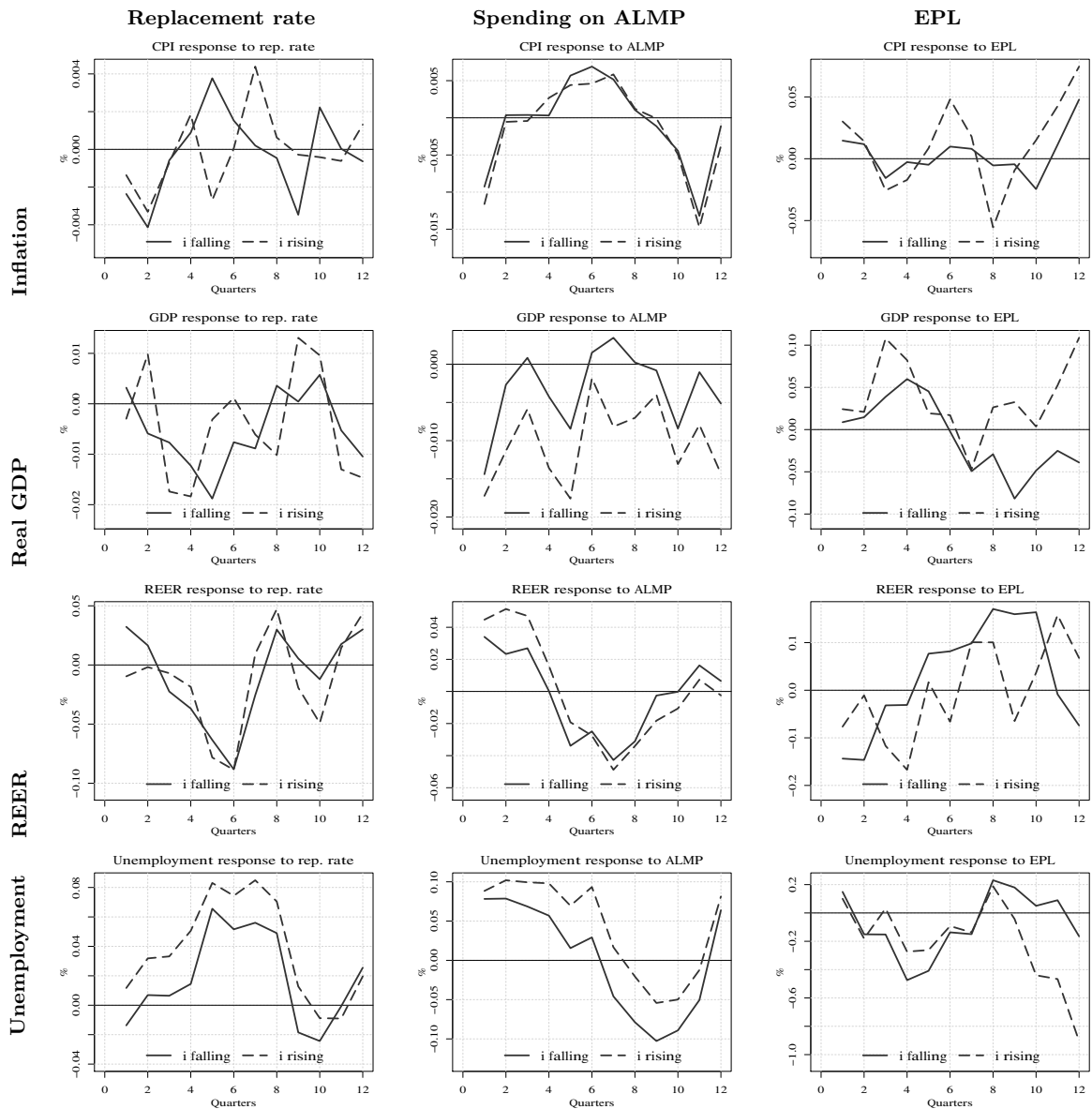

Figure 3.9: Changes in macroeconomic variables due to the 1% increase in labor market policies (replacement rate, ALMP and EPL), 1985-2010, model D

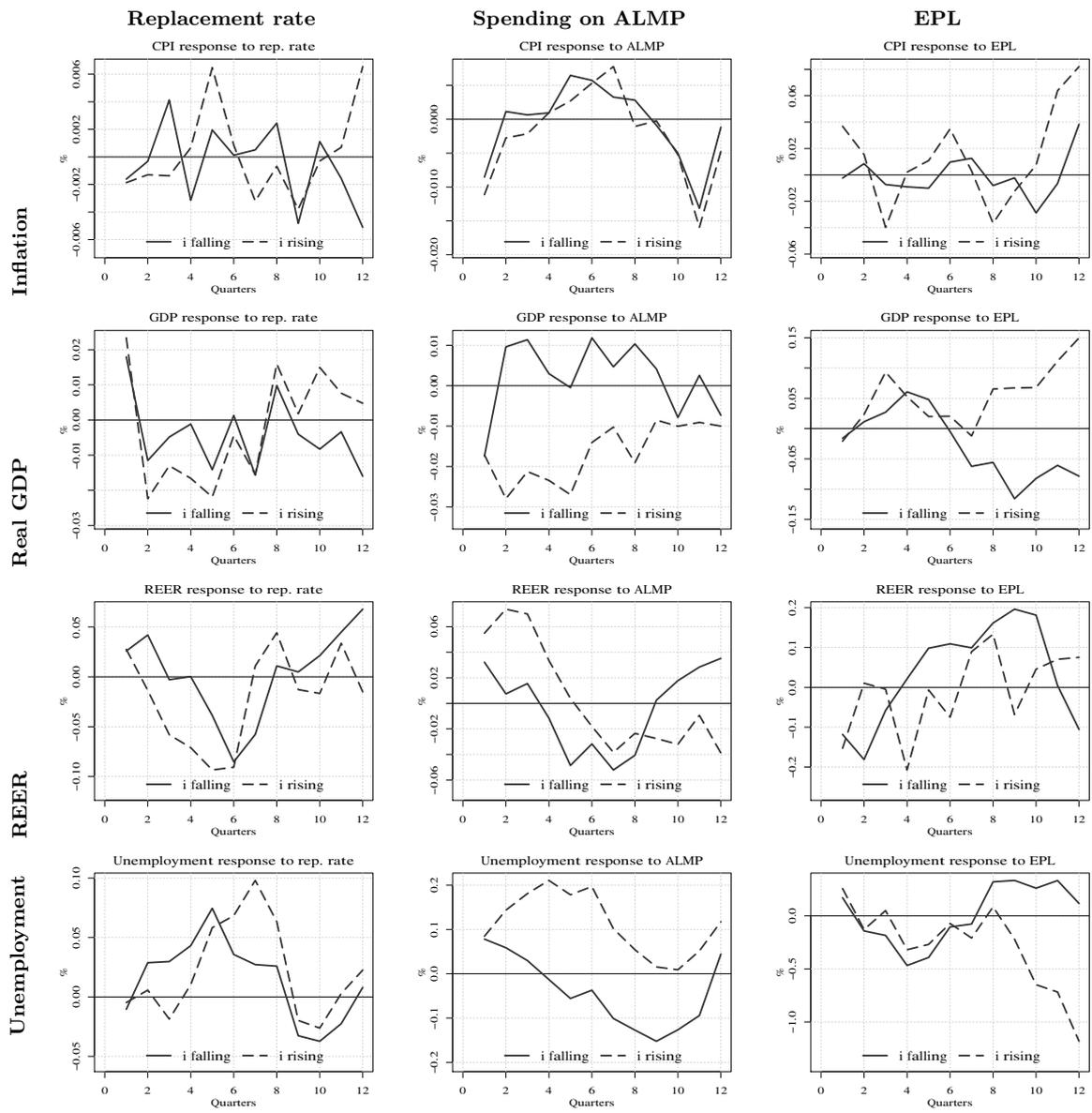

Figure 3.10: Changes in macroeconomic variables due to the 1% increase in labor market policies (replacement rate, ALMP and EPL), 1985-2010, model E

Asymptotic unbiasedness does not guarantee good behavior in small samples, nor does it establish asymptotic efficiency. For the former, however, Monte-Carlo small-sample evidence in Hansen (2008, 2010) is encouraging for one and multi-step forecasts; as for the latter, Cheng and Hansen (2015); Hansen (2008) conjecture that the result in Ing and Wei (2005) where Mallow's criterion leads to asymptotic optimality given autoregressive process with IID errors may be extended to more general cases. We leave a fully fledged simulation exercise for future research, but our framework seems to build on encouraging grounds, established in the literature.

## 4.1 Weights

We move to reporting weights, calculated following criterion (4.3). Figure 4.1 is particularly revealing: weights to combine impulse responses vary substantially across macro variables, policy variables and horizons. Our methodology easily admits this heterogeneity. For instance, model D is particularly well suited for modeling responses in inflation and unemployment (with model B being its major rival) but totally marginal when it comes to the real effective exchange rate where the choice of models is quite erratic across horizons and labor market policies. Interestingly, data prefer delayed or averaged effects of monetary policy with the most recent changes carrying lower weight. We combine those flexible weights, reported in Figure (4.3), and move to the averaged responses.

## 4.2 Average Local Projections

Since economic agents might respond only to the fact that the policy rate changed or rate's change or longer rates or interactions, we proceed to the average model that captures all these potential channels. A combination of the local projections is visualized in Figure 4.2. Both the replacement rate and the ALMP reduce inflation on impact, as opposed to the employment protection. Prices do not seem to depend on the monetary policy environment in any fundamental way, except for the EPL, when tightening monetary policy delivers more inflationary result. Real GDP increases with the replacement rate on impact if the monetary policy was tightening but, independent of the monetary policy stance, the ultimate effect is contractionary. The aggregate economy (real GDP) reacts more positively to changes in the ALMP if expenditure is increased when a loosening monetary policy is in place. An increase in labor market rigidity can be supported if implemented during a tightening monetary policy as it is associated with an increase in real GDP over the longer run. Similarly to inflation, the real effective exchange rate goes up on impact and depreciation takes time to realize for the replacement rate and ALMP, unlike EPL, which depreciates on impact and mimics a J curve over time. Long-run effects differ depending on the monetary policy stance once labor policy has been initiated.

The replacement rate seems to be increasing unemployment, as predicted by the standard economic theory, but the negative effect fades out if an increase in unemployment benefits is implemented in the loosening monetary policy environment. Similarly, ALMP seems to help to reduce

---

each response variable and each policy variable), we leave these computational exercises for future explorations on robustness and optimality.

[25]Notice that projection is over observed variables, unlike, for instance, practices in recursive multi-step forecasting when predicted values are used to construct future values. As noted by Chevillon (2007), model mis-specification (in particular, mis-specified unit roots, neglected serial correlation and location shifts) is better handled when different forecasting models are used for each forecast horizon (the so-called direct multi-step forecasting). Impulse responses are nothing but functions of multi-step forecasts.

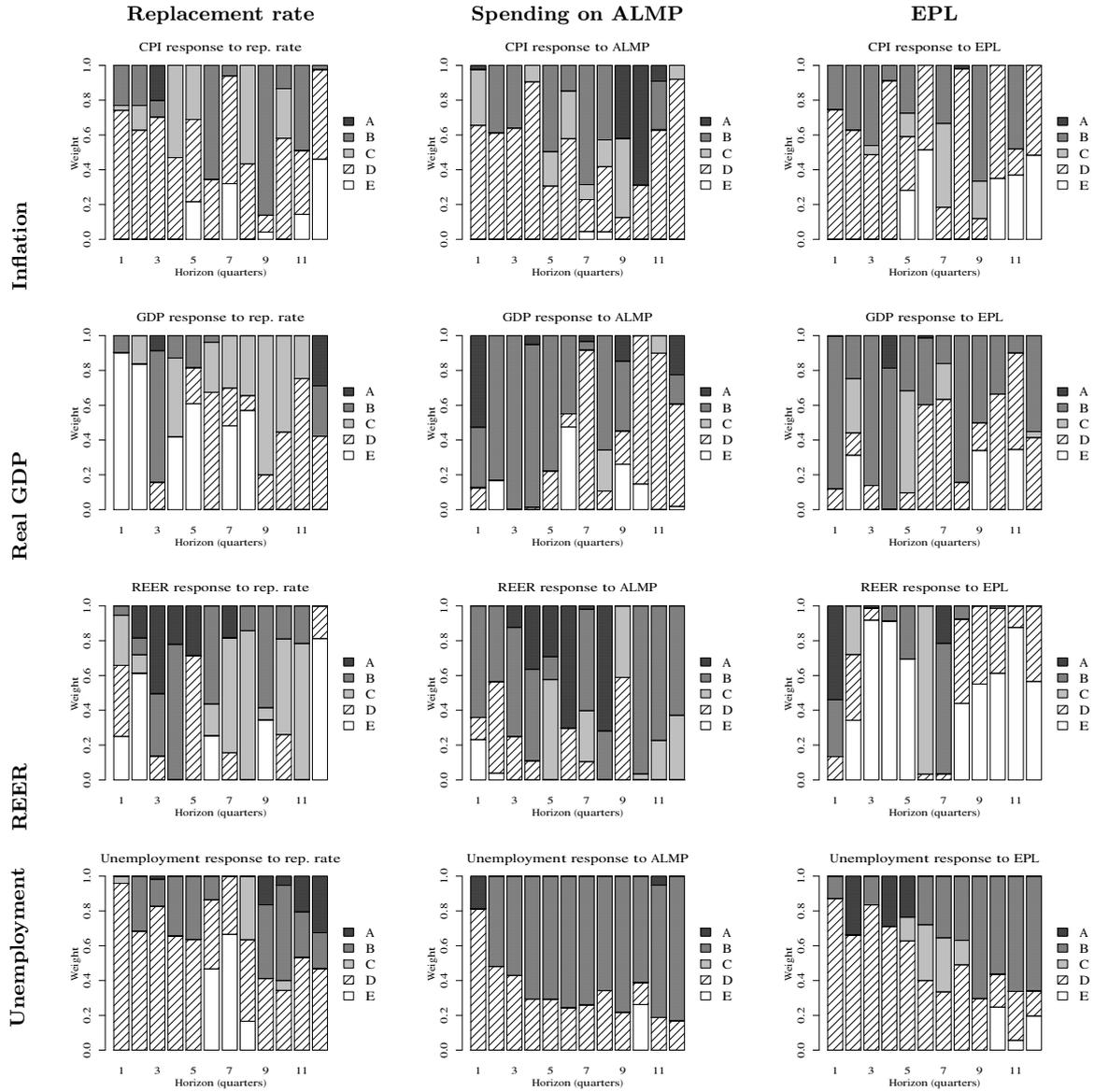

Figure 4.1: Average model weights for each horizon, impulse and response variables (Mallow's criterion)

unemployment if monetary policy is loose. This effect may also shed light on why the previously quoted empirical literature fails to find a desirable aggregate effect of ALMP when it does not differentiate between other events, such as the monetary policy regime. Finally, employment protection does not deliver an increase in unemployment if initiated simultaneously with a tightening monetary policy. An increase in replacement rate and ALMP requires financial inputs, whereas EPL is usually conducted by changing law and financial requirement, if any, is experienced only later. It thus seems that expenditure-increasing policies work during an economic downturn when loose monetary policy is in place whereas institutional changes are favored during economic booms, in line with the current, IMF or ECB, suggestions to use fiscal expansions during recessions and conduct labor market reforms during (and thus capitalize on) good times (IMF, 2018, ECB, 2017).

We summarize our key findings by these takeaways:

**Takeaway 6. Impact on Inflation and Real GDP** The replacement rate and ALMP initially reduce inflation, contrasting with EPL, which can be more inflationary under tightening monetary policy. While replacement rates can boost real GDP in the short term under tightening monetary policy, both replacement rates and ALMP tend to have a contractionary effect on the economy in the longer run.

**Takeaway 7. Monetary Policy Environment** The effectiveness of labor market policies, including their impact on inflation, real GDP, and unemployment, varies significantly with the prevailing monetary policy stance. Loosening monetary policy enhances the unemployment-reducing effects of ALMP and the replacement rate, whereas tightening aligns with the efficacy of EPL in boosting real GDP over time.

**Takeaway 8. Policy Implementation Timing** Expenditure-increasing policies like ALMP and adjustments in the replacement rate are more effective during economic downturns with loose monetary policy. In contrast, institutional changes like EPL are preferred during economic upturns.

### 4.3 Test on the Importance of Monetary Policy

Though differences in reactions are captured visually, we also devise a statistical test to evaluate whether macroeconomic responses to labor market policies react to monetary policy or not. We test whether $H_0 : \tilde{\delta}_k - \delta_k = 0$ against a two-sided alternative. Though the test can distinguish whether the Models A-E add *additional information* about labor market effects on macroeconomic aggregates, it does not tell us whether the benchmark model is correct or not. With that caveat in mind, we report a fraction of occasions when t-test has been accepted for the Mallow's averaged model in Table 4.1, where the chosen p-value was 0.1 (so we report a fraction for which p-value are larger than 0.1). For instance, the ALMP parameter in the equation of the real GDP cannot be differentiated from the baseline model when none of the extensions in Models A-E are included for the first period after a shock but only 31% of the cases are registered for the second period. The test is horizon-specific and distinguishes between informational advantages of an inclusion of monetary policy variables for the *entire* dynamic path of responses.

#### 4.3.1 Horizon-specific Results

To ease reading of the results, we also depict proportions for each horizon (quarter) graphically in Figure 4.3. If the parameters were identical to the specification with no monetary policy effects, we would expect the proportion to be close to 1 (or if, say, 10 times out of 100 equality occurs by chance, the no-effect conclusion, or acceptance of the null hypothesis, would be made for the cases when proportion is equal to or larger than 0.9). Many cases fall below 0.9 threshold, indicating important interactions and channels that make labor markets affect the macroeconomy differently, depending on the monetary policy stance. In particular, ALMP interactions with monetary policy are crucial for real GDP and unemployment whereas changes in unemployment benefits deliver different results on international prices (REER) and domestic inflation, depending on the monetary policy stance.

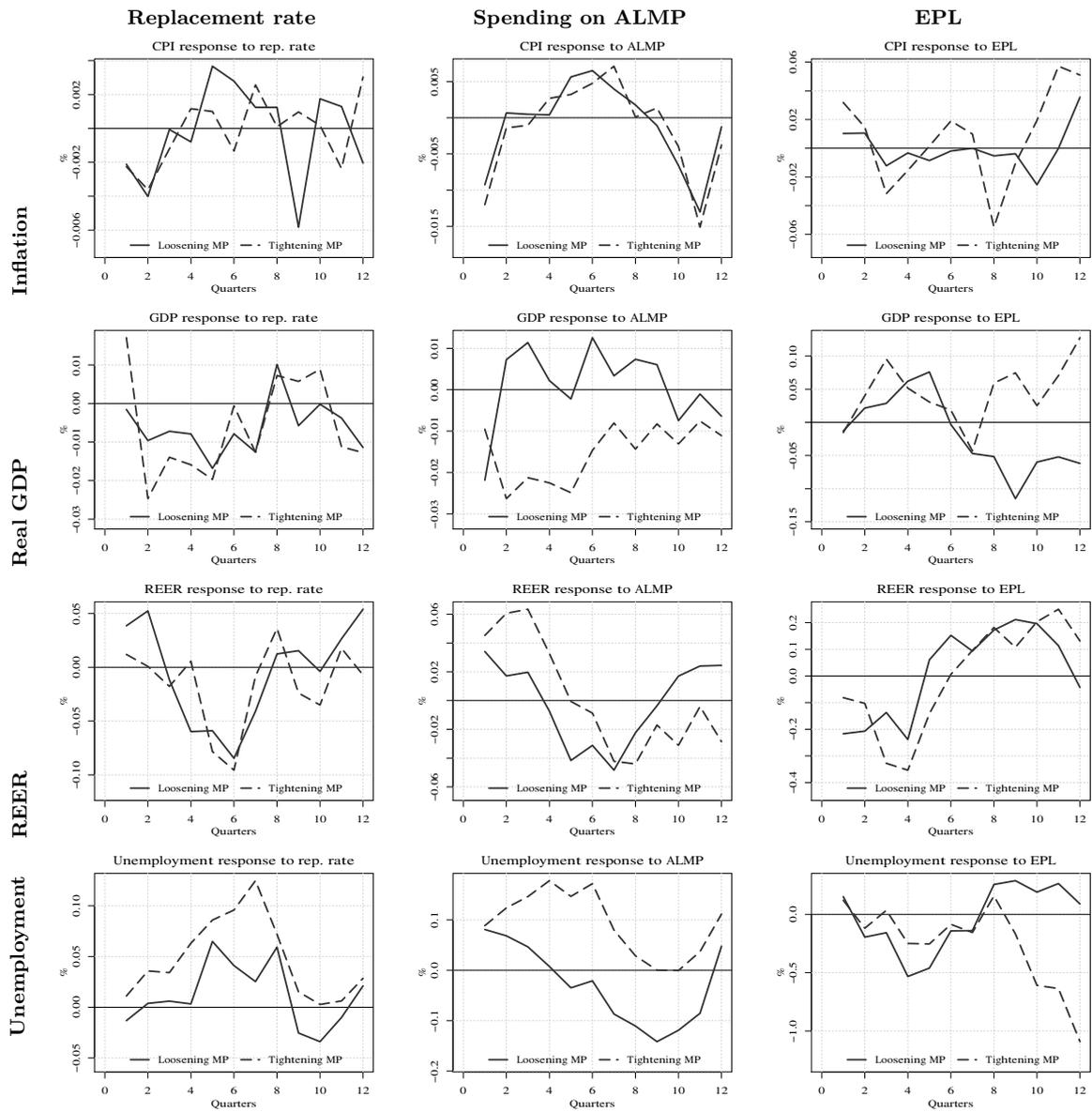

Figure 4.2: Changes in macroeconomic variables due to the 1% increase in labor market policies (replacement rate, ALMP and EPL), weighted (Mallow's averaging) impulse responses (local projections)

Table 4.1: Proportion of accepted t-tests

|  |  | 1 | 2 | 3 | 4 | 5 | 6 | 7 | 8 | 9 | 10 | 11 | 12 |
|---|---|---|---|---|---|---|---|---|---|---|---|---|---|
| Real GDP | ALMP | 1 | 0.31 | 1 | 0.01 | 0.96 | 0.67 | 0.96 | 0.52 | 0.99 | 0.99 | 0.94 | 0.96 |
|  | Rep. rate | 0.15 | 0.42 | 0.99 | 0.6 | 0.66 | 0.82 | 0.81 | 0.48 | 0.62 | 0.71 | 0.93 | 1 |
|  | EPL | 1 | 0.9 | 0.96 | 1 | 0.92 | 0.99 | 1 | 0.68 | 0.46 | 0.35 | 0.95 | 0.66 |
| Inflation | ALMP | 0.98 | 0.85 | 1 | 0.96 | 0.99 | 0.99 | 0.98 | 0.99 | 0.62 | 1 | 0.85 | 0.98 |
|  | Rep. rate | 1 | 0.99 | 1 | 0.89 | 0.79 | 0.27 | 0.67 | 0.8 | 0 | 0.85 | 0.26 | 0.26 |
|  | EPL | 0.97 | 1 | 0.97 | 0.82 | 0.96 | 0.79 | 0.96 | 0.71 | 1 | 0.53 | 0.97 | 0.97 |
| Unemployment | ALMP | 1 | 1 | 0.93 | 0 | 0 | 0 | 0.01 | 0.32 | 0.54 | 0.55 | 0.01 | 1 |
|  | Rep. rate | 0.62 | 0.37 | 0.74 | 0.59 | 0.99 | 0.99 | 0.49 | 0.96 | 0.97 | 1 | 1 | 1 |
|  | EPL | 1 | 1 | 0.93 | 0.99 | 0.97 | 1 | 1 | 1 | 0.99 | 0.42 | 0.48 | 0.7 |
| REER | ALMP | 0.99 | 0.91 | 0.87 | 0.27 | 0.85 | 1 | 1 | 1 | 0.96 | 1 | 1 | 0.62 |
|  | Rep. rate | 0.52 | 0.15 | 0.65 | 0 | 0.45 | 0.92 | 0.51 | 0.72 | 0.37 | 0.57 | 0.82 | 0.03 |
|  | EPL | 0.99 | 0.89 | 0.39 | 0.47 | 0.78 | 0.84 | 1 | 0.94 | 0.82 | 0.91 | 0.47 | 0.67 |

Acceptance proportions for the real GDP indicate no significant difference for the ALMP on impact but this is no longer true later, substantial difference for the policy of the replacement rates, and differences for the EPL over the longer run, but not initially (Table 4.1). Inflation seems to display quite erratic patterns – consistent differences are registered for the replacement rate from the third period onwards. Interestingly, the ALMP starts displaying different results, depending on the monetary policy, for unemployment, which is the ultimate objective of this policy measure. This finding connects to the puzzle of ALMP that has been found in the literature (and discussed in our literature review): ALMP does not display consistent results on unemployment and varies substantially across countries and time periods, with rather limited evidence on its efficacy. We thus demonstrate one additional channel that can rationalize such varying results, and call for the use of confounding events, such as monetary policy interventions, to uncover the true effects of labor market policies.

The replacement rate has different effects on unemployment in the beginning, whereas the EPL has such effects at a very late stage, depending on the interest rate environment. The replacement rate is also channeled to REER differently in tightening and loosening interest rate environments. The EPL seems to also affect the REER differently, especially from the second to the sixth, and from the ninth to the twelfth periods. The ALMP varies across a few horizons but has limited support for consistently different effects.

**Takeaway 9. Horizon-Specific Monetary Policy Impact** The statistical test shows that ALMP's impact on real GDP is initially insignificant but becomes non-ignorable over time, while replacement rates and EPL have delayed effects on inflation and unemployment. This highlights the importance of considering the timing of monetary policy interactions to fully understand labor market policy outcomes.

### 4.3.2 Multiple Testing

The test above has considered each horizon separately. We now take into account the fact that each variable and each policy measure require twelve tests (one for each horizon), thus requiring us to control the family wise error rate (the probability of at least one type I error) or the false discovery rate (where some false positives are allowed for). We choose three results – a standard Bonferroni adjustment, the Holm's method, which is valid under quite arbitrary assumptions, and the Benjamini and Yekutieli method, which controls the expected proportion of false findings among the rejected hypotheses.[26] Ultimately, we take all the p-values for all horizons (separately for each macroeconomic variables and labor market policy), and compute adjusted p-values (overall, there are 12 p-values in each such testing for each time period, over which we will be averaging). For instance, a standard Bonferroni method, reported in the first column of Table 4.2, postulated a new significance level 0.1/12=0.0083 for the null hypothesis to be rejected for each time period.[27] We recomputed proportions for the entire impulse response function: if the null is rejected for the majority of horizons, the response function is considered different once monetary policy is taken into account.

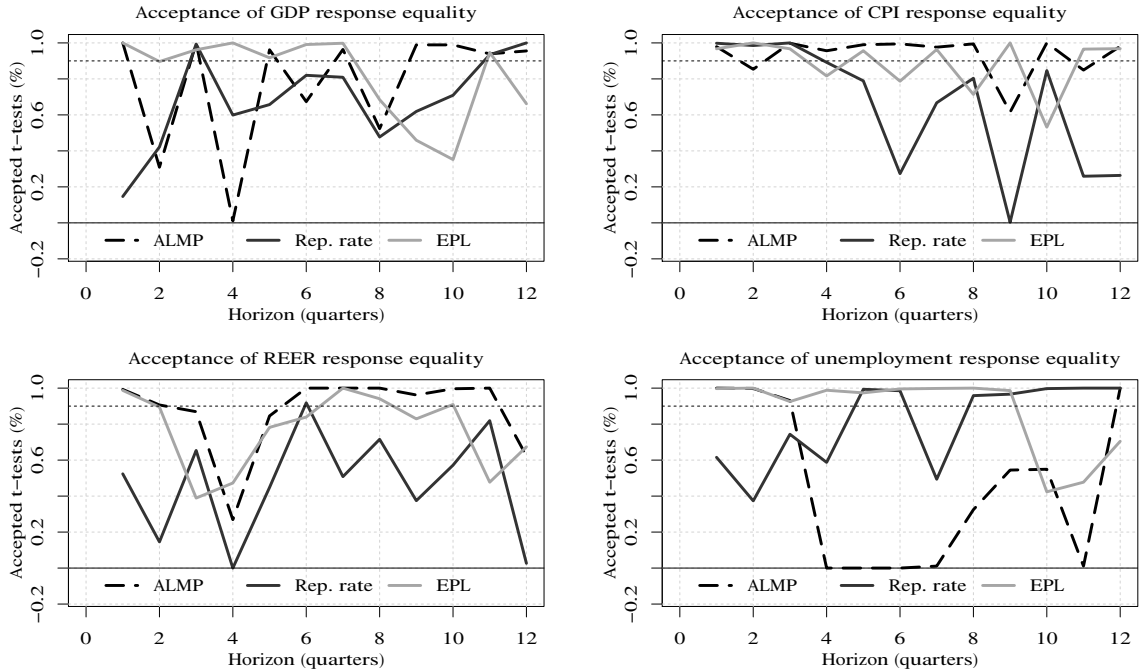

**Figure 4.3: Acceptance regions of $H_0 : \tilde{\delta}_k - \delta_k = 0$ for the replacement rate, ALMP and EPL in models for inflation, real GDP, REER and unemployment**

It is clear that using a threshold of at least one significant difference, all macro variables and all policy reforms pass it. Sticking to the 10% tolerance level, we find that monetary policy delivers different impulse response functions if replacement rate policies are evaluated for the real GDP, inflation and REER. When it comes to unemployment, however, differences are significant for the active labor market policies. Hence, if the *entire dynamic path* was of interest about the real GDP, monetary policy did matter least for the ALMP reforms (similarly to responses in inflation and REER). The most significant result is with regards to the replacement rate (unemployment benefits) policies and competitiveness, that is, REER. Despite horizon adjustments, our results remain largely intact compared to the horizon-specific tests above. Finally, if we pooled over $h$ and $t$, the second part of the Table 4.2, the most robust result for the entire period is obtained with regards to unemployment benefits and real GDP as well as REER. Importantly, the monetary policy effect of ALMP on unemployment remains even after double averaging in the case of the

---

[26]Though we concern ourselves with the probability of the entire impulse response function, we refer to Lütkepohl et al. (2015), who argue that Bonferroni method may be useful to construct bands around estimated impulse response functions, based on the estimated parameters and their empirical distributions. We see an interesting research direction to develop more clear guidance to uncover optimal *joint* confidence bands of different horizons.

[27]Note that changes in interest rates depend on time; therefore, the t-tests are conducted for each time period (in fact, the tests are conducted for all $t$ time periods and all $i$ countries, so reported tests are averaged over both these dimensions).

Table 4.2: Proportion of the Equality Rejection with Adjusted *p*-values (left: averaging over $h$, right: averaging over $h$ and $t$)

|  |  | Bonferroni | Holm | Benjamini & Yekutieli | Bonferroni | Holm | Benjamini & and Yekutieli |
|---|---|---|---|---|---|---|---|
| Real GDP | ALMP | 0.99 | 0.99 | 0.99 | 1 | 1 | 1 |
|  | Rep. rate | 0.79 | 0.78 | 0.79 | 0.9 | 0.9 | 0.83 |
|  | EPL | 0.92 | 0.92 | 0.93 | 0.99 | 0.99 | 0.97 |
| Inflation (CPI) | ALMP | 0.99 | 0.99 | 0.99 | 1 | 1 | 1 |
|  | Rep. rate | 0.89 | 0.88 | 0.9 | 0.97 | 0.97 | 0.94 |
|  | EPL | 0.96 | 0.96 | 0.97 | 0.99 | 0.99 | 0.98 |
| Unemployment | ALMP | 0.81 | 0.8 | 0.81 | 0.99 | 0.99 | 0.89 |
|  | Rep. rate | 0.94 | 0.93 | 0.94 | 0.98 | 0.98 | 0.97 |
|  | EPL | 0.95 | 0.95 | 0.96 | 1 | 1 | 1 |
| REER | ALMP | 0.99 | 0.99 | 0.99 | 1 | 1 | 1 |
|  | Rep. rate | 0.65 | 0.62 | 0.62 | 0.77 | 0.77 | 0.68 |
|  | EPL | 0.92 | 0.92 | 0.92 | 0.96 | 0.96 | 0.96 |

Benjamini and Yekutieli method.

Our analysis underscores the importance of considering the entire dynamic path and monetary policy stance when evaluating the effectiveness of labor market reforms. In conclusion:

**Takeaway 10. Multi-horizon approach and multiple testing of the monetary policy role** Monetary policy significantly influences the macroeconomic impact of labor market policies, in particular, how replacement rates affect real GDP, inflation, and REER, and how ALMPs impact unemployment outcomes.

## 5 Extensions

To conserve space, we report 90% confidence intervals in Section D.1 with three sets of the results (sub-sample 1985-1998 in Section D.1.1, sub-sample 1999-2010 in Section D.1.2 and the overall sample in Section D.1.3). Unlike the main text, where we tested $H_0 : \tilde{\delta}_k - \delta_k = 0$, we also test $H_0 : \tilde{\delta}_k = 0$ against a two-sided alternative. We report probability values from all the models separately in Appendices, Section D.2. All the models and all the extra parameters, in addition to the first two lags of the labor market policies, are visualized for each horizon in Figures D.16-D.17. We also cover an extension of the multi-factor error structure in Appendix C, but here, we focus on two essential extensions: additional controls for anticipation effects and aggregate demand fluctuations and an expanded sample.

### 5.1 Anticipation Effects and Aggregate Demand Fluctuations

Even though we control for the history of macroeconomic dynamics, past policies, and aggregate shocks, one may nevertheless be concerned that pre-determinedness of controls in equations (2.1) may not hold due to the anticipated component. Following Auerbach and Gorodnichenko (2012), Duval and Furceri (2018), and Lastauskas and Stakenas (2020a), we introduce the OECD forecast for yeat $t$ GDP growth, made at $t-1$, as an additional variable that controls for agents' expectations about the evolution of an economy. We have manually extracted projections from the OECD Economic Outlook June and December editions. Note that due to the quarterly nature of our exercise, the construction of anticipated controls is more nuanced in our case. In the first and the second quarters, we included the current year's GDP forecast published in the previous year's OECD's December edition, whereas in the third and the fourth quarters we included next year's

GDP forecast published in the current year's June edition. Conditional on past information set, we now control for any expectations about the economic environment that may be correlated with the policy reforms. In addition to anticipation effects, and even though we are dealing with changes in variables (growth rates if levels are measured in logarithms), we also introduce an output gap (constructed using a standard HP filter)[28] to control for the unemployment and other macro effects of aggregate demand fluctuations over the business cycle.[29]

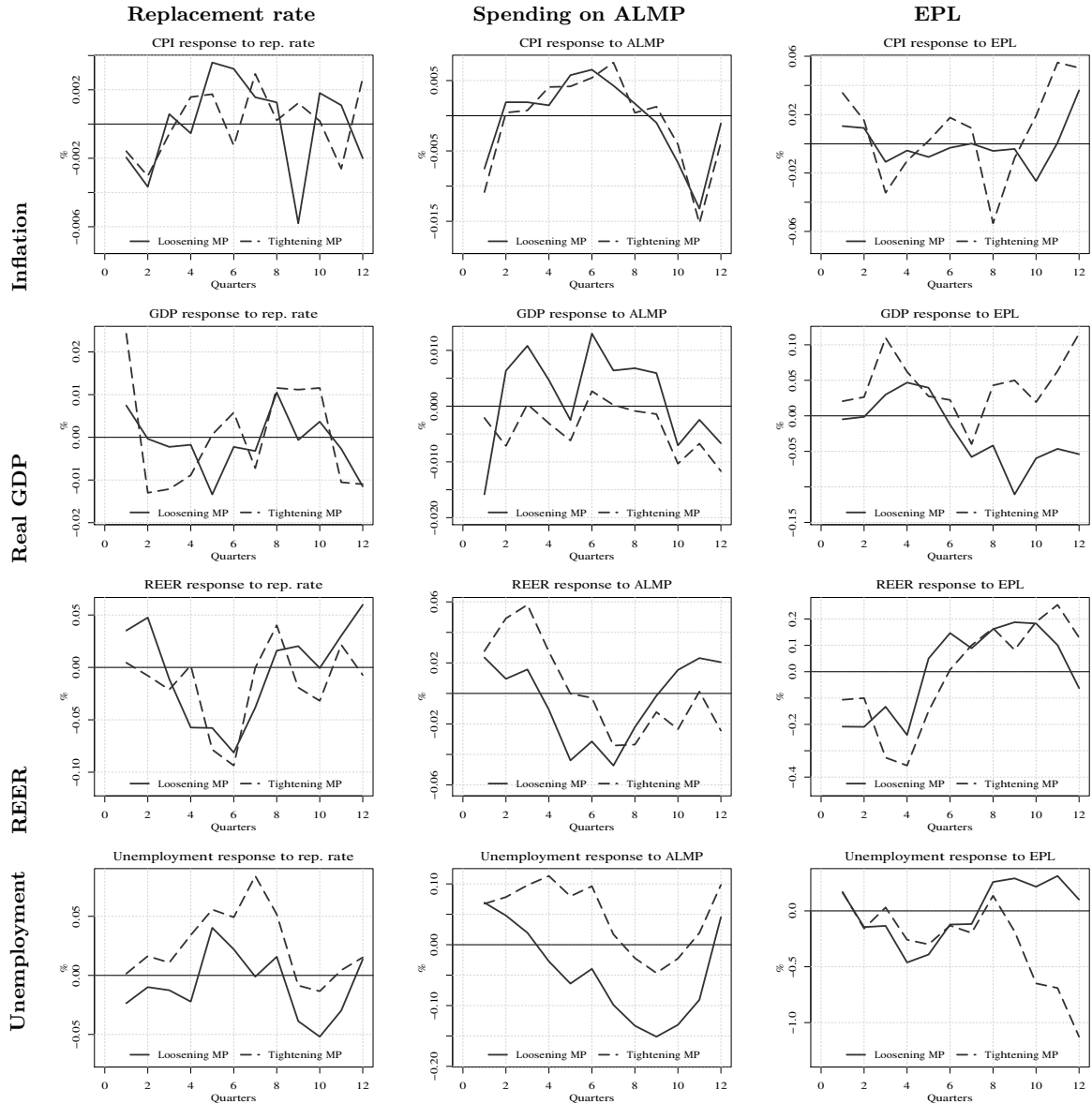

Figure 5.1: **Changes in macroeconomic variables due to the 1% increase in labor market policies (replacement rate, ALMP and EPL), weighted (Mallow's averaging) impulse responses (local projections) with OECD GDP forecasts and HP-filtered output gaps**

---

[28] We use $\lambda = 1600$ for our quarterly data as originally suggested by Hodrick and Prescott (1997).

[29] To alleviate notation, these additional variables are not accounted within $\boldsymbol{X}_{it}$ in the model (2.1).

Figure 5.1 collects weighted local projections with additional controls for anticipation and output gaps. As before, the replacement rate and the ALMP tend to reduce whereas EPL tends to increase inflation on impact. A more inflationary result may be achieved by increasing labor market rigidity in the tightening monetary policy environment. Unemployment benefits may have expansionary effect only in the short run. ALMP supports aggregate economy when a loosening monetary policy is in place; in contrast, EPL is more inflationary and expansionary under increasing interest rates. Changes in the real effective exchange rate resemble results from before with a J curve dynamics for the EPL. Finally, unemployment suffers less from an increase in the replacement rate and goes down due to the ALMP expenditure if monetary policy is loose. We find the same narrative for the EPL which supports the macroeconomy when interest rates are higher. The story about the expenditure-increasing policies delivering during an economic downturn when loose monetary policy is in place whereas institutional changes being favored during economic booms is supported by accounting for anticipation effects and aggregate demand fluctuations.

**Takeaway 11. Monetary Policy and Labor Market Rigidity Allowing for Anticipation and Demand Fluctuations** The replacement rate and ALMP reduce inflation on impact, while EPL increases it. ALMP benefits the economy under loosening monetary policy, whereas EPL is more inflationary and expansionary with tightening rates. Unemployment benefits have a short-term expansionary effect, and EPL supports the macroeconomy during high-interest periods, aligning with the narrative that expenditure-increasing policies are effective during downturns with loose monetary policy.

## 5.2 Expanded Sample

The last robustness check we conducted was expanding the time and country coverage. In the baseline, we analyzed data until 2010, when conventional monetary policy prevailed in the euro area. We extended the sample up to and including Q4 2020 and included Luxembourg and new euro area members: Estonia, Latvia, Lithuania, Slovakia, and Slovenia. Though data coverage varies across variables and countries, the maximum time coverage is 1985-2020 for 17 euro area countries in the expanded sample (please refer to Appendix A for data description and summary statistics).

We replicate the analysis and report local projections for each model A-E in Appendix E.1.1, covering evidence from the most recent period with joint monetary policy in the euro area (1999-2020), whereas Appendix E.1.2 covers a full expanded sample. Compared to the counterparts of our smaller sample (refer to Section 3.1 and Appendix D.1.3), we find several qualitative differences in reactions across the models. For instance, in the expanded sample, the replacement rate is less recessionary and barely impacts REER. At the same time, EPL is less impactful on real GDP and unemployment in the tightening interest rate environment (models A and B). This illustrates the role of time-varying model uncertainty as some models deliver differences that matter and others that do not.

When we produce Mallow's weights, reported in Appendix E.2 for baseline and extended models using an expanded sample (to be compared with Figure 4.1),we also find a few differences. They are unsurprising since models A-E delivered observable differences for the expanded sample. For instance, model E started to dominate the average model for EPL's impact on real GDP and unemployment, though, for a smaller sample, model B played a relatively more prominent role. Recall that model E is the most exhaustive, whereas model B relies on periods with annual changes in the interest rate.

Despite the weight differences, what actually matters is the average impulse response, which is the ultimate object of interest. Figure E.13 reports average projections for the baseline model, and Figure E.14 for an extension with output gaps and announcements being controlled for. Comparing the former to Figure 4.2, we find that real variables (real GDP and unemployment) display considerable differences, whereas REER and inflation responses are quite similar for shorter and expanded samples. As mentioned, monetary policy changed after 2010, so the updated sample reflects (though not fully captured as new tools, beyond interest rates, were widely used). Another reason for differences might be that the baseline model still fails to capture demand fluctuations and anticipation, which may play an even larger role post-2010.

And, indeed, we find that having accounted for them, Figures 5.1 and E.14 deliver very similar dynamic responses and thus the same qualitative messages, as already summarized in Takeaway 11. As before, EPL reduction (less protection) and ALMP spending are better implemented under the looser monetary policy as they reduce unemployment more and can boost international competitiveness, whereas more labor protection is warranted under the tighter monetary policy. We conclude with the following Takeaway:

**Takeaway 12. Robustness with Expanded Sample** Expanding the analysis to include data up to 2020 and additional euro area countries reveals some qualitative differences in the impacts of labor market policies. Notably, the replacement rate becomes less recessionary, and EPL has a lessened effect on real GDP and unemployment under tightening monetary policy. However, after accounting for demand fluctuations and anticipation effects—which are more pronounced post-2010 due to changes in monetary policy—the main results remain consistent. This reinforces that reducing EPL and increasing active labor market policy (ALMP) spending are more effective under looser monetary policy, enhancing unemployment reduction and international competitiveness, while increased labor protection is more beneficial under tighter monetary policy environments.

## 6 Conclusions

We considered the heterogeneous effects of labor market institutions on the macroeconomy in different interest rate environments. We built a baseline model with five potential alternatives how monetary policy environment can change labor market policies' impact on the macroeconomy. We found that replacement rates and ALMP initially reduce inflation, but EPL increases it, with their effects on GDP, unemployment, and inflation varying significantly with monetary policy stance and timing. Loosening policies enhance the unemployment-reducing effects of ALMP and replacement rates, whereas tightening policies improve EPL's impact on GDP.

In other words, our hypothesis that the effects of labor policy changes initiated in periods of loose monetary policy differ from the effects of those initiated when monetary tightening prevails was largely confirmed. We developed a test that shed light both on significant differences due to monetary policy over any chosen horizon *and* the entire impulse response function. We found that monetary policy carries unequal significance for different horizons but is particularly important for the effect of the replacement rate across responses in macroeconomic variables and active labor market policies on unemployment. This contributes to the solution of the empirical puzzle, found in the labor literature, that the activation policies may not have aggregate effects on the reduction of unemployment whenever monetary policy regimes or business cycle conditions are ignored.

In practical terms, this means that governments should consider the monetary policy stance and business cycle position when implementing labor market interventions, as macroeconomic outcomes may vary. The welfare-improving (unemployment-reducing) effects of activation schemes are justified when monetary policy is loose. Similarly, liberalization (lowering) of employment protection may be expansionary when interest rates decrease. These findings highlight the need for closer coordination between government (executive) and monetary policy authorities to ensure that labor market interventions are supported by appropriate monetary policy conditions.

Methodologically, we proposed to apply frequentist tools for model averaging; in particular we averaged popular local projections using the Mallow's $C_p$ criterion. Our framework does not require knowledge of the exact functional form (one can specify many alternatives), is thus robust to misspecification, admits non-linearities, and addresses uncertainty regarding interactions between labor policies and the macroeconomy. Econometrically, the method is easy to implement as it effectively relies on ordinary least squares. Policymakers are most interested in the dynamic path of a policy (the short and medium term policy effects); this is our target, too. Due to the local nature of estimation, each period after a shock carries a specific parameter. Its applicability thus goes well beyond our application, in particular, where heterogeneous responses over time are suspected.

A number of promising research directions remain. On the methodology side, we see developments on the ways to calculate robust standard errors and optimal joint confidence bands of different horizons particularly useful for empirical applications. On the literature of labor market policies side, we foresee applications of our methodology on an even larger set of countries (a major obstacle so far remains data availability), exploring even more layers of heterogeneity and interactions. For instance, political cycles might be incorporated into our framework to explore how they affect labor policies in countries with different levels of central bank independence. Our framework, which jointly models dynamic paths of policies and model uncertainty, is particularly well suited for multi-country policy interventions at the business cycle frequency. Additionally, more granular data applications would be of interest as well as a larger set of policies, such as goods or financial markets. More theoretical contributions that combine business cycle fluctuations with labor market policy reforms and monetary policy stance would also be a fruitful avenue of study.

# Online Appendix for "Labor Market Policies in High- and Low-Interest Rate Environments: Evidence from the Euro Area"

Povilas Lastauskas
Homerton College, University of Cambridge, and Vilnius University

Julius Stakėnas
Vilnius University

**Abstract**

This Online Appendix collects supplementary material, namely: summary statistics for an expanded sample, a description of the dataset and its sources; a theoretical argument to use Mallow's weights to produce robust local projections in the presence of stationary, dependent data and even multi-factor error structure; additional empirical results and robustness checks. An additional Supplementary Material is posted online on the corresponding author's website.

# A  Data

## A.1  Summary Statistics for an Expanded Sample

Table A.1: Summary Statistics Using Expanded Sample

| Country | GDP | | Inflation (CPI) | | Unemployment | | REER | | Short interest rate | | ALMP | | UB | | EPL | |
|---|---|---|---|---|---|---|---|---|---|---|---|---|---|---|---|---|
| | mean | sd | mean | sd | mean | sd | mean | sd | mean | sd | mean | sd | mean | sd | mean | sd |
| Austria | 0.45 | 1.60 | -0.00 | 0.33 | 0.27 | 5.81 | 0.06 | 1.06 | -0.05 | 0.43 | 0.38 | 5.49 | 0.15 | 1.65 | -0.10 | 0.77 |
| Belgium | 0.40 | 1.53 | -0.01 | 0.42 | -0.42 | 4.81 | 0.15 | 1.12 | -0.08 | 0.49 | 0.88 | 3.18 | -0.08 | 0.51 | 0.11 | 1.88 |
| Finland | 0.43 | 1.42 | -0.01 | 0.41 | 0.33 | 5.68 | -0.20 | 2.72 | -0.10 | 0.78 | -0.15 | 5.15 | 0.02 | 0.82 | -0.18 | 0.84 |
| France | 0.36 | 1.94 | -0.01 | 0.33 | -0.18 | 2.66 | 0.00 | 1.19 | -0.08 | 0.55 | 0.22 | 3.01 | 0.10 | 0.78 | -0.01 | 0.61 |
| Germany | 0.40 | 1.45 | -0.00 | 0.41 | -0.50 | 3.17 | 0.02 | 1.67 | -0.05 | 0.40 | 0.11 | 5.13 | -0.21 | 0.93 | -0.12 | 0.49 |
| Greece | 0.19 | 2.49 | -0.03 | 0.72 | 0.47 | 3.66 | -0.00 | 2.02 | -0.12 | 2.74 | 0.07 | 12.42 | 1.15 | 2.98 | -0.30 | 2.07 |
| Ireland | 1.24 | 2.71 | -0.01 | 0.45 | -0.70 | 5.94 | -0.24 | 2.48 | -0.11 | 1.22 | 0.89 | 8.51 | 0.17 | 1.42 | 0.00 | 1.13 |
| Italy | 0.20 | 1.83 | -0.02 | 0.28 | 0.10 | 3.35 | -0.00 | 2.34 | -0.12 | 0.70 | 0.18 | 5.20 | 3.72 | 12.18 | -0.19 | 1.50 |
| Netherlands | 0.45 | 1.16 | -0.00 | 0.36 | -0.46 | 4.87 | 0.09 | 1.66 | -0.05 | 0.40 | 1.36 | 8.48 | -0.40 | 1.32 | 0.01 | 0.75 |
| Portugal | 0.45 | 2.08 | -0.04 | 0.52 | -0.17 | 5.71 | 0.31 | 1.34 | -0.16 | 0.85 | 0.97 | 7.08 | 0.55 | 1.77 | -0.37 | 1.54 |
| Spain | 0.48 | 2.19 | -0.01 | 0.52 | -0.16 | 3.91 | 0.24 | 1.78 | -0.09 | 0.81 | 1.13 | 7.05 | -0.11 | 0.93 | -0.37 | 2.11 |
| Estonia | 0.95 | 2.06 | -0.01 | 0.80 | -0.31 | 13.22 | 1.14 | 1.85 | -0.09 | 1.31 | 3.36 | 10.86 | 0.02 | 1.23 | -0.88 | 6.06 |
| Latvia | 0.85 | 2.41 | -0.41 | 9.15 | -0.62 | 8.54 | 0.74 | 2.58 | -0.06 | 1.58 | 0.38 | 12.28 | 0.37 | 2.99 | 0.00 | 0.00 |
| Lithuania | 1.00 | 1.99 | -0.03 | 1.47 | -0.39 | 8.37 | 1.40 | 2.95 | -0.14 | 1.58 | 2.64 | 19.66 | 0.59 | 3.32 | -0.92 | 4.40 |
| Luxembourg | 0.78 | 2.05 | -0.01 | 0.46 | 0.99 | 12.84 | 0.30 | 0.66 | -0.08 | 0.49 | 0.21 | 5.72 | 0.02 | 0.29 | 0.00 | 0.00 |
| Slovenia | 0.60 | 1.91 | -0.05 | 0.67 | -0.32 | 6.14 | 0.17 | 1.14 | -0.15 | 0.58 | -0.01 | 7.56 | 0.44 | 3.31 | -0.52 | 3.31 |
| Slovakia | 0.89 | 2.10 | -0.06 | 1.49 | -0.59 | 4.27 | 0.53 | 1.92 | -0.16 | 1.19 | 0.07 | 12.47 | -0.90 | 4.80 | -0.21 | 2.02 |

## A.2 Description and Sources

**Table A.2: Data definitions and sources**

| Variable | Definition | Measure | Source | Adjustments |
|---|---|---|---|---|
| Public expenditure on active labor market policies | Expenditure on public interventions, which are explicitly targeted at groups of persons with difficulties in the labor market: the unemployed, the employed at risk of involuntary job loss and inactive persons who would like to enter the labor market. Total expenditure on active measures can be broken down into 7 categories, which include labor market policy (LMP) services (category 1) and LMP measures (categories 2-7). LMP measures cover activation measures for the unemployed and other target groups including the categories of training, job rotation and job sharing, employment incentives, supported employment and rehabilitation, direct job creation, and start-up incentives. | Expenditure per number of unemployed (we take the number of unemployed in the previous year) divided by GDP per capita | OECD, Eurostat | The main data source is Eurostat. For missing data entries, OECD data were used. Annual data were interpolated using Denton's method. The data were also logarithmically transformed. |
| Population | Resident population, i.e. all persons, regardless of citizenship, who have a permanent place of residence in the country. | 1000s | OECD | |
| Real effective exchange rate | Weighted average of a country's currency relative to an index or basket of other major currencies adjusted for variations in relative prices using unit labor costs in manufacturing (for Greece, we used HICP). The weights are determined by comparing the relative trade balances, in terms of one country's currency, with each other country within the index. | Index (2000=100) | OECD, Eurostat (for Greece) | The data were logarithmically transformed. |
| Output gap | Cyclical component of real gross domestic product filtered with the Hodrick Prescott filter ($\lambda = 1600$) | Source and definition of GDP as in this Table | Own calculations | |
| Short term interest rate | 3-month money market rates | Percentage | Eurostat | |

| Variable | Definition | Measure | Source | Adjustments |
|---|---|---|---|---|
| Real gross domestic product | Value of all final goods and services produced within an economy per quarter/year, taking into account changes in the general price level. | Index (2000=100) | GVAR database, IMF International Financial Statistics database, OECD National Accounts database, Eurostat | The primary source is the GVAR 2013 Vintage database. The IMF International Financial Statistics (IFS) database was also consulted for Portugal and Denmark, GDP Volume (2005=100). Data for Denmark were seasonally adjusted using the U.S. Census Bureau's ARIMA X12. Statistics for Luxembourg and Ireland, were obtained from the OECD National Accounts database. The most extensive adjustments were made to data pertinent to Greece. Data for 1980-1994 were obtained from the OECD National Accounts database. The base year was adjusted (2000=100) with a backward extrapolation until Q1 1980 using quarterly growth rates based on the OECD estimates. Data from 1995 onwards were extracted from Eurostat and seasonally adjusted with ARIMA X12. The data were also logarithmically transformed. |
| GDP forecast | The variable is used as an expectation measure for future GDP growth. Constructing the time series, in quarters 1 and 2, the measure is equal to the annual GDP growth forecast of the running year published in the previous year (last year's Economic Outlook issue 2). In quarters 3 and 4, the measure is equal to annual GDP growth forecast of the next year as projected in the running year's Economic Outlook issue 1. | Annual GDP change | Historical OECD Economic Outlook Issues 1 and 2 | |
| EPL (employment protection legislation index) | A measure of procedures and costs involved in dismissing individuals or groups of workers and the procedures involved in hiring workers on fixed-term or temporary work agency contracts. | Index | OECD, own calculations | Separate EPL indices for regular and temporary workers were averaged according to shares of temporary/regular contract workers in the sample for a specific country. |
| Inflation (CPI) | Q-o-Q difference of consumer price index for all items | Index | OECD, own calculations | 2010=100, seasonally adjusted using arima x12. The index was logarithmically transformed. |

| Variable | Definition | Measure | Source | Adjustments |
|---|---|---|---|---|
| Unemployment rate | Number of unemployed persons as a percentage of the labor force with a seasonal adjustment. | Percentage | Eurostat Labour Market database, IMF International Financial Statistics database, OECD Labour database | Q1 1983 - Q2 1998 data for Greece was interpolated from OECD annual data using Denton's interpolation method. For Austria, Finland and Germany, where Eurostat quarterly unemployment data were not available, it was constructed from annual IMF data using Chow-Lin interpolation method (quarterly indicator series was constructed from quarterly registered unemployed series and interpolated annual labor force series). In the case of Spain, as Eurostat and OECD annual data for this country exhibit certain differences, Eurostat annual data were extrapolated using annual OECD data and then the latter annual series was interpolated using Chow-Lin method (quarterly indicator series was constructed from quarterly registered unemployed series and interpolated annual labor force series). The data were also logarithmically transformed. |
| Unemployment replacement rate | Proportion of net in-work income that is maintained when unemployed. The OECD summary measure is defined as the average of the gross unemployment benefit replacement rates for two earnings levels, three family situations and three durations of unemployment. For further details, see OECD (1994), The OECD Jobs Study (chapter 8) and Martin J. (1996), 'Measures of Replacement Rates for the Purpose of International Comparisons: A Note', OECD Economic Studies, No. 26. Pre-2003 data have been revised. | Values between 0 and 1 (before log transformation) | OECD, Eurostat, van Vliet and Caminada (2012) | Biannual data were interpolated to quarterly frequency using Denton's method. Data were spliced from two OECD measurements, keeping original data until 2005 and rescaling after 2005. The data were also logarithmically transformed. Data for the Baltic States for 2005-2014 were taken from the European Commission's Tax and benefits indicators database. Prior data come from van Vliet and Caminada (2012). Data were spliced keeping the original data from the European Commission and constructing earlier data using percentage point changes observed in the van Vliet and Caminada (2012) dataset; Annual data were interpolated using Denton's method. |

## B  Discussion

**Proposition.** *Local projection, weighted by Mallow's weights $C_T(w)$, is asymptotically unbiased in the sense that such averaging delivers asymptotically unbiased estimator of the mean squared (forecast) error in the presence of stationary, dependent data and even multi-factor error structure.*

Taken separately, results on the optimality of both, local projections and Mallow's averaging, are already known in the literature. Our contribution is to combine them as well as to admit a common factor in the data generating process. We shall start with the assumptions about the data generating process: let the information set (filtration) be defined as $\Omega_t = \sigma(y_t, \{z_t\}, y_{t-1}, \{z_{t-1}\}, \ldots)$, where $\{z_t\}$ denotes a set already used in the equation (4.1) and defined in (4.2). Usual orthogonality condition $\mathbb{E}(u_{t+h}|\Omega_t) = 0$, strict stationarity and ergodicity of data, finite fourth moments ($\mathbb{E} \|z_t\|^4 \leq C$, $\mathbb{E} \|u_t\|^4 \leq C$ for a generic constant $C$), (semi-)positive definiteness $\mathbb{E} z_t z_t' \geq 0$, and weak dependence ($T^{-1/2} \sum_{t=1-h}^{T-h} z_t u_{t+h} \xrightarrow{d} N(0, \Sigma)$, such that $\Sigma = \sum_{|j|<h} \mathbb{E}\left(z_t z_{t-j}' u_{t+h} u_{t+h-j}\right)$) are assumed to hold. For the factor-dependent data, we admit $\{X_{it}\}$ and factors to explain $y_t$, also let factors be correlated with the regresssors. An extreme of such dependence (see, for instance, Pesaran (2006))is the assumption that $\{X_{it}\}$ are generated by $\{f_t\}$ and idiosyncratic components. We shall not deviate from the literature or seek the smallest set of assumptions to justify the method; rather, our aim is to demonstrate how the proposed method fits within the existing body of knowledge and how to implement it.

Let a vector with factors $\{f_t\}$ be a zero mean covariance stationary process with absolutely summable autocovariances, distributed independently of individual specific errors across cross-sectional units and time. Finite moments condition $\mathbb{E} \|f_t\|^4 \leq C$, $\sum_{t=1}^T f_t f_t' \xrightarrow{p} \Sigma_f > 0$ and deterministic factor loadings (or finite fourth moments) are assumed to hold. Though the method can be applied for non-stationary factors, we deal with the stationary setup (differenced variables) as it readily complies to the theory on the model averaging.

By (4.1),
$$y_{t+k}(m) = z_t(m)' a(m) + u_{t+k}(m),$$

and its least squares forecast is
$$\hat{y}_{T+k|T}(m) = \tilde{z}_T(m)' \hat{a}(m),$$

where $\tilde{z}_T(m)$ has unobserved factors replaced by their estimates $\tilde{f}_t$. The combination of forecasts is given by
$$\hat{y}_{T+k|T}(w) = \sum_{m=1}^M w(m) \hat{y}_{T+k|T}(m),$$

and the error term of the combination of projections is given by $u_{t+k}(w) = \sum_{m=1}^M w(m) \hat{u}_{T+k}(m)$. The mean squared forecast error then follows almost immediately:

$$MSFE_T(w) = \mathbb{E}\left(y_{T+k} - \hat{y}_{T+k|T}(w)\right)^2$$
$$= \mathbb{E}\left(u_{T+k} - \sum_{m=1}^M w(m)\left(\tilde{z}_T(m)' \hat{a}(m) - z_T(m)' a(m)\right)\right)^2$$
$$\simeq \mathbb{E} u_{t+k}^2 + \mathbb{E}\left(\sum_{m=1}^M w(m)\left(z_t(m)' a(m)\right) - \tilde{z}_t(m)' \hat{a}(m)\right)^2,$$

where the result follows from the error term's orthogonality and stationarity (as well the fact that factors span the space of the same regressors, assumed to be stationary). Under homoskedasticity, $\mathbb{E}\left(u_{t+k}^2 | \Omega_t\right) = \sigma^2$, we have that

$$MSFE_T(w) = \sigma^2 + \mathbb{E}\left(\sum_{m=1}^M w(m)\left(z_t(m)' a(m)\right) - \tilde{z}_t(m)' \hat{a}(m)\right)^2$$
$$= \sigma^2 + \mathbb{E}\left(\sum_{m=1}^M w(m)\left(u_t - \hat{u}_t(w)\right)\right)^2 = \sigma^2 + \mathbb{E} L_T(w),$$

where the second equality follows from a decomposition of $y_t = \hat{y}_t + \hat{u}_t$ and $L_T(w) = \frac{1}{T} \sum_{t=1}^T \left(u_t - \hat{u}_t(w)\right)^2$, an in-sample squared error. The difference can be further simplified by invoking a projection operator $P$ (where $\tilde{P}$ denotes its estimated counterpart):

$$\hat{u}_t(w) = u_t + \sum_{m=1}^M w(m) z_t(m)' a(m) - \sum_{m=1}^M w(m) \tilde{z}_t(m)' \hat{a}(m)$$
$$= u_t + \sum_{m=1}^M w(m) z_t(m)' a(m) - \sum_{m=1}^M w(m) \tilde{P}(m) z_t(m)' a(m) - \sum_{m=1}^M w(m) \tilde{P}(m) u_t,$$

which, using matrix notation, can be written as
$$\hat{u}(w) = u + \left(I - \tilde{P}(w)\right) z' a - \tilde{P}(w) u,$$

such that $\tilde{P}(w) = \sum_{m=1}^{M} w(m) \tilde{P}(m)$ and $\tilde{P}(m) = \tilde{z}(m) \left(\tilde{z}(m)' \tilde{z}(m)\right)^{-1} \tilde{z}(m)'$. We thus have

$$\hat{u}(w) = u + \left(I - \tilde{P}(w)\right) z(m)' a(m) - \tilde{P}(w) u.$$

Hence,

$$\frac{1}{T}\hat{u}_t(w)' \hat{u}_t(w)$$
$$= \tfrac{1}{T}u'u + \tfrac{1}{T}\left(\left(I - \tilde{P}(w)\right) z(m)' a(m) - \tilde{P}(w) u\right)' u + \tfrac{1}{T}u'\left(\left(I - \tilde{P}(w)\right) z(m)' a(m) - \tilde{P}(w) u\right)$$
$$+ \tfrac{1}{T}\left(\left(I - \tilde{P}(w)\right) z(m)' a(m) - \tilde{P}(w) u\right)' \left(\left(I - \tilde{P}(w)\right) z(m)' a(m) - \tilde{P}(w) u\right)$$
$$= \tfrac{1}{T}u'u + \tfrac{1}{T}\left(\left(I - \tilde{P}(w)\right) z(m)' a(m) - \tilde{P}(w) u\right)' u + \tfrac{1}{T}u'\left(\left(I - \tilde{P}(w)\right) z(m)' a(m) - \tilde{P}(w) u\right)$$
$$+ \tfrac{1}{T}\left(\left(\left(I - \tilde{P}(w)\right) z(m)' a(m)\right)' \left(I - \tilde{P}(w)\right) z(m)' a(m) - \left(\tilde{P}(w) u\right)' \left(I - \tilde{P}(w)\right) z(m)' a(m)\right)$$
$$- \tfrac{1}{T}\left(\left(\left(I - \tilde{P}(w)\right) z(m)' a(m)\right)' \tilde{P}(w) u + \left(\tilde{P}(w) u\right)' \tilde{P}(w) u\right)$$
$$= \tfrac{1}{T}u'u + \tfrac{1}{T}\left(\left(I - \tilde{P}(w)\right) z(m)' a(m) - \tilde{P}(w) u\right)' u + \tfrac{1}{T}u'\left(I - \tilde{P}(w)\right) z(m)' a(m) - \tfrac{1}{T}u'\tilde{P}(w) u$$
$$+ L_T(w) - \tfrac{1}{T}\left(\tilde{P}(w) u\right)' \left(I - \tilde{P}(w)\right) z(m)' a(m) - \tfrac{1}{T}\left(\left(I - \tilde{P}(w)\right) z(m)' a(m)\right)' \tilde{P}(w) u$$
$$= \tfrac{1}{T}u'u + 2\tfrac{1}{\sqrt{T}}\left(\tfrac{1}{\sqrt{T}}\left(z(m)' a(m)\right)' \left(I - \tilde{P}(w)\right)' u\right) - 2\tfrac{1}{T}u'\tilde{P}(w) u + L_T(w),$$

from the standard properties of the projection matrix (symmetry and idempotence). Following Cheng and Hansen (2015), call $r_{1T}(w) = \tfrac{1}{\sqrt{T}}\left(z(m)' a(m)\right)' \left(I - \tilde{P}(w)\right)' u$ and $r_{2T}(w) = u'\tilde{P}(w) u - \hat{\sigma}_T^2 \sum_{m=1} w(m) \dim(z_t(m))$ (refer to the criterion in the equation (4.3)), it then follows that

$$C_T(w) = \tfrac{1}{T}u'u + 2\tfrac{1}{\sqrt{T}} r_{1T}(w) - 2\tfrac{1}{T} r_{2T}(w) + L_T(w).$$

What Cheng and Hansen (2015) prove is the convergence in distribution of $r_{1T}(w)$ and $r_{2T}(w)$ to mean-zero random variables. Using results in Bai and Ng (2006), estimated projection using unobserved factors may be approximated by the true projection and the estimated factors span column space of the true factors, $u'\tilde{P}(w) u = u' P(w) u + o_p(1)$. Clearly, $r_{1T}(w) = O_p\left(T^{-1/2}\right)$ and $r_{2T}(w) = O_p\left(T^{-1}\right)$, whereas $L_T(w)$ is non-zero for any $w$ as long as we rule out a degenerate case with a true model receiving a full weight.

Two differences arise in our application. First, like proposed by Jordà (2005), we consider $y_{T+k}$ and its projection onto the space, generated by the regressors in (4.2). Hence, the quality of the result depends on the choice of $k$, which limits the number of effective observations used for estimation. This is because for all projections the historical set remains the same. Unlike model averaging literature, we are interested in each horizon, one-by-one, with the horizon-specific impulse responses. This is not an issue for the point or well-defined forecasting exercises but is the main objective for the policy evaluation exercise for short, medium and long terms, as results of economic reforms usually deliver very different impacts over different time horizons (Cacciatore et al., 2016b). Another difference lies in the use of estimated unobserved factors, not considered in Jordà (2005). We follow Bada and Kneip (2014) in applying parameter cascading strategy to arrive at what the authors called entirely updated estimators. Unlike Cheng and Hansen (2015), we do not require to know the number of factors in empirical applications. It has been shown that the method by Bada and Kneip (2014) achieves more efficient estimates in terms of mean squared error than Bai et al. (2009) with an externally selected factor dimension; hence, it renders theoretical assumption on the known factor number closer to being true. We report the results from the panel criteria, as proposed by Bai and Ng (2002), which is integrated into a global estimation that alternates between an inner iteration of parameters, factors and their loadings, as functions of the factor dimension, and an outer iteration to select the optimal dimension. We implement all the procedures in R (codes will be made available on the corresponding author's website).

## C   Cross-Sectional Dependence

In addition to taking care of a rich set of observables, one could argue that, despite taking past values of macroeconomic variables, some factor structure (cross-sectional dependence) still remains in the error term. To ensure that our results are robust to such data generating process, we follow Bai (2009) and Bada and Kneip (2014), and let the error term to be subject to the multi-factor structure:

$$\begin{aligned}
\triangle y_{i,t+k} = \alpha_i &+ \boldsymbol{\beta}' \boldsymbol{X}_{it} + \boldsymbol{\gamma}' X_{i,t-1} + \delta_1 \triangle \ln LMP_{it} + \delta_2 \triangle \ln LMP_{i,t-1} \quad && k=1,\ldots,12, \\
&+ \delta_3 \triangle \ln LMP_{it} \times \mathbb{I}_{\triangle i_{it}<0} + \sum_\ell \lambda_{i\ell} f_{\ell t} + u_{i,t+k} && \text{Model A,} \\
&+ \delta_3 \triangle \ln LMP_{it} \times \mathbb{I}_{\triangle^a i_{it}} + \sum_\ell \lambda_{i\ell} f_{\ell t} + u_{i,t+k} && \text{Model B,} \\
&+ \delta_3 \triangle \ln LMP_{it} \times \triangle i_{it} + \sum_\ell \lambda_{i\ell} f_{\ell t} + u_{i,t+k} && \text{Model C,} \\
&+ \delta_3 \triangle \ln LMP_{it} \times \triangle^a i_{it} + \sum_\ell \lambda_{i\ell} f_{\ell t} + u_{i,t+k} && \text{Model D,} \\
+ \delta_3 \triangle \ln LMP_{it} &\times \mathbb{I}_{\triangle i_{it}<0} + \delta_4 \triangle \ln LMP_{it} \times \triangle i_{it} + \\
\delta_5 \triangle \ln LMP_{it} \times \triangle i_{it} &\times \mathbb{I}_{\triangle i_{it}<0} + \sum_\ell \lambda_{i\ell} f_{\ell t} + u_{i,t+k} && \text{Model E.}
\end{aligned} \quad (C.1)$$

As before, let us rewrite the model in (C.1) as $\triangle y_{t+k}(m) = z_t(m)' a(m) + u_{t+k}(m)$, where all variables along with the factors are combined into

$$\begin{aligned}
z_t(m) = (1,\ & X_t,\ X_{t-1},\ f_t(m),\ldots,\ f_{t-p_{max}}(m), \\
& \triangle \ln LMP_t,\ \triangle \ln LMP_{t-1},\ g(\triangle \ln LMP_t(m)),\ldots)',
\end{aligned} \quad (C.2)$$

and $m$ denotes one of the models (A-E) that are covered in specifications (2.1) (zero factors) and (C.1), where error factor structure applies. Notice that the models may include a number of factors or none at all; also, there is no requirement for the models to be nested. To make the procedure feasible, one can substitute unobserved factors with their principal components, control for their existence by cross-sectional averages or estimate them along with other parameters. None of the above addresses simultaneously parameter identification and a number of factors, something of crucial importance in establishing an average model. We follow Bada and Kneip (2014) and estimate each model with an integrated penalty term in the objective function with an iterative procedure to avoid under- or over-parameterization. Effectively, the fitting procedure is a penalized least squares method with iterations to establish an optimal dimension of the factor structure.[30] The factors are estimated by the first eigenvectors that correspond to the first largest eigenvalues, an exact number determined during the estimation procedure.

Resorting to Proposition 1 and Cheng and Hansen (2015), we know that, since factors are generated from the same variables, Mallow's optimality condition can be extended to models with factor structure (recall that Mallow's criterion is directly applicable to any context where fitted values are a linear function of the dependent variable). Accounting for the factor structure arguably makes homoskedasticity assumption, used in proving optimality of Mallow's averaging, hold in more cases than otherwise. Dynamic regression, as is ours, poses some challenges since $\boldsymbol{X}$ includes lags of the dependent variable, thus invalidating linearity assumption. Fortunately, Cheng and Hansen (2015) demonstrate that Mallow's averaging remains valid for dynamic models, too. Despite a multi-factor error structure, the least squares methodology still applies; it is just conducted iteratively with the penalization for the uncertainty about a number of factors (also refer to discussion in Appendix B). This opens up vistas to apply a least squares penalization as is done in the Mallow's combination weights criterion, something missing in the policy evaluation literature (Gobillon and Magnac (2016) consider policy evaluation with the factor structure, yet they abstract from model uncertainty or locally robust dynamic paths of policy changes).

---

[30]The procedures by Bada and Kneip (2014) are rooted in the parameter cascading strategy, put forward by Cao and Ramsay (2010) to estimate models with multi-level parameters.

## C.1 Results for the Baseline Model

Figure C.1 visualizes Mallow's weights for the impulse response averaging, once error factor structure is incorporated into the weight-selection algorithm. Compared to Figure 4.1, factor structure makes weights somewhat less dispersed across horizons and models. Inflation, however, still requires us to apply all the models, across labor market policies, though Model D dominates, as in Figure 4.1. The average response of real GDP to the replacement rate first relies on Model D and moves to Model B, though averaging is more dispersed for other labor market policies. The most stark difference, compared to Figure 4.1, relates to the REER: Model D clearly dominates across all the macro responses and labor variables. Unemployment also prefers Model D, thus creating a difference to the situation when factor structure was ignored as then Model B was preferred for the shock in the ALMP. Data favor specifications with annual rather than quarterly changes or signs of an annual change, supporting our interpretation of the monetary policy trend. Yet, convergence to one particular model across response and shock variables, also horizons, indicates that the prior substantial heterogeneity can be attributed to the existence of unobserved factors, which drive response variables and potentially correlate with the policy variables.

Average responses are depicted in Figure C.2. The most important difference is the reduced importance of monetary policy stance for the macroeconomic effect of labor policies. Though the role of monetary policy does not disappear, it seems that the growth rates across open and integrated European economies are driven by common factors, and they potentially correlate with the joint labor and monetary policy actions. In other words, interactions between labor and monetary policies are driven substantially more by time-varying component than by an idiosyncratic component. Since the euro was first introduced in 1999, a substantial portion of our sample actually reflects a joint monetary policy. The only question, then, remains about whether changes in labor market policies happened to be alike. If the answer was positive, we would expect that parameters $\delta_1$ and $\delta_2$ in the equation (C.1) would also be affected (and not only the interaction terms). In fact, the scales in Figures 4.2 and C.2 are substantially different, thus suggesting that a great deal of variation in labor market policies are actually attributable to a common though unobserved factor. We interpret this result as a confirmation that not only monetary policy became euro area specific but also changes in labor market policies tend to be driven more by a common cyclical factor rather than country-specific components.

**Takeaway 13. Baseline Model and Factor Structure** Incorporating error factor structure into Mallow's weights reveals that common cyclical factors, rather than country-specific components, drive the responses of macroeconomic variables to labor market policies. This suggests that both labor and monetary policies in the euro area are influenced by *shared* economic conditions.

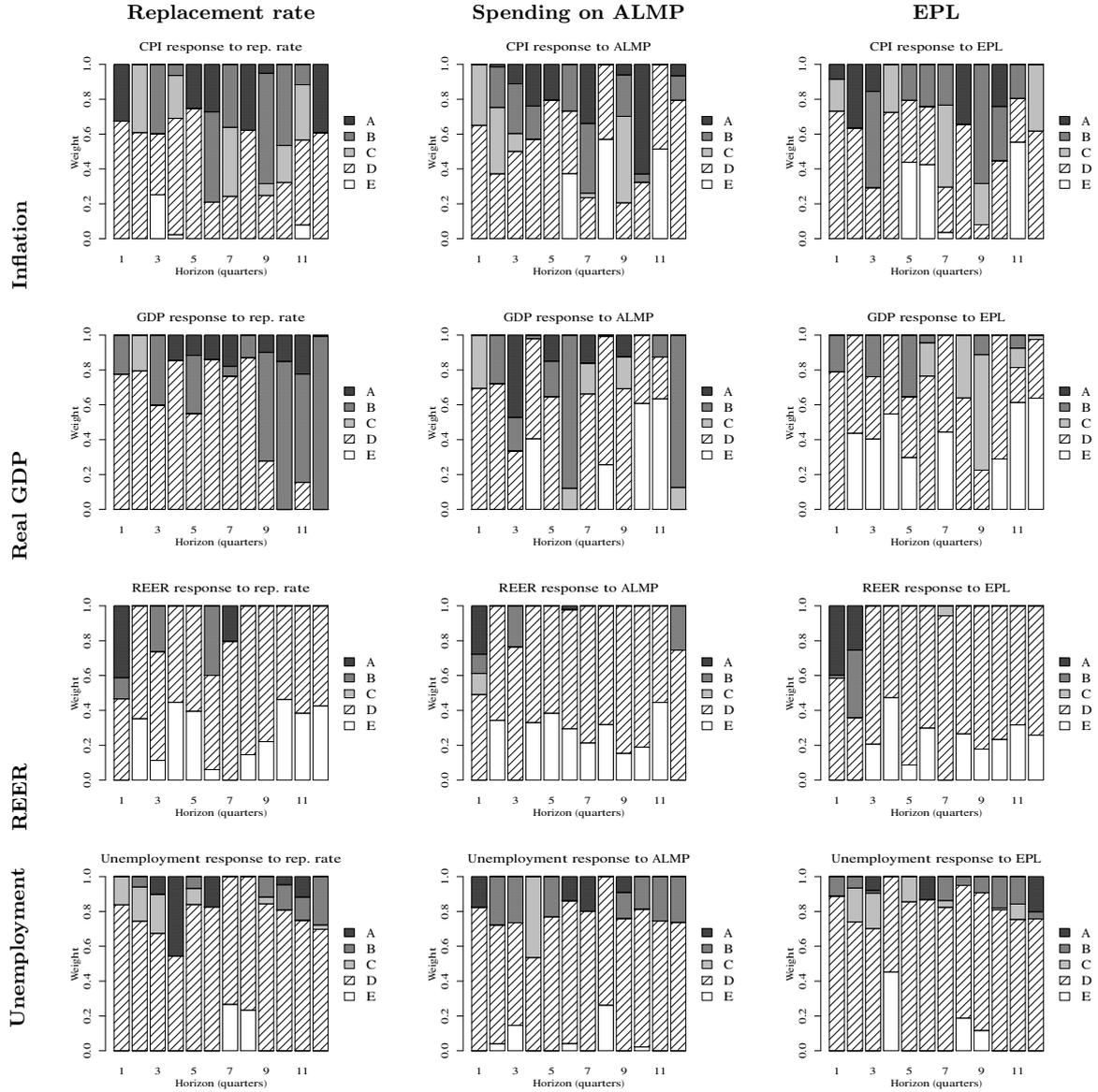

Figure C.1: Average model weights for each horizon, impulse and response variables (Mallow's criterion with multi-factor error structure)

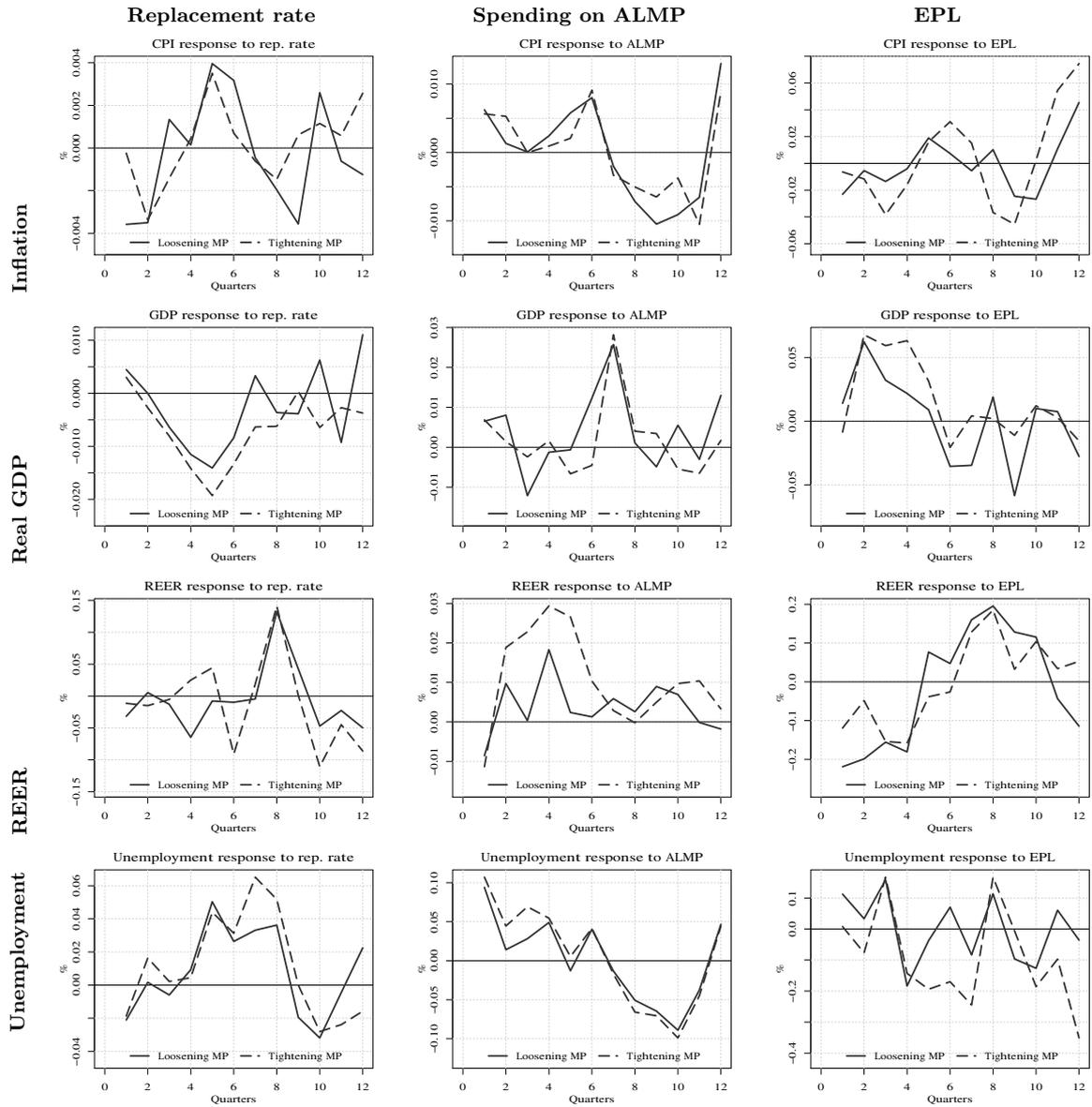

Figure C.2: Changes in macroeconomic variables due to the 1% increase in labor market policies (replacement rate, ALMP and EPL), weighted (Mallow's averaging and multi-factor error structure) impulse responses (local projections)

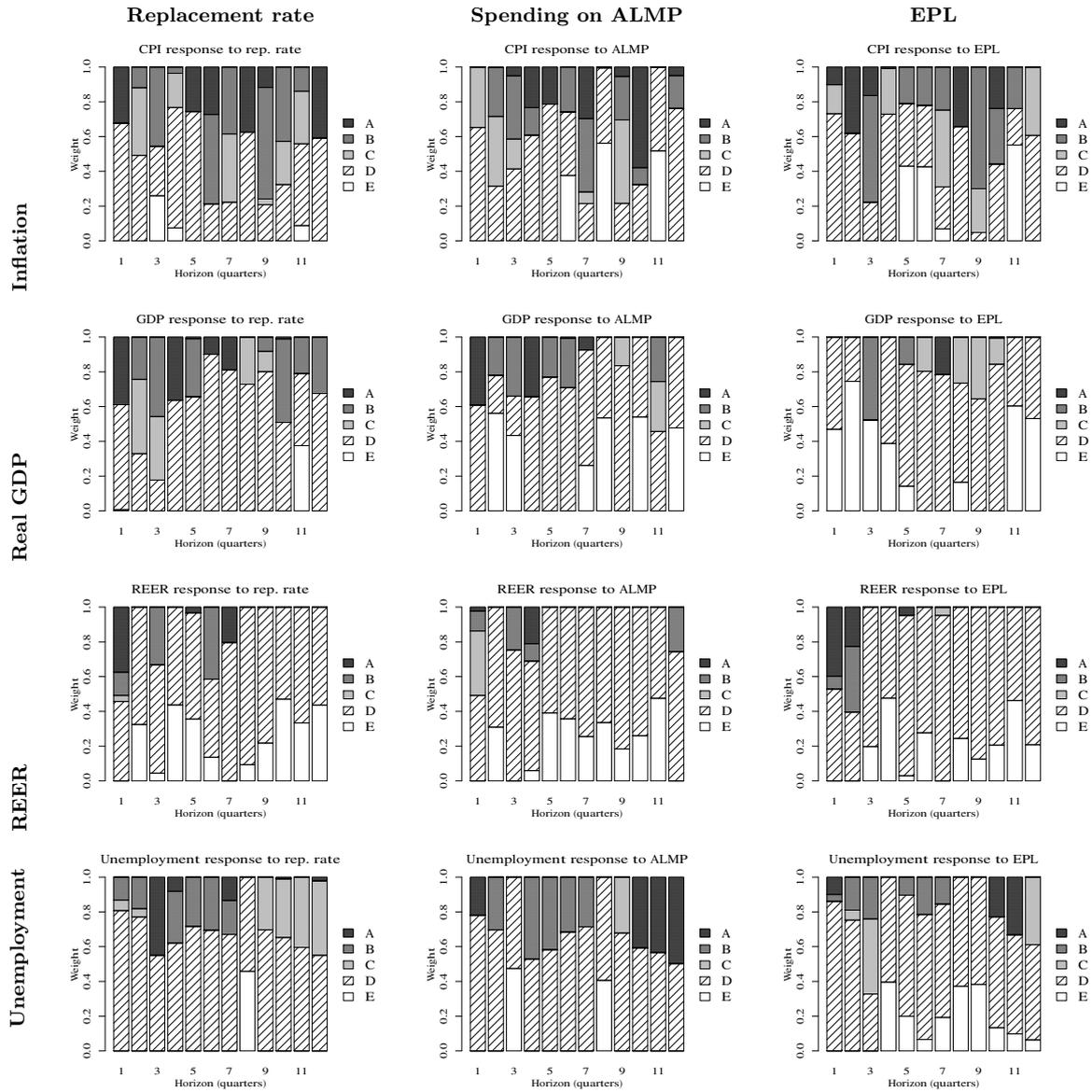

Figure C.3: Average model weights for each horizon, impulse and response variables (Mallow's criterion with multi-factor error structure and additional controls)

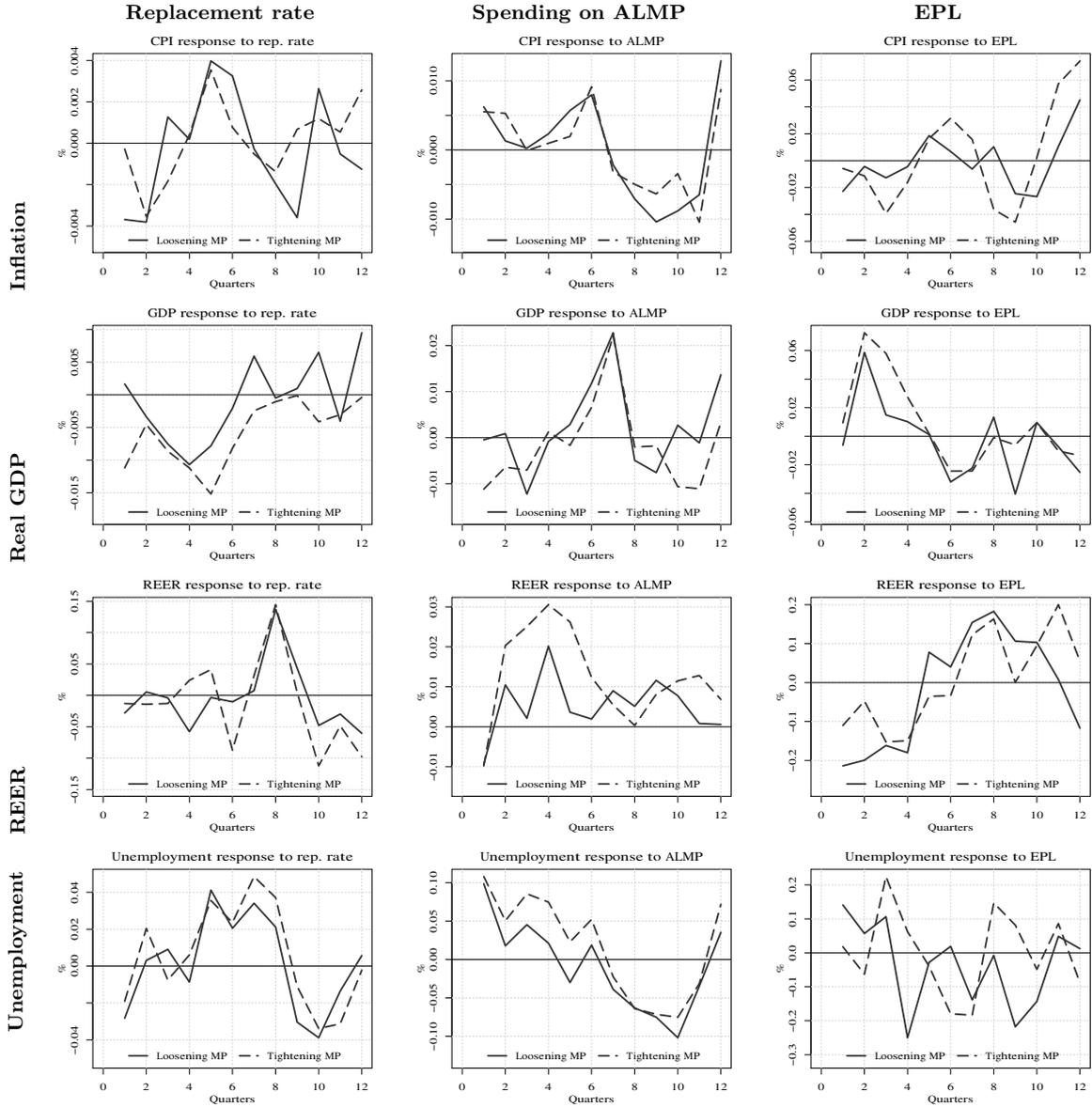

Figure C.4: Changes in macroeconomic variables due to the 1% increase in labor market policies (replacement rate, ALMP and EPL), weighted (Mallow's averaging, additional controls and multi-factor error structure) impulse responses (local projections)

## C.2 Results for the Extended Model

In this robustness check, we extend the set of controls to include GDP forecasts as well as output gaps, in addition to the multi-factor error structure. Figure C.3 is largely comparable to Figure C.1 when additional controls where not taken into account. visualizes Mallow's weights for the impulse response averaging, once error factor structure is incorporated into the weight-selection algorithm. Model D carries most weight for all variables of interest, though dispersion is quite vast and is clearly horizon, policy and outcome variable specific. Nevertheless, data clearly favor specifications the monetary policy trend rather than with more volatile alternatives.

The reported weights are used to construct average response functions, visualized in Figure C.4. Unsurprisingly, since weights remained quite comparable, so are the responses (refer to Figure C.2). Though much of variation in labor market policies is indeed attributable to a common factor, some visual differences remain. For instance, replacement rate is less detrimental whereas unemployment would decline faster and more strongly after an increase in ALMP if initiated during a loosening monetary policy period.

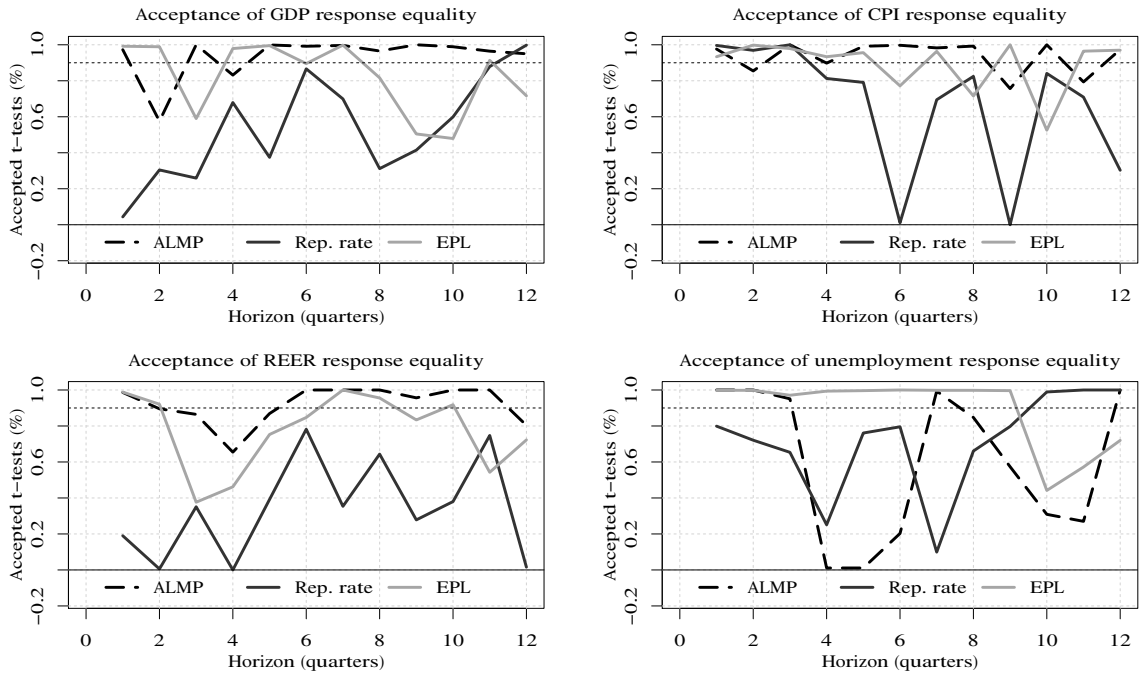

Figure C.5: Acceptance regions of $H_0 : \tilde{\delta}_k - \delta_k = 0$ for the replacement rate, ALMP and EPL in models for inflation, real GDP, REER and unemployment (with additional controls)

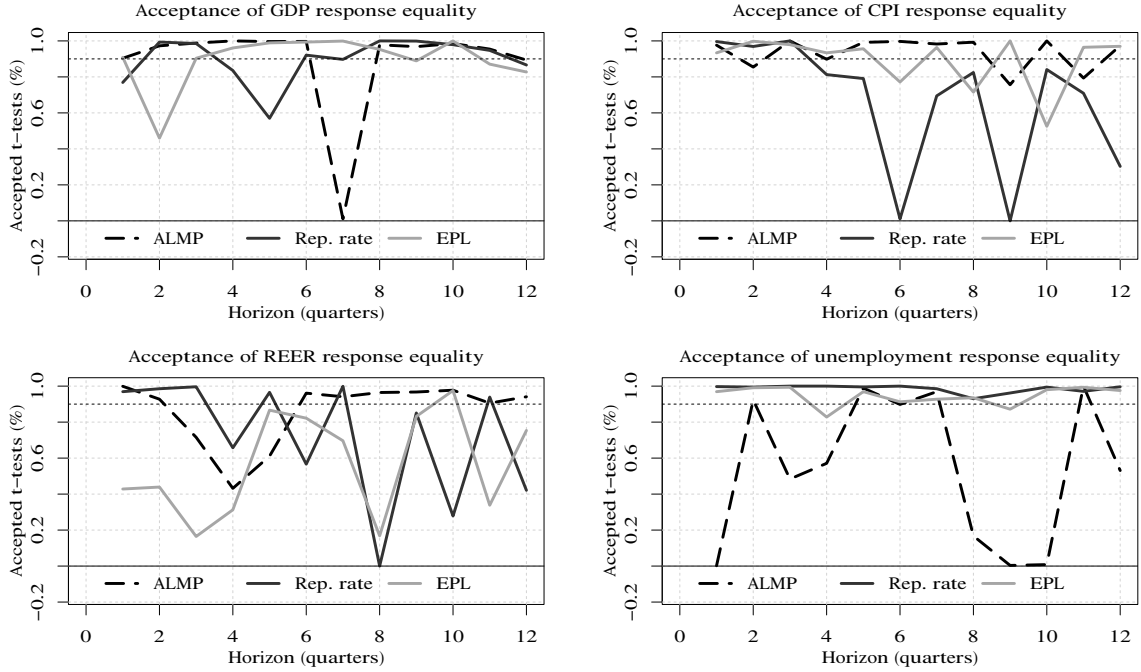

**Figure C.6:** Acceptance regions of $H_0 : \tilde{\delta}_k - \delta_k = 0$ for the replacement rate, ALMP and EPL in models for inflation, real GDP, REER and unemployment (with additional controls and the multi-factor error structure)

**Table C.1:** Proportion of the Equality Rejection with Adjusted *p*-values (left: averaging over $h$; right: averaging over $h$ and $t$)

|  |  | Bonferroni | Holm | Benjamini & Yekutieli | Bonferroni | Holm | Benjamini & Yekutieli |
|---|---|---|---|---|---|---|---|
| Real GDP | ALMP | 0.98 | 0.98 | 0.99 | 1 | 1 | 1 |
|  | Rep. rate | 0.66 | 0.64 | 0.64 | 0.77 | 0.77 | 0.68 |
|  | EPL | 0.94 | 0.94 | 0.94 | 0.99 | 0.99 | 0.98 |
| Inflation (CPI) | ALMP | 0.99 | 0.99 | 0.99 | 1 | 1 | 1 |
|  | Rep. rate | 0.86 | 0.85 | 0.88 | 0.98 | 0.98 | 0.95 |
|  | EPL | 0.96 | 0.96 | 0.97 | 0.99 | 0.99 | 0.98 |
| Unemployment | ALMP | 0.92 | 0.92 | 0.96 | 1 | 1 | 1 |
|  | Rep. rate | 0.84 | 0.82 | 0.83 | 0.96 | 0.96 | 0.89 |
|  | EPL | 0.97 | 0.96 | 0.98 | 1 | 1 | 1 |
| REER | ALMP | 0.99 | 0.99 | 0.99 | 1 | 1 | 1 |
|  | Rep. rate | 0.49 | 0.45 | 0.45 | 0.67 | 0.67 | 0.51 |
|  | EPL | 0.92 | 0.92 | 0.93 | 0.96 | 0.96 | 0.96 |

Figures C.5 and C.6 plot acceptance regions for models with additional controls and without and with the multi-factor error structure. They document, as in the baseline case, many cases falling below 0.9 threshold. A notable difference in Figure C.5 is in the effect on GDP, which is no clearly different for the replacement rate, unlike previous dominance by ALMP. Other effects remain comparable: monetary policy plays a role for the impact of the replacement rate on inflation and real effective exchange rate, and ALMP on unemployment (for the horizon specific results also refer to Supplementary Material A.1). Extending analysis to the multiple testing case, Table C.1

documents substantial effects from the monetary policy for the replacement rate across all outcome variables dynamic paths and, to a lesser extent, for the EPL and a path of REER as well as real GDP. Pooling over $h$ and $t$ points to significant effects for the replacement rate and real GDP as well as REER. As before, there is evidence for the different unemployment reactions to ALMP if one of the multiple testing methods was used.

**Takeaway 14. Robustness of Monetary Policy Impact** Including GDP forecasts and output gaps confirms that Model D consistently carries the most weight, highlighting the strong influence of monetary policy trends. The robustness check shows that replacement rates and ALMP have horizon- and monetary policy-specific impacts on GDP, unemployment, inflation, and REER, whereas pooled test confirms that the changes in replacement rate delivers different results depending on the monetary policy stance.

To save space, we report other results in Supplementary Material A.[31] We leave it to future research to delve into the precise nature and relationship of common factors in macroeconomic and policy variables.

---

[31]Reader is referred to online Supplementary Material for results from the extended model, including probability values from all the models, all the models and all the extra parameters, in addition to the first two lags of the labor market policies.

# D Baseline Model

## D.1 90% Confidence Intervals

### D.1.1 Sub-sample 1985-1998

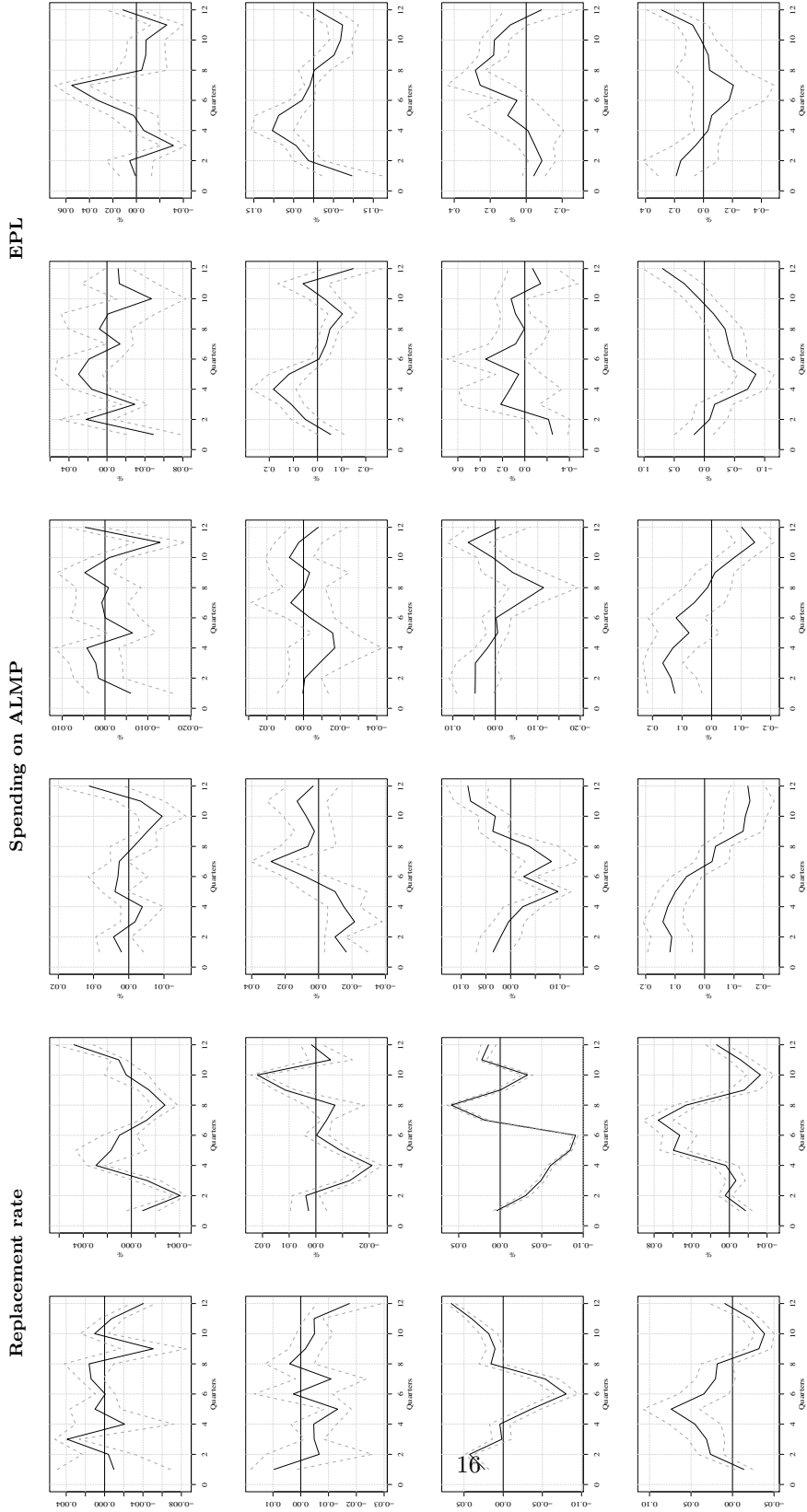

Figure D.1: Changes in macroeconomic variables due to the 1% increase in labor market policies (replacement rate, ALMP and EPL), falling interest rate (left) and rising interest rate (right), model A

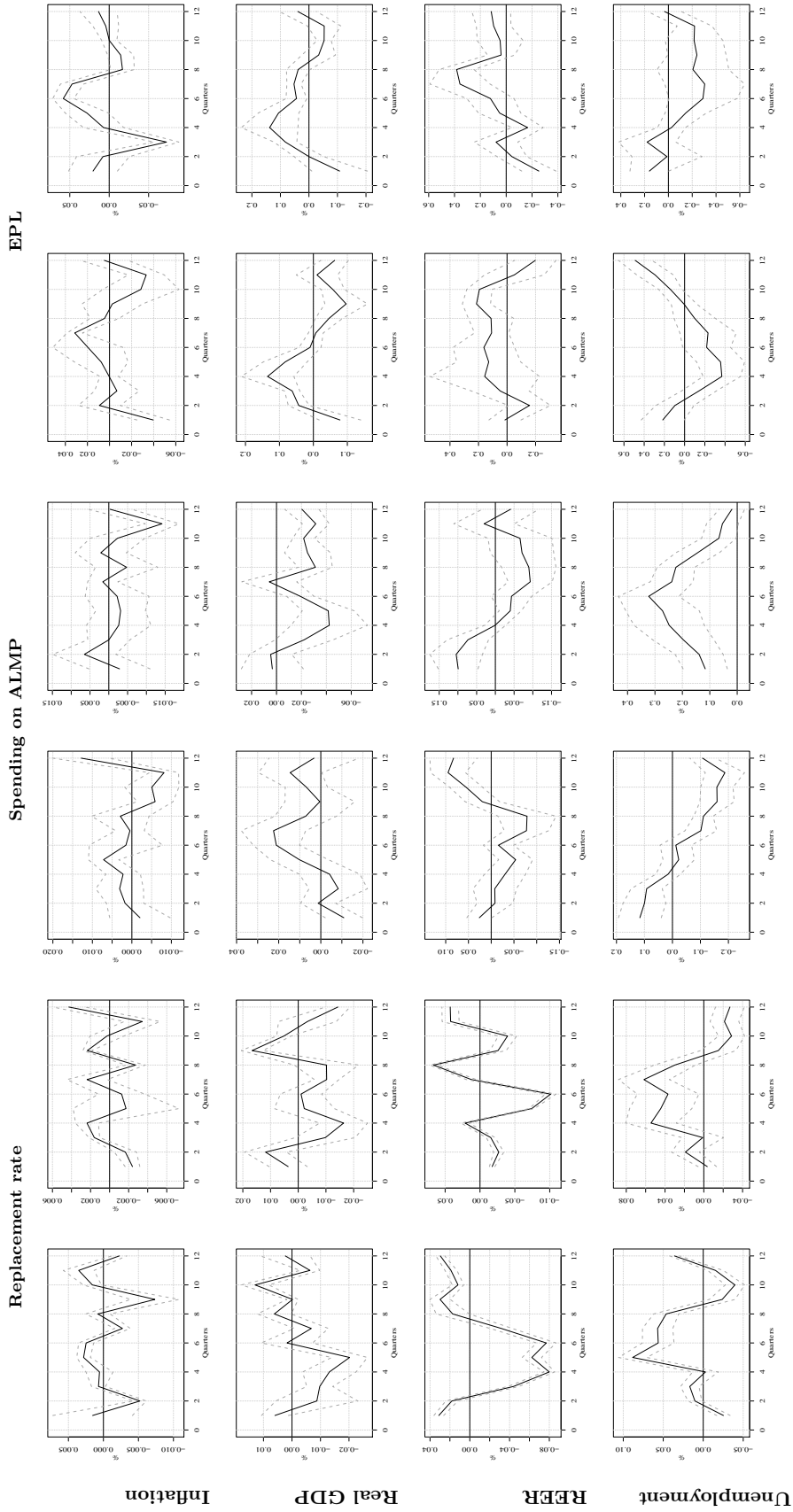

Figure D.2: Changes in macroeconomic variables due to the 1% increase in labor market policies (replacement rate, ALMP and EPL), falling interest rate (left) and rising interest rate (right), model B

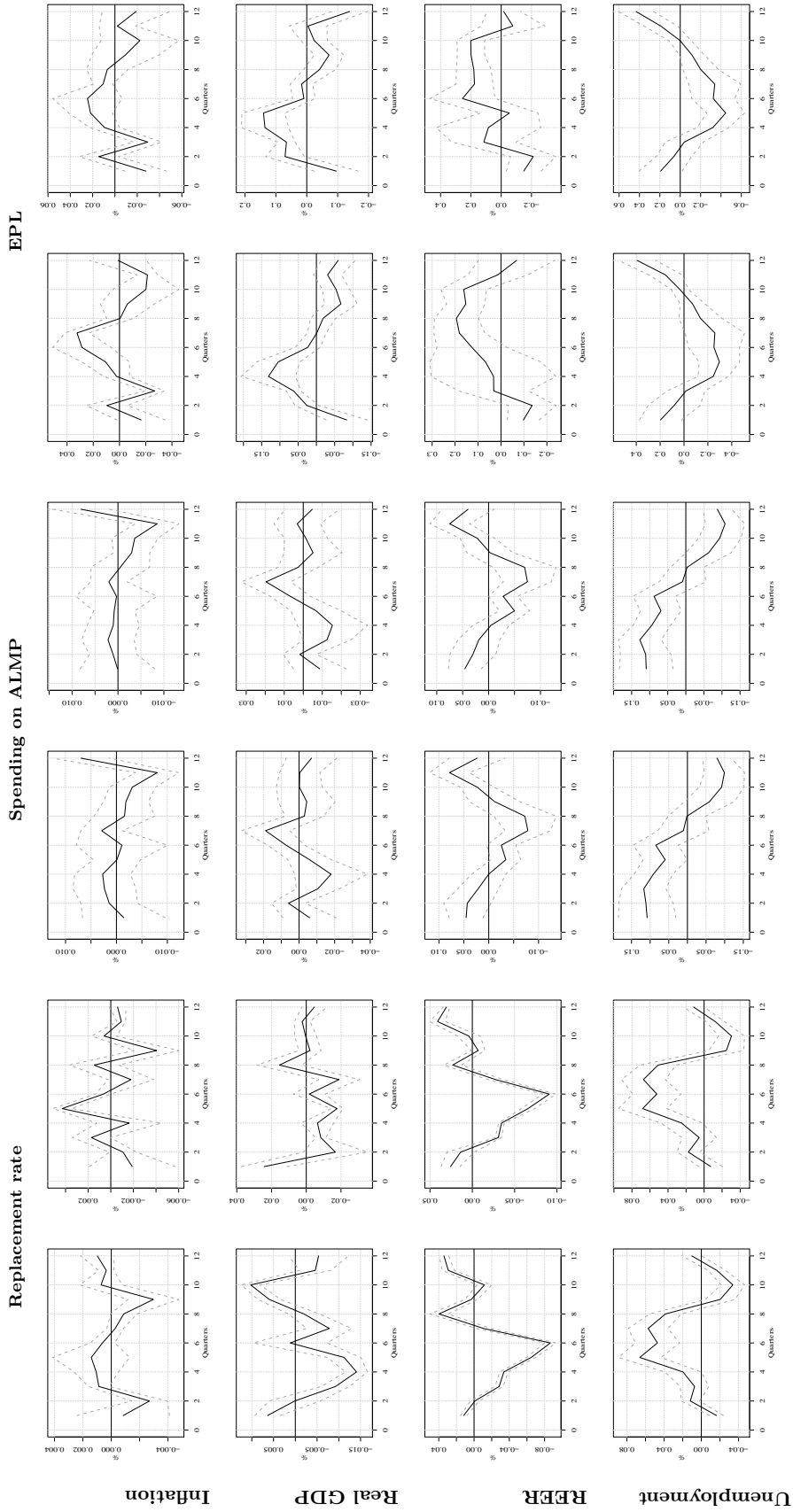

Figure D.3: Changes in macroeconomic variables due to the 1% increase in labor market policies (replacement rate, ALMP and EPL), first quarterly quartile (left) and third quarterly quartile (right), model C

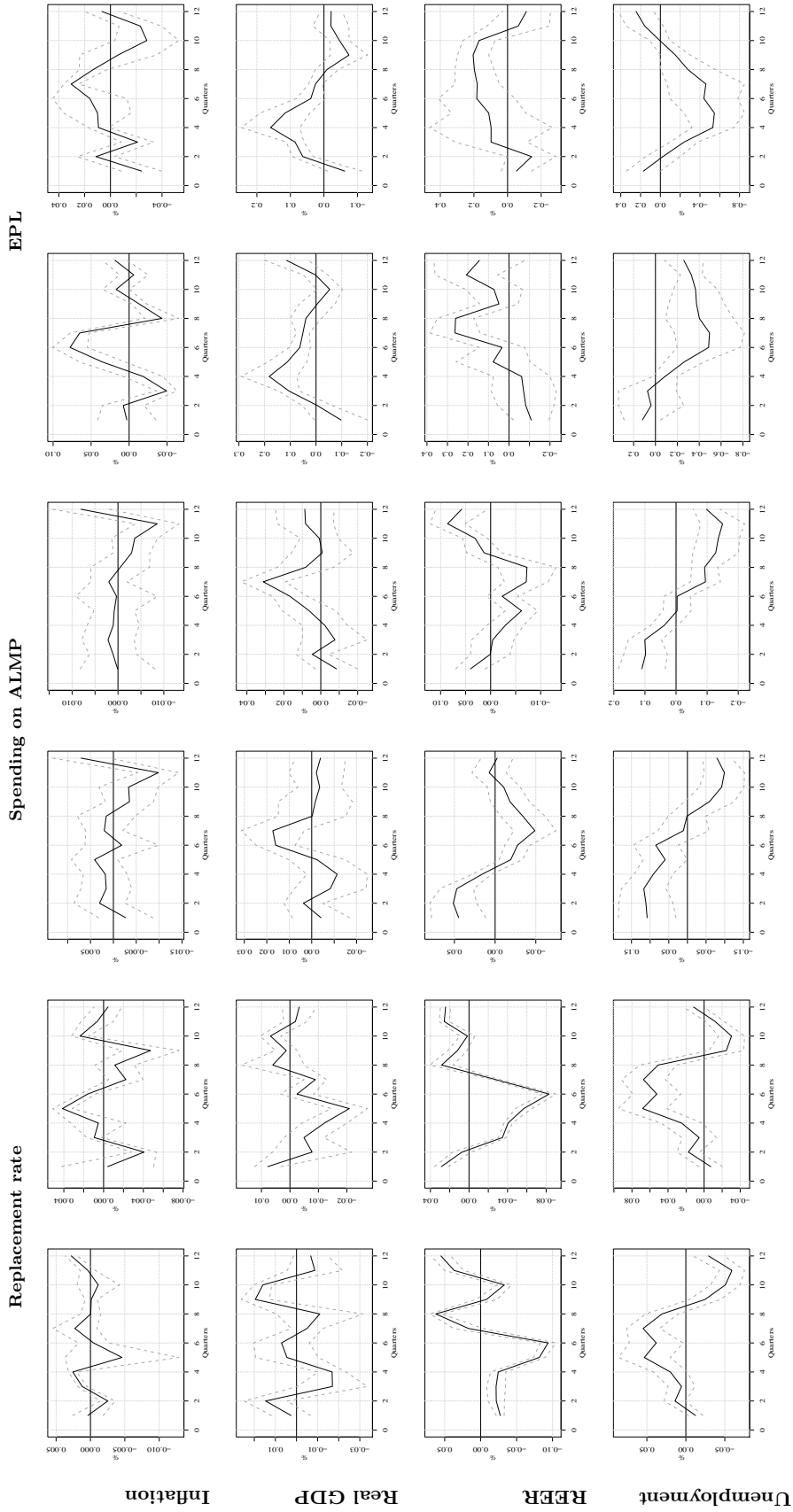

Figure D.4: Changes in macroeconomic variables due to the 1% increase in labor market policies (replacement rate, ALMP and EPL), first annual quartile (left) and third annual quartile (right), model D

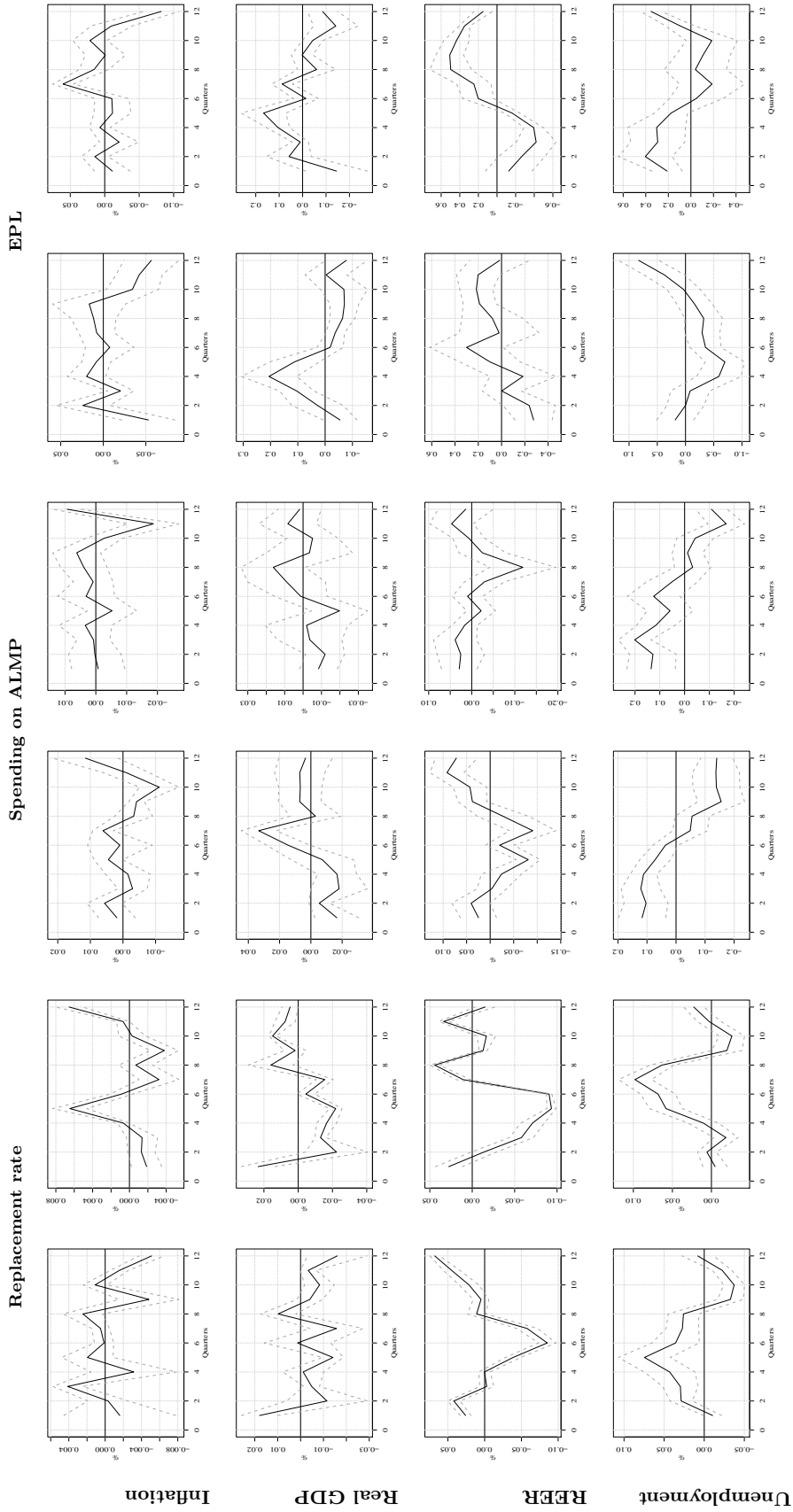

Figure D.5: Changes in macroeconomic variables due to the 1% increase in labor market policies (replacement rate, ALMP and EPL), falling interest rate (left) and rising interest rate (right), model E

## D.1.2 Sub-sample 1999-2010

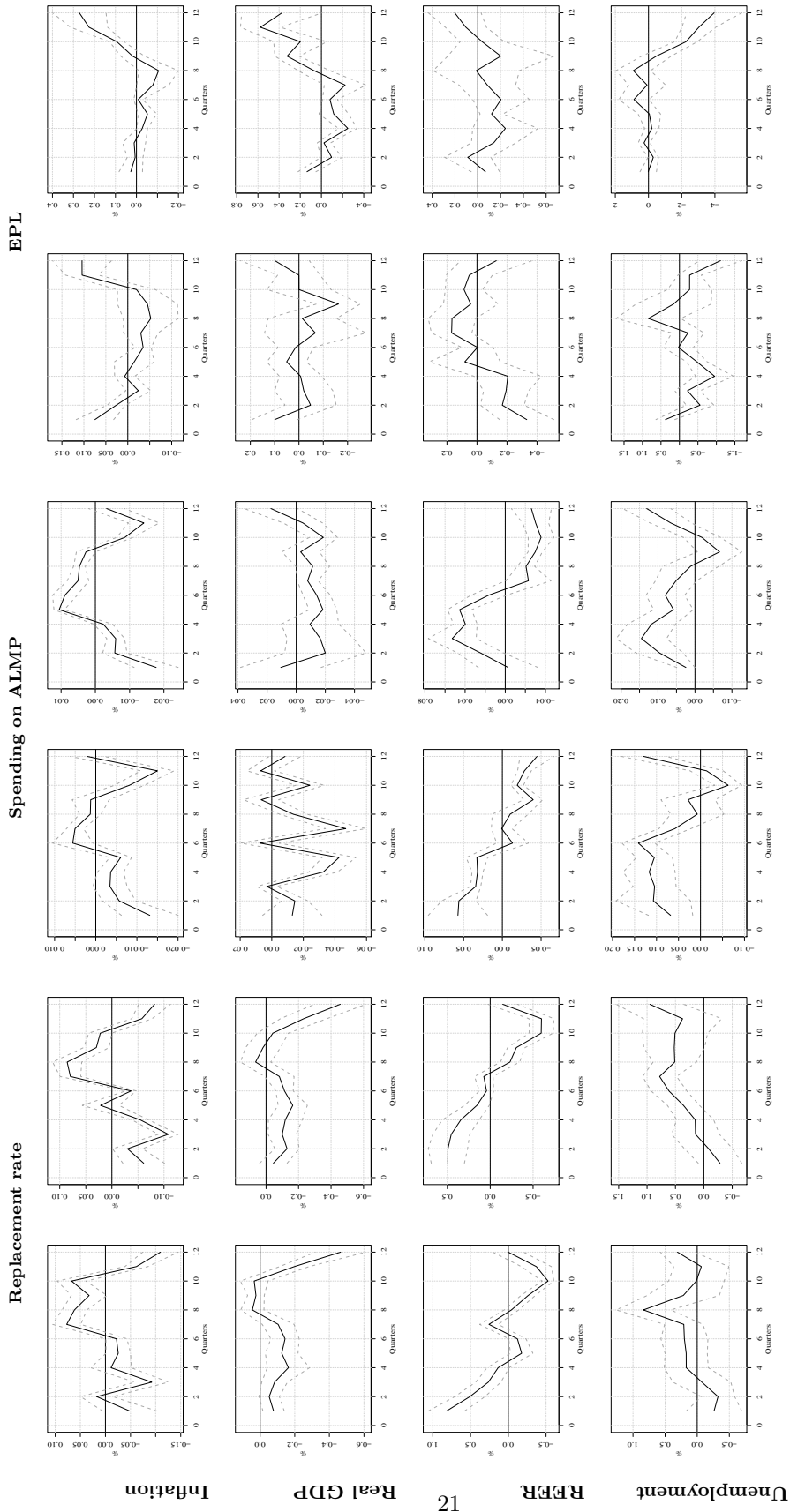

Figure D.6: Changes in macroeconomic variables due to the 1% increase in labor market policies (replacement rate, ALMP and EPL), falling interest rate (left) and rising interest rate (right), model A

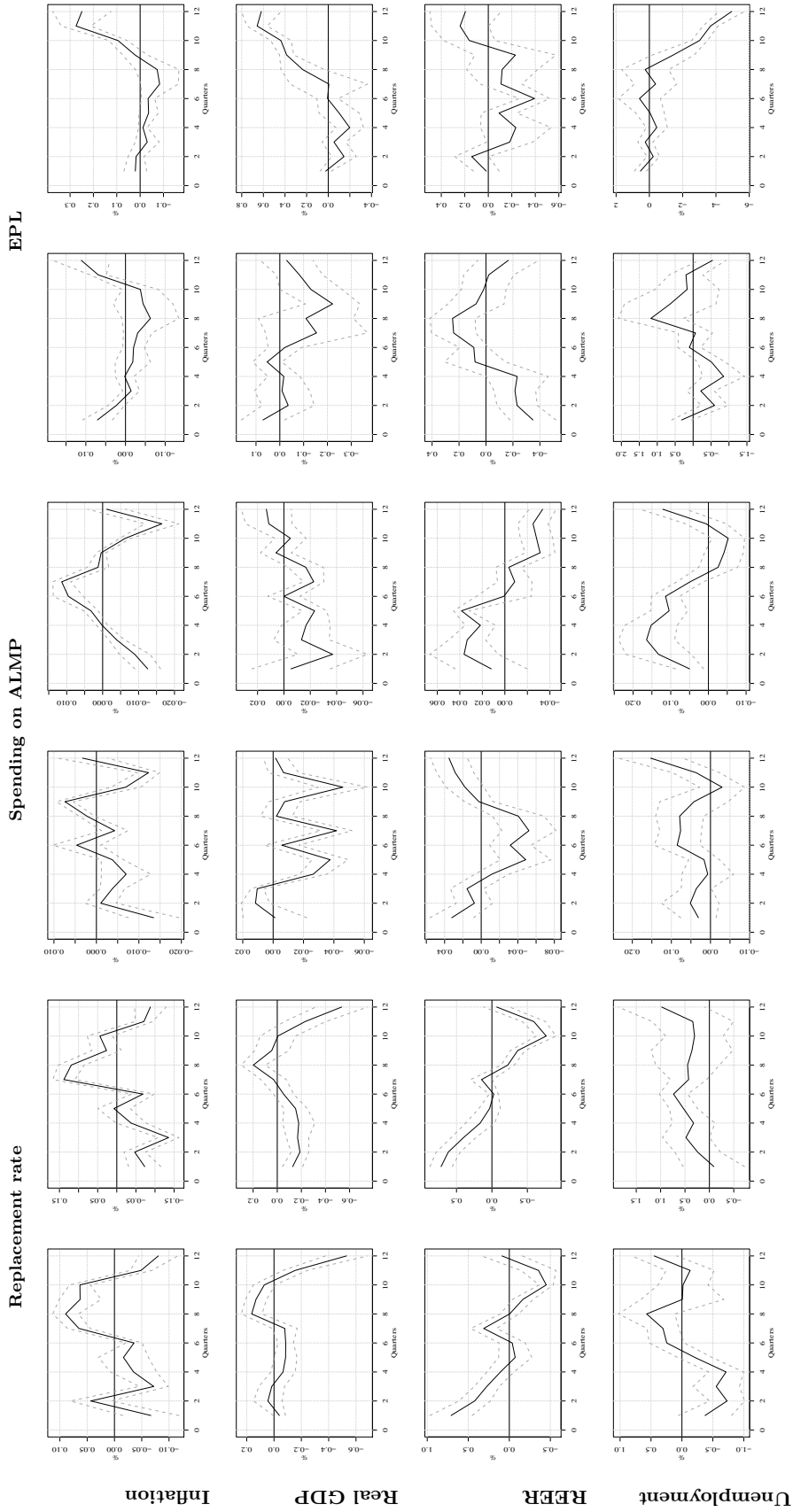

Figure D.7: Changes in macroeconomic variables due to the 1% increase in labor market policies (replacement rate, ALMP and EPL), falling interest rate (left) and rising interest rate (right), model B

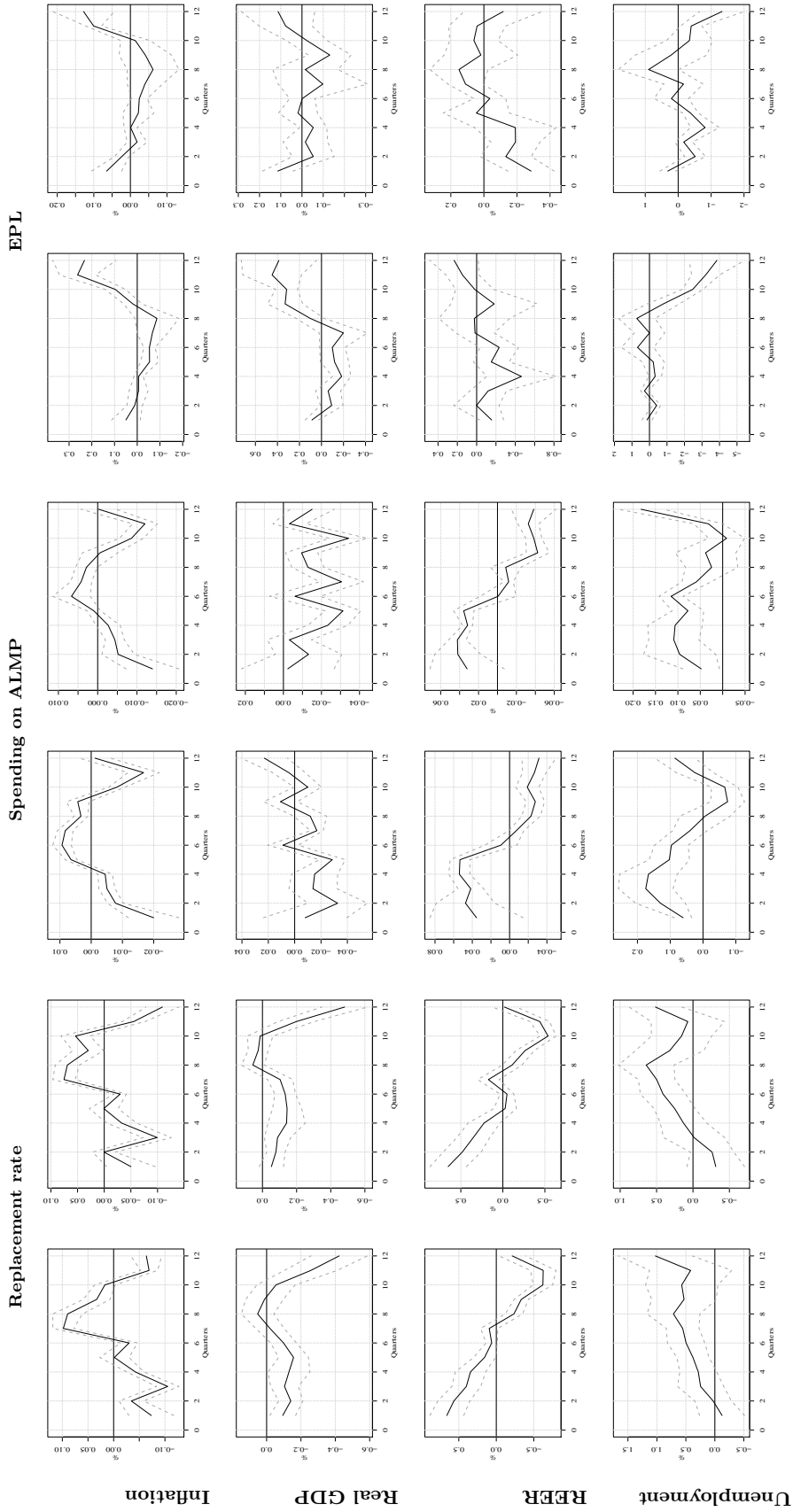

Figure D.8: Changes in macroeconomic variables due to the 1% increase in labor market policies (replacement rate, ALMP and EPL), first quarterly quartile (left) and third quarterly quartile (right), model C

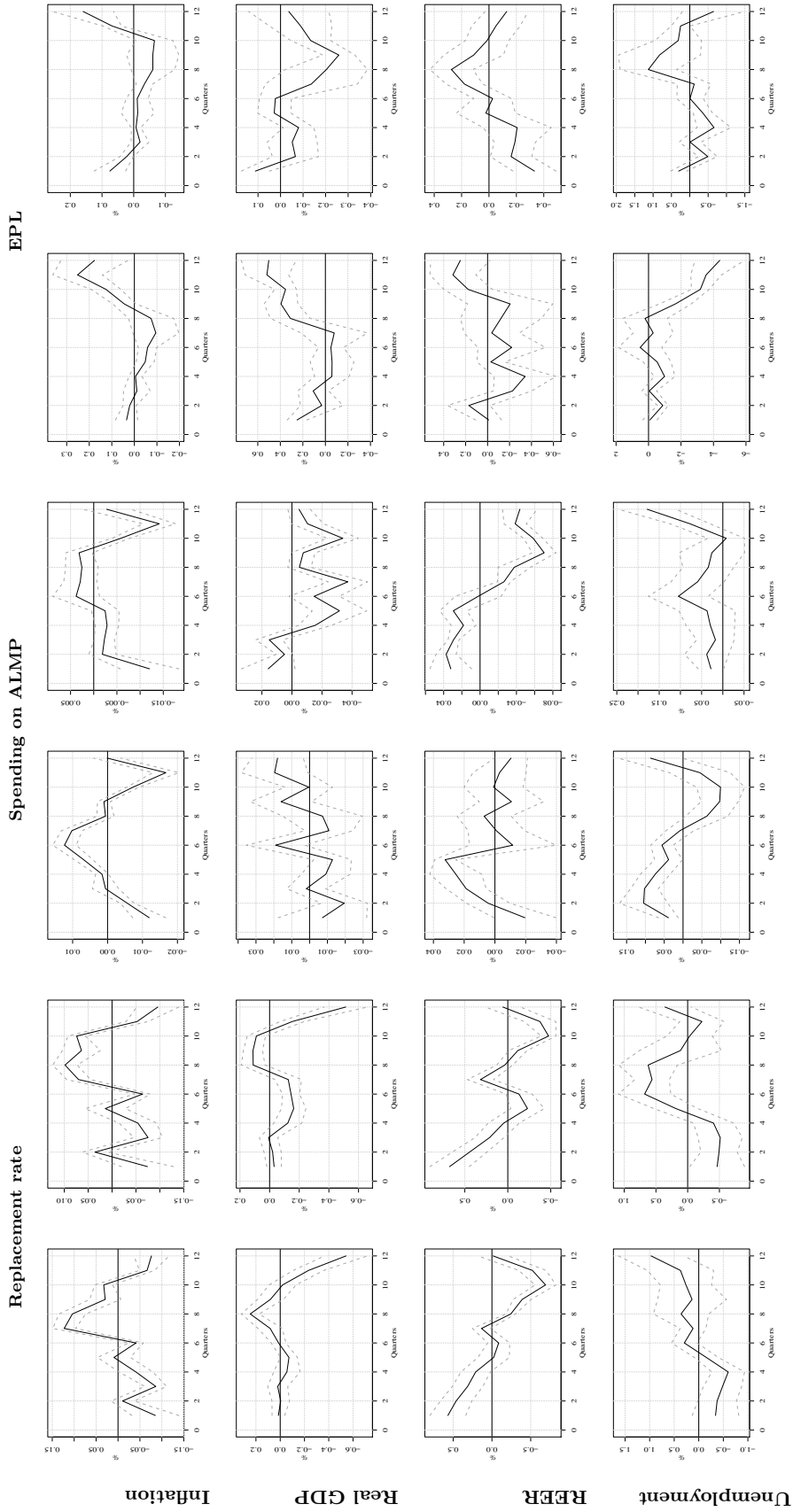

Figure D.9: Changes in macroeconomic variables due to the 1% increase in labor market policies (replacement rate, ALMP and EPL), first annual quartile (left) and third annual quartile (right), model D

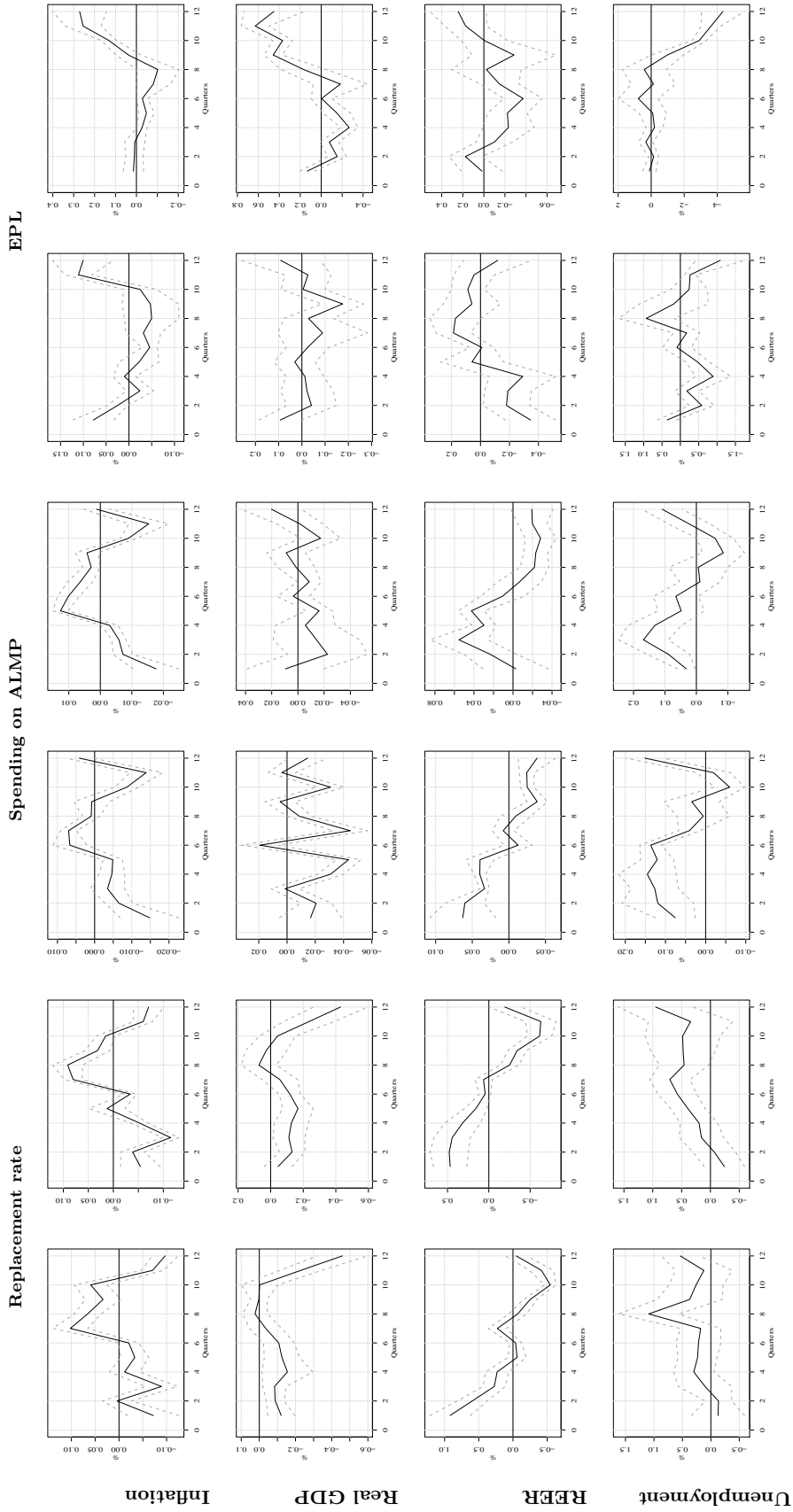

Figure D.10: Changes in macroeconomic variables due to the 1% increase in labor market policies (replacement rate, ALMP and EPL), falling interest rate (left) and rising interest rate (right), model E

### D.1.3 Full Sample

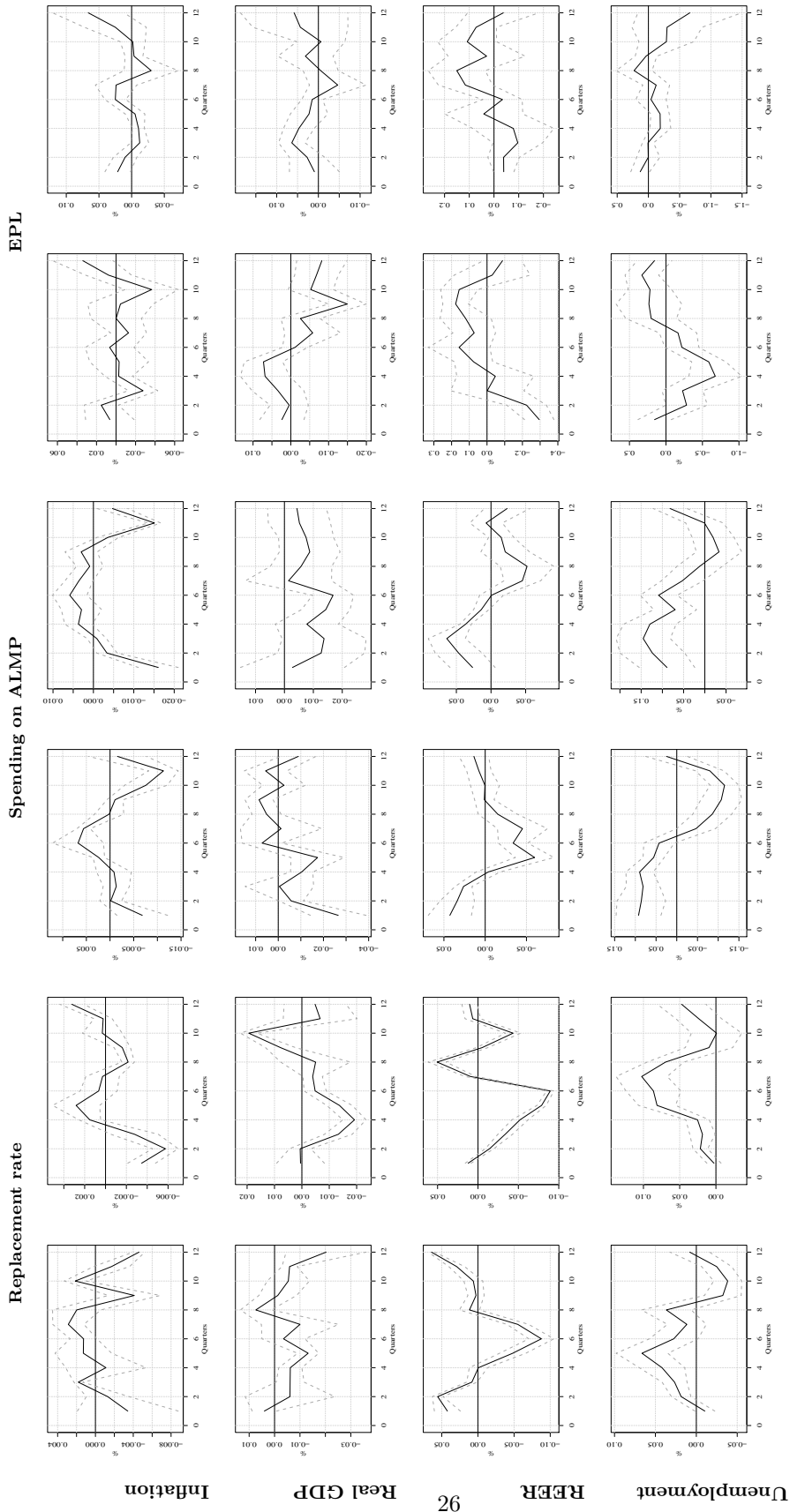

Figure D.11: Changes in macroeconomic variables due to the 1% increase in labor market policies (replacement rate, ALMP and EPL), falling interest rate (left) and rising interest rate (right), model A

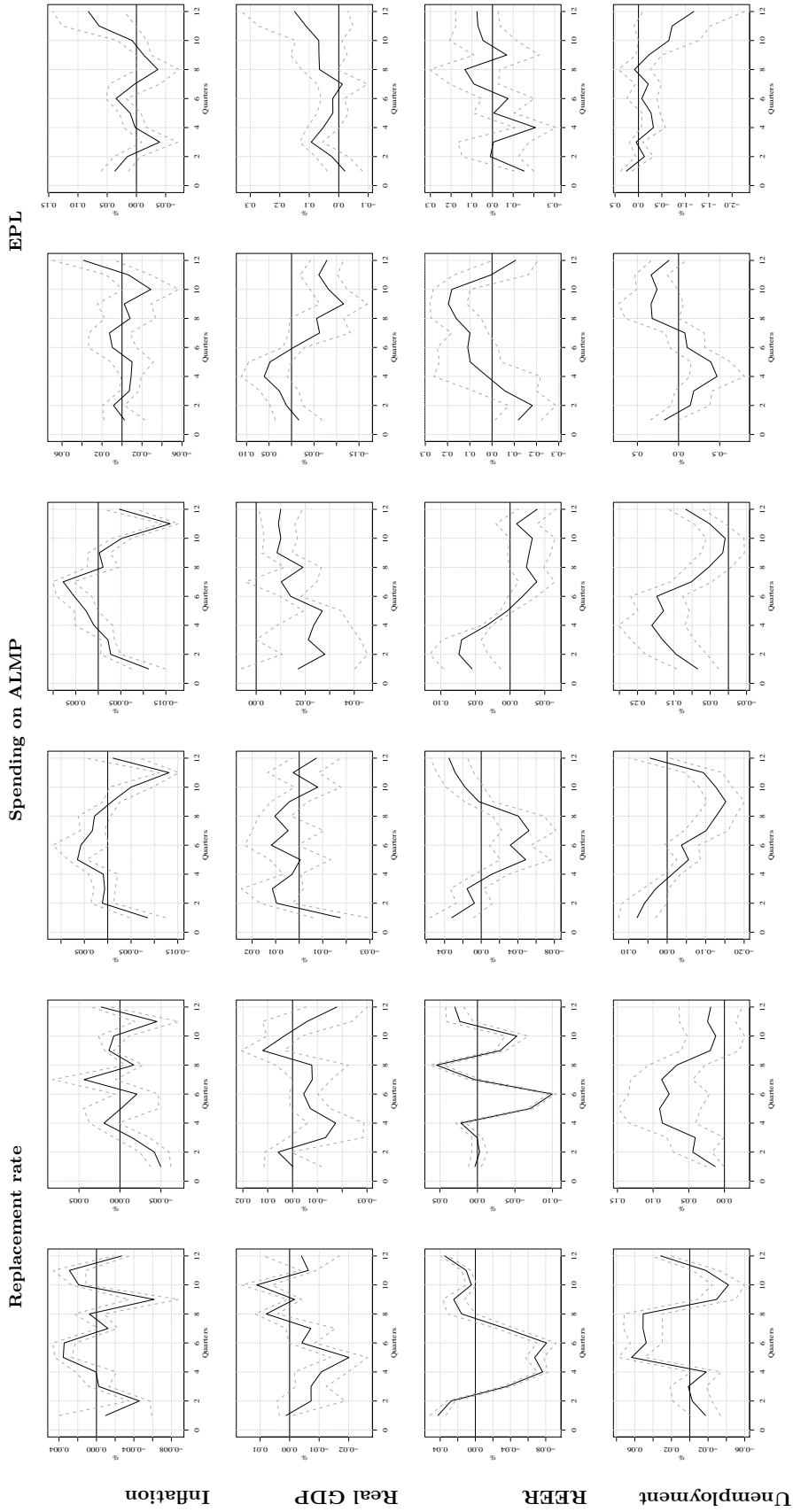

Figure D.12: Changes in macroeconomic variables due to the 1% increase in labor market policies (replacement rate, ALMP and EPL), falling interest rate (left) and rising interest rate (right), model B

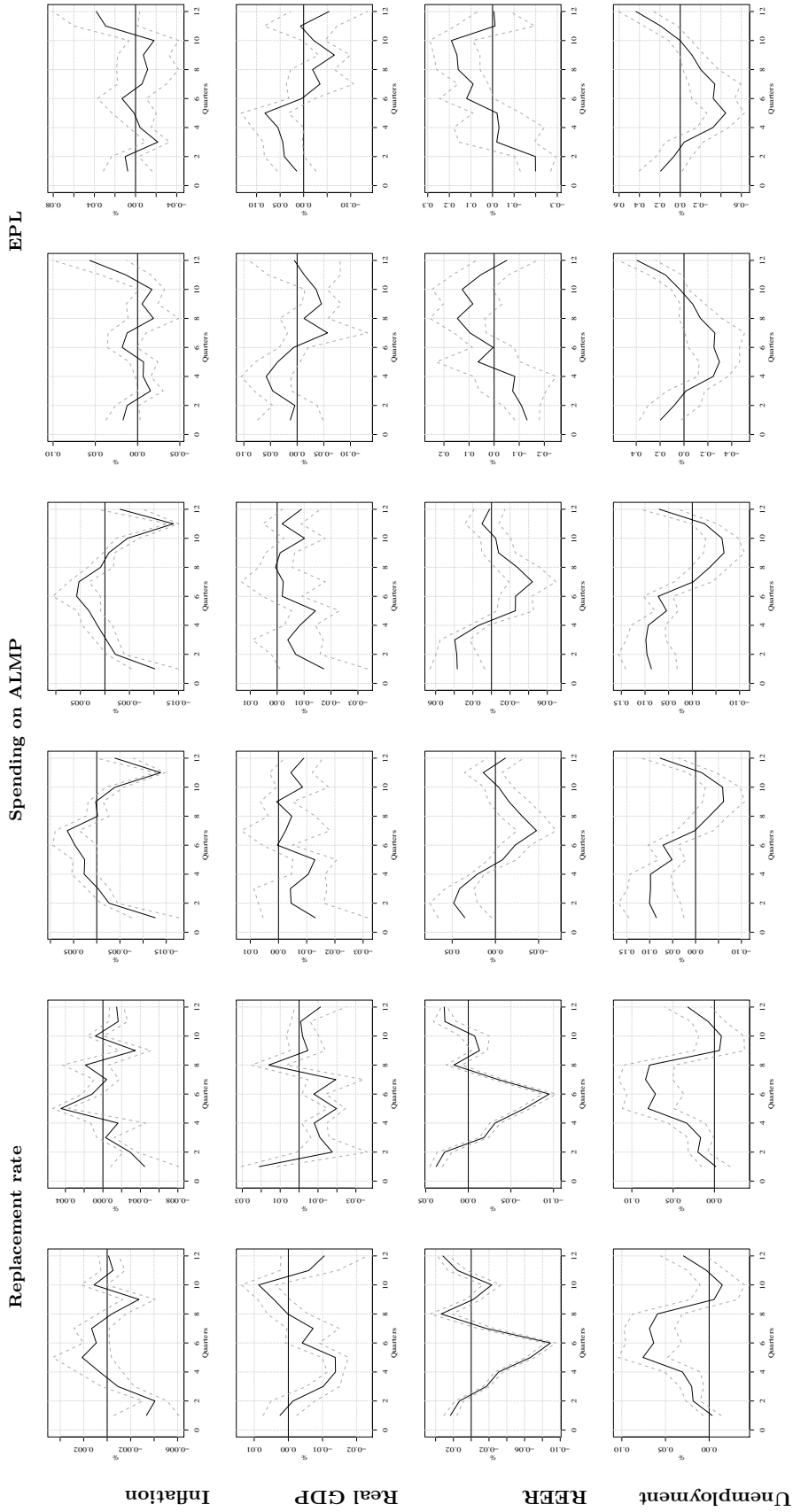

Figure D.13: Changes in macroeconomic variables due to the 1% increase in labor market policies (replacement rate, ALMP and EPL), first quarterly quartile (left) and third quarterly quartile (right), model C

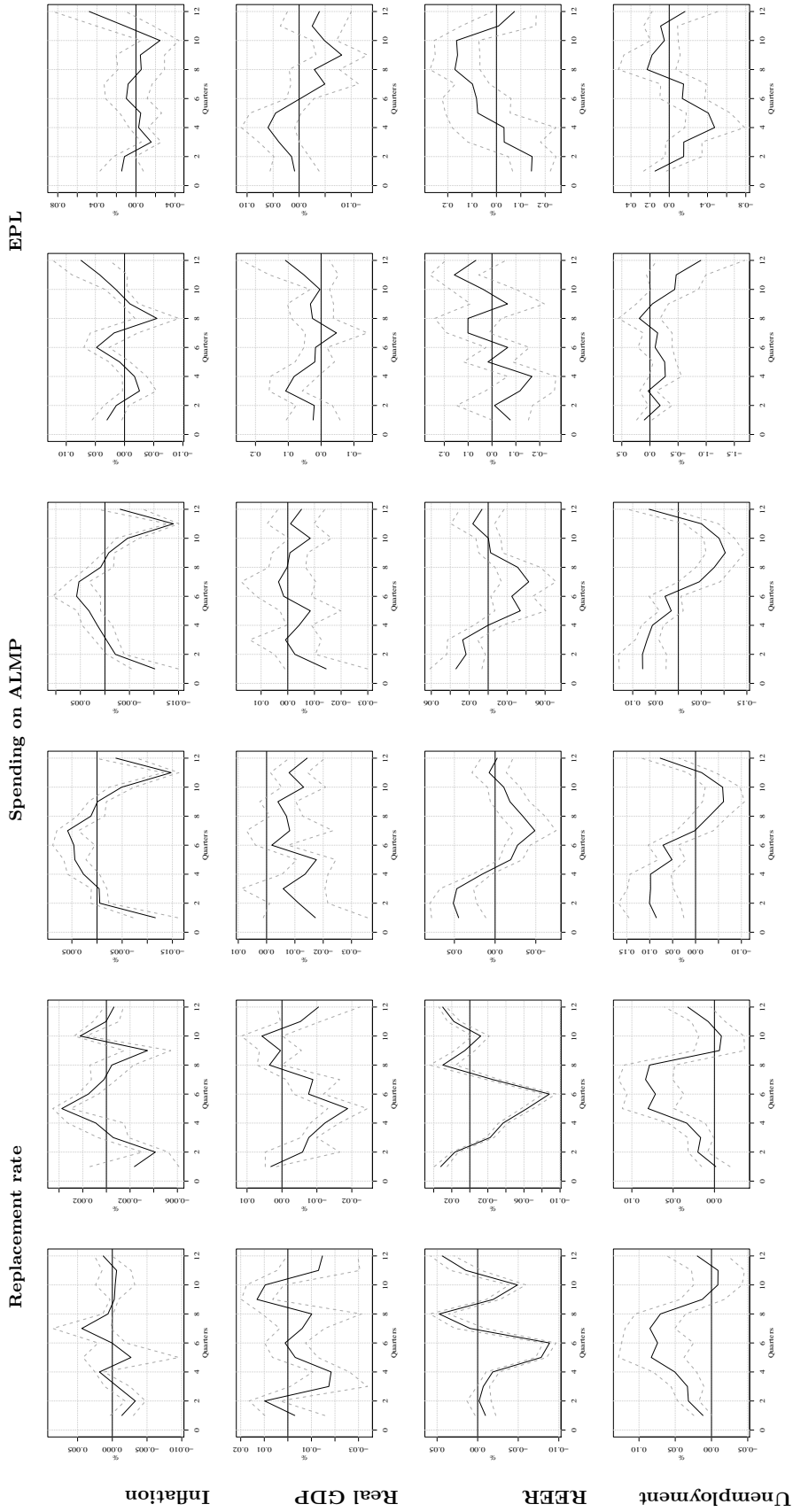

Figure D.14: Changes in macroeconomic variables due to the 1% increase in labor market policies (replacement rate, ALMP and EPL), first annual quartile (left) and third annual quartile (right), model D

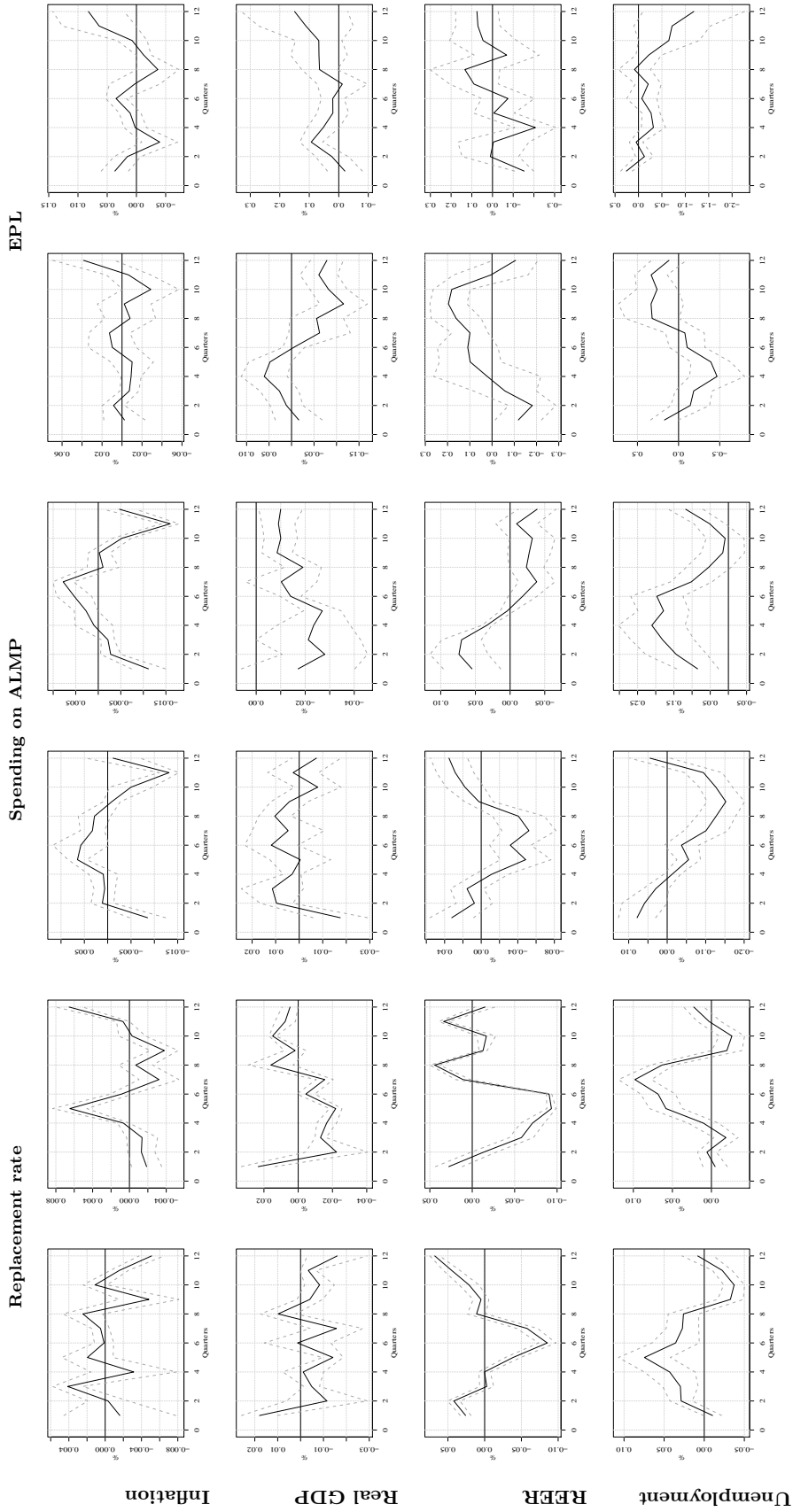

Figure D.15: Changes in macroeconomic variables due to the 1% increase in labor market policies (replacement rate, ALMP and EPL), falling interest rate (left) and rising interest rate (right), model E

## D.2 Probability Values

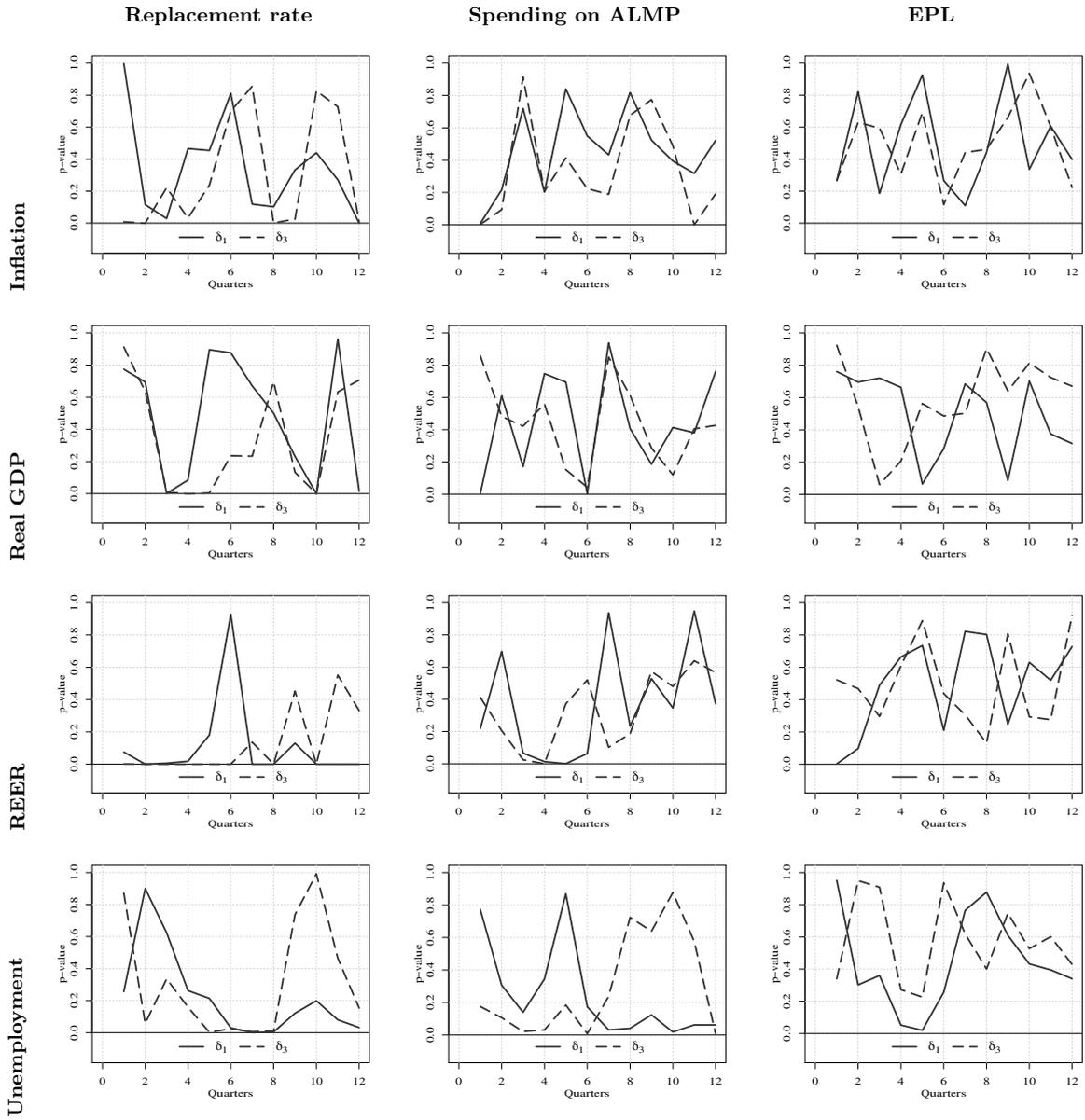

Figure D.16: P-values of the parameter equality, 1985-2010, model A

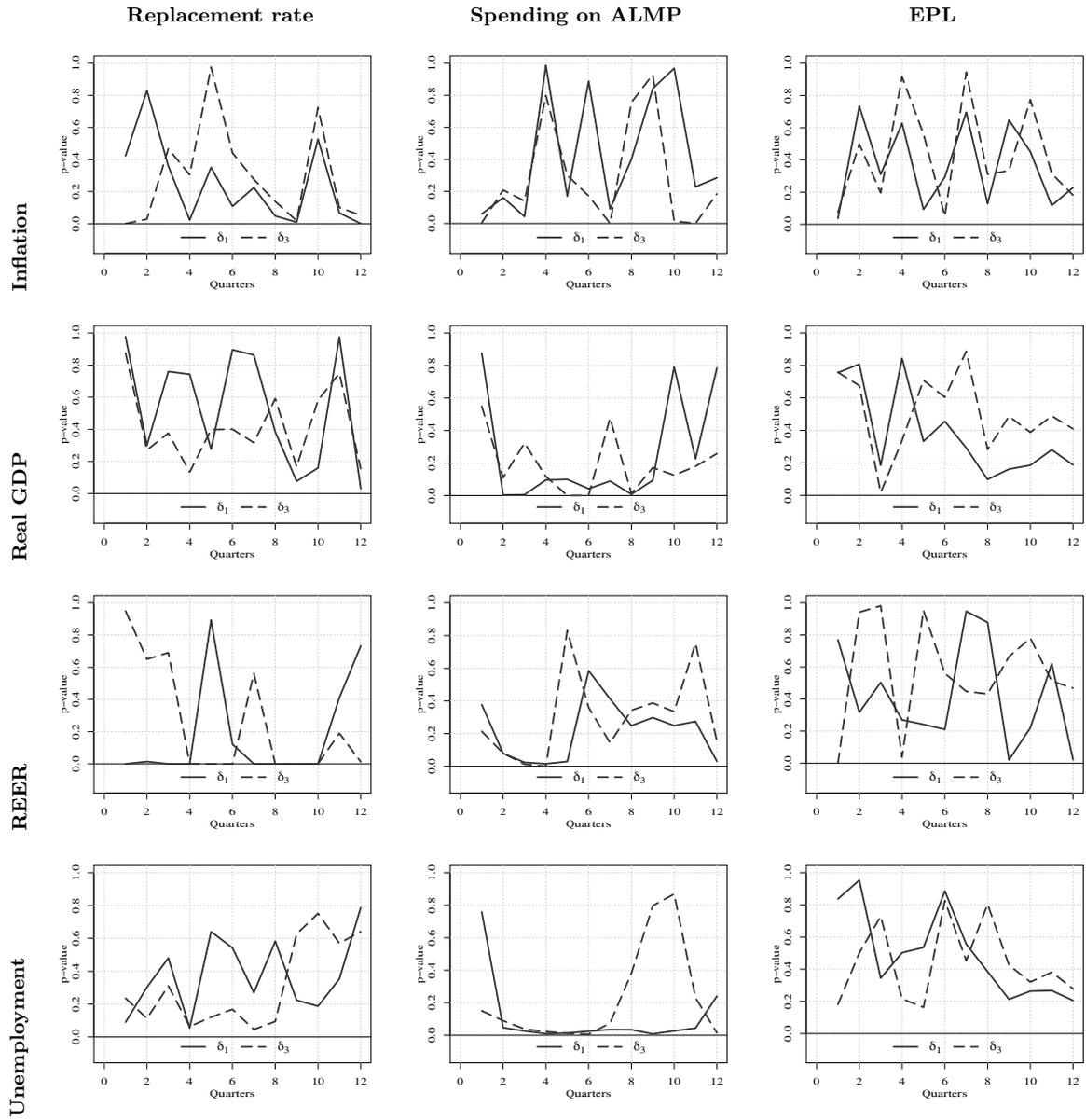

Figure D.17: P-values of the parameter equality, 1985-2010, model B

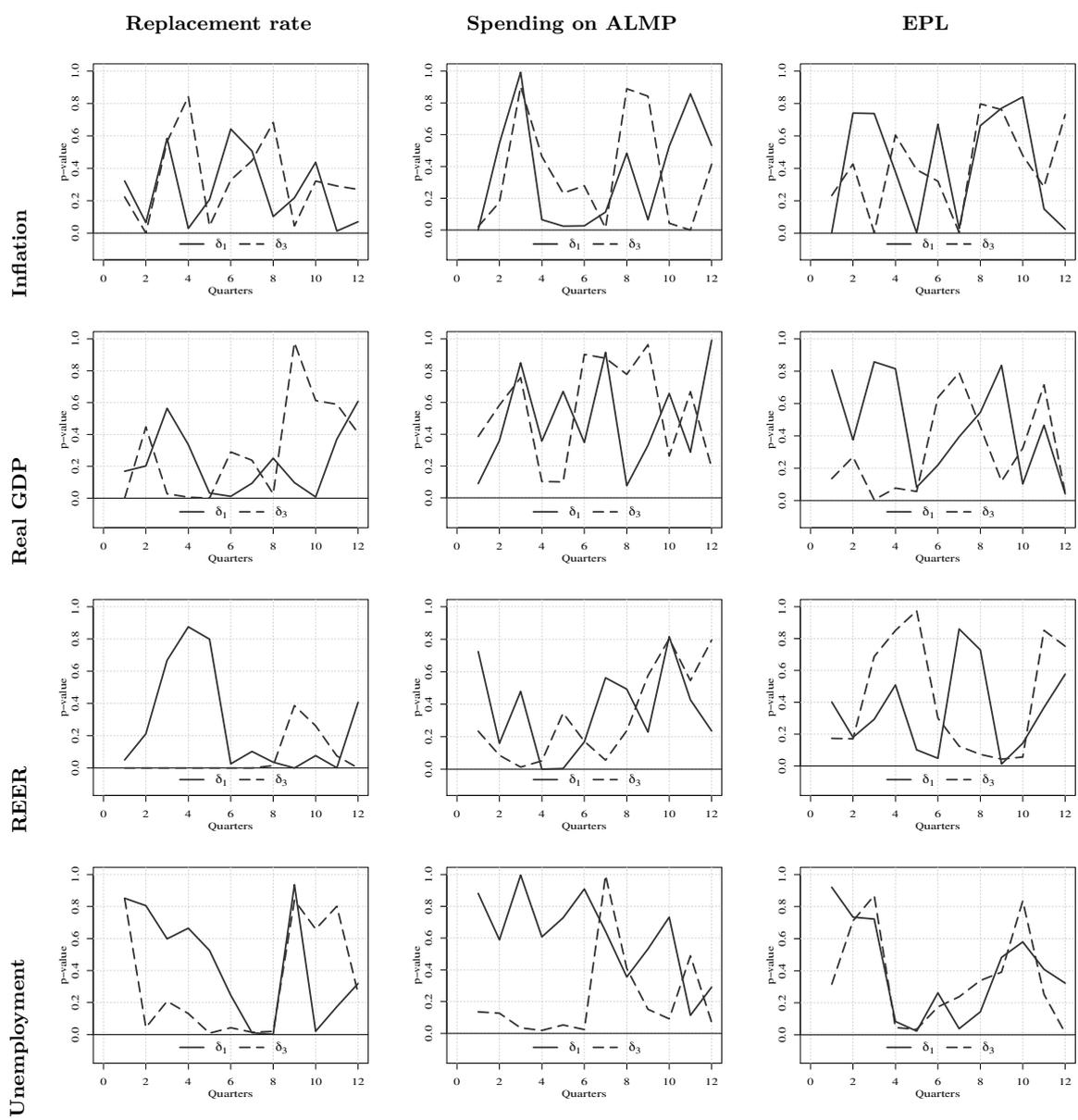

Figure D.18: P-values of the parameter equality, 1985-2010, model C

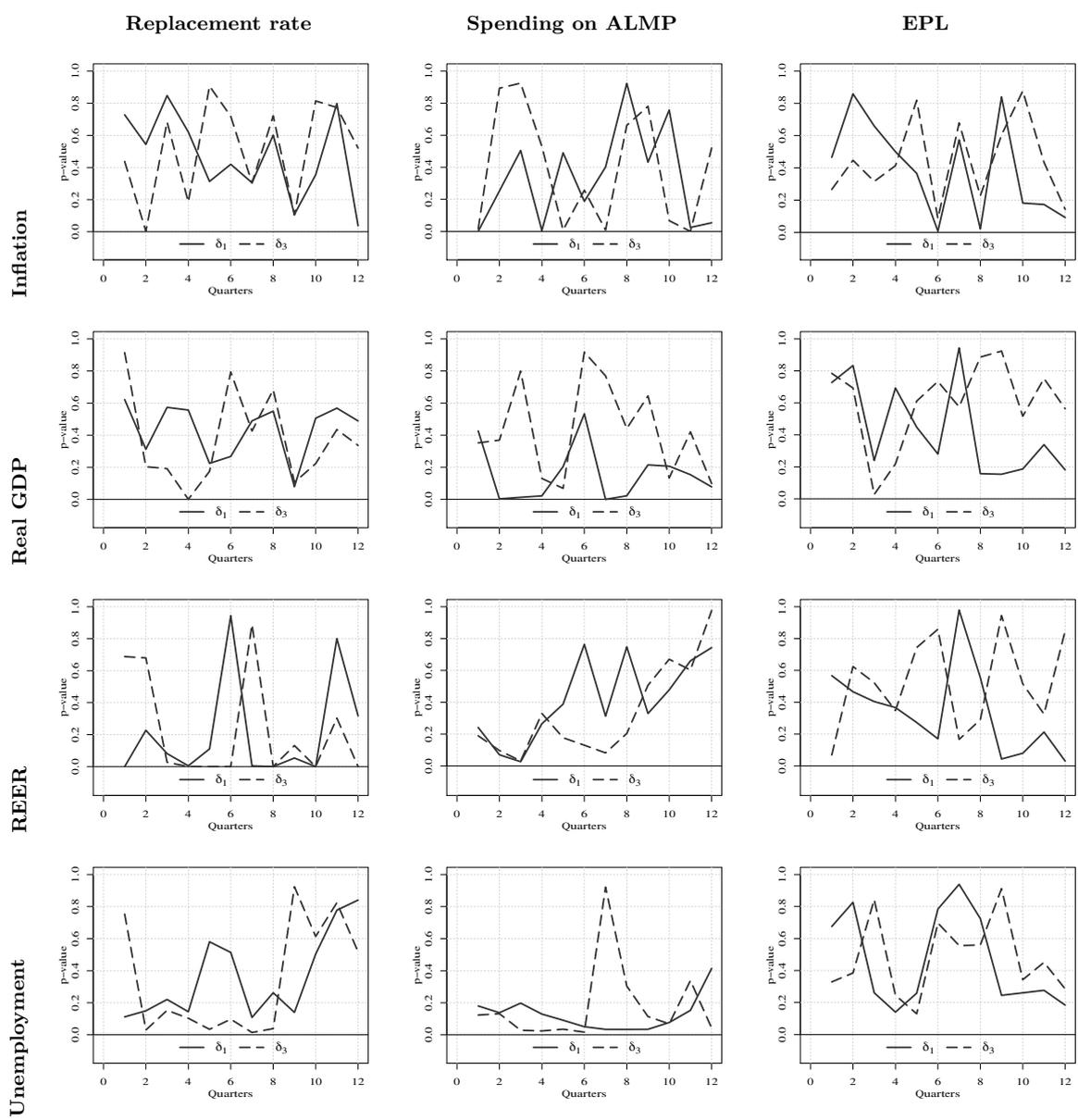

Figure D.19: P-values of the parameter equality, 1985-2010, model D

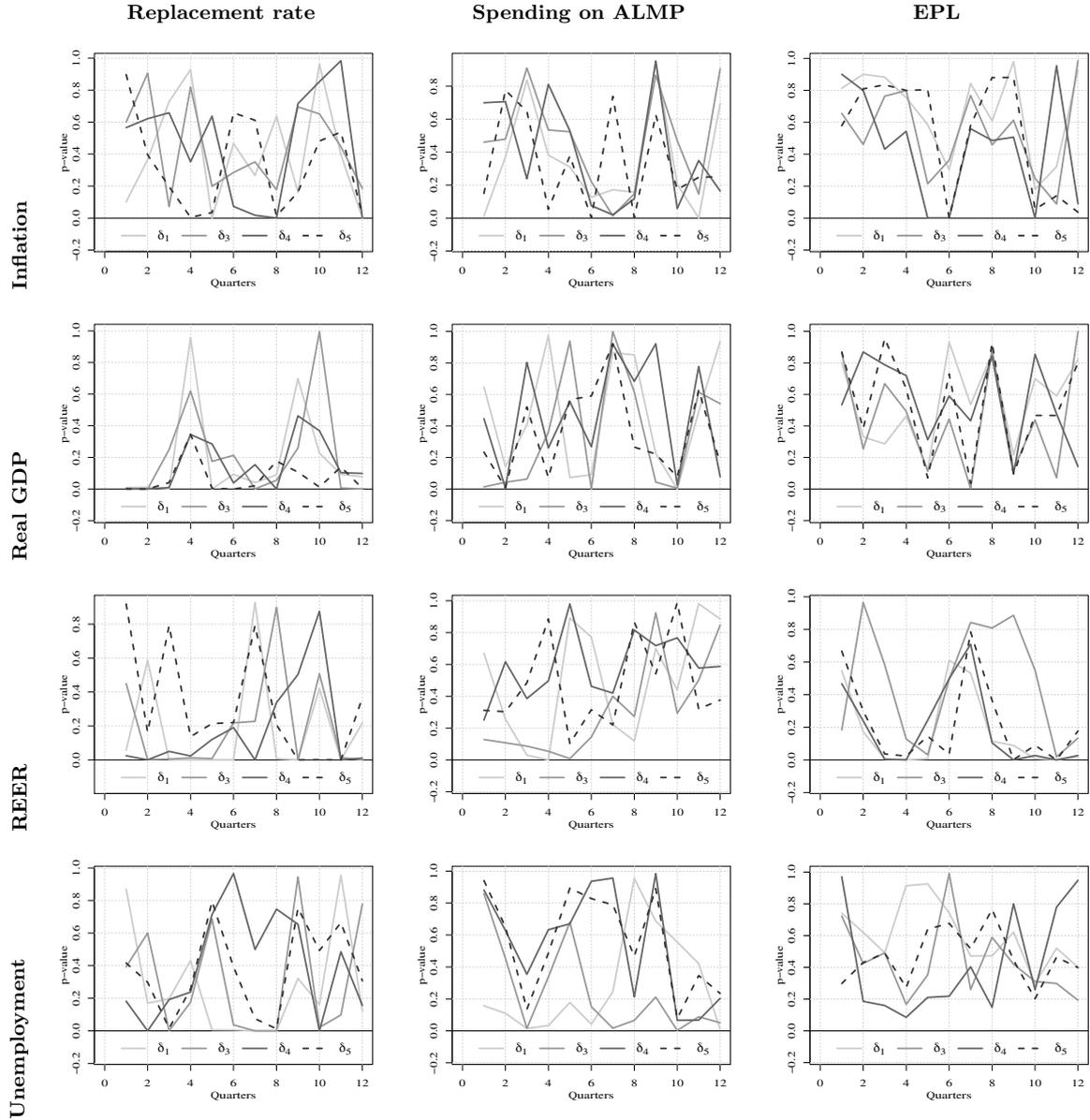

**Figure D.20: P-values of the parameter equality, 1985-2010, model E**

# E  Time and Country Coverage Extension

In this section, we replicate the main results while expanding the time and country coverage. Most data extend up to and include Q4 2020, with a few exceptions (e.g., OECD EPL data is available only until 2019). We have also expanded the country coverage to include Luxembourg and new euro area members: Estonia, Latvia, Lithuania, Slovakia, and Slovenia. The temporal coverage for these new states varies considerably and depends on the specific variable. Generally, the maximum time coverage is 1985-2020 for 17 euro area countries.

All data series, except for output gap, were extended keeping the original data and using the percentage changes observed in the new data to prolong the data series. Output gap data was replaced entirely with a new estimate based on a longer sample. This was done in order to avoid breaks in the data as the filtered estimate becomes less reliable around the edges of the sample, so new and old data cannot be joined properly.

Due to changes in OECD data collection, extended data for the unemployment replacement rate is based on net replacement rates instead of gross rates as it was in the original series. We replicated the original definition of the composite replacement rate indicator computing average replacement rate for two earnings levels, three family situations and three durations of unemployment.

## E.1 Confidence Intervals: An Expanded Sample

### E.1.1 Sub-sample 1999-2020

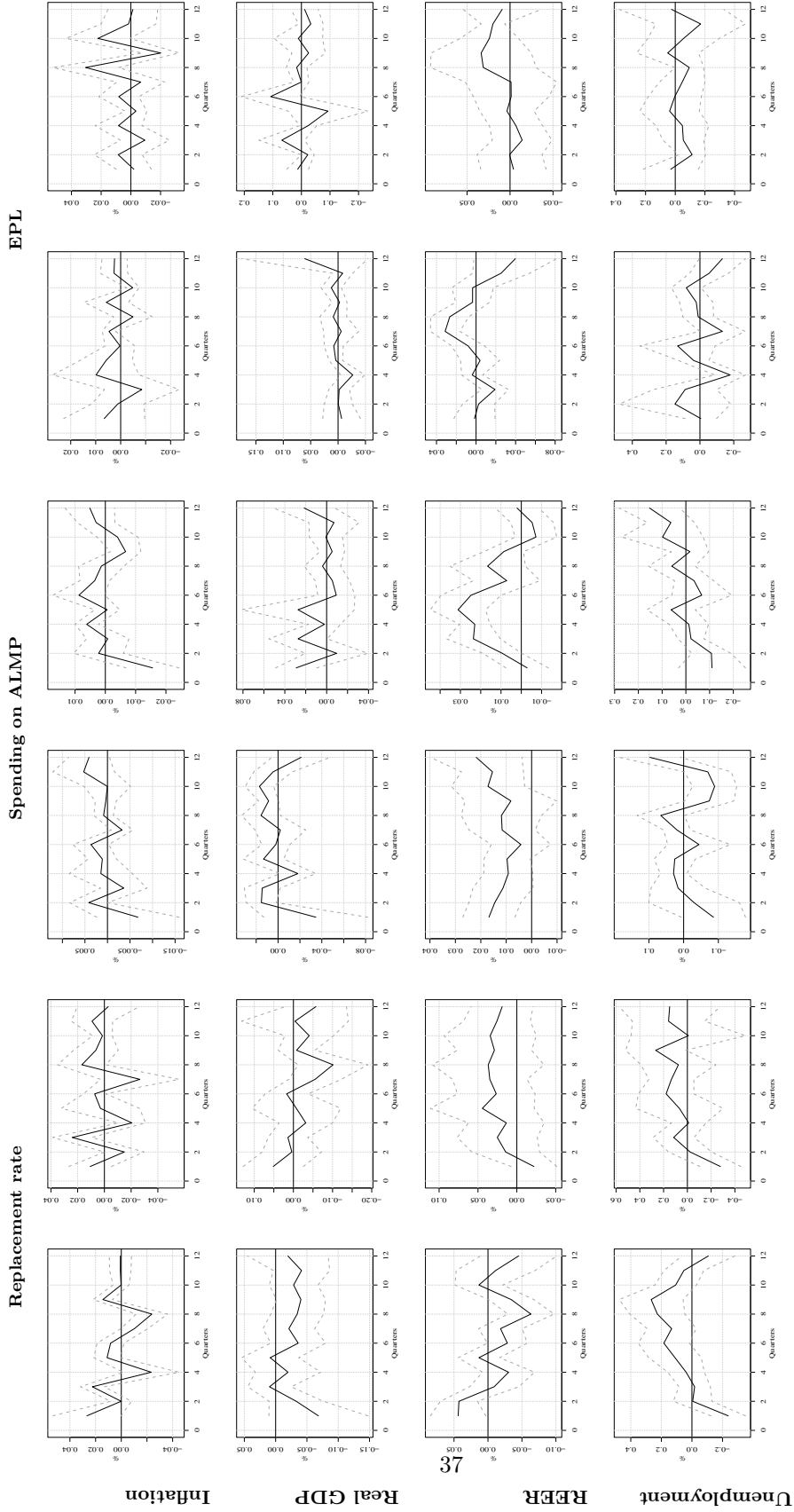

Figure E.1: Changes in macroeconomic variables due to the 1% increase in labor market policies (replacement rate, ALMP and EPL), falling interest rate (left) and rising interest rate (right), an expanded sample, model A

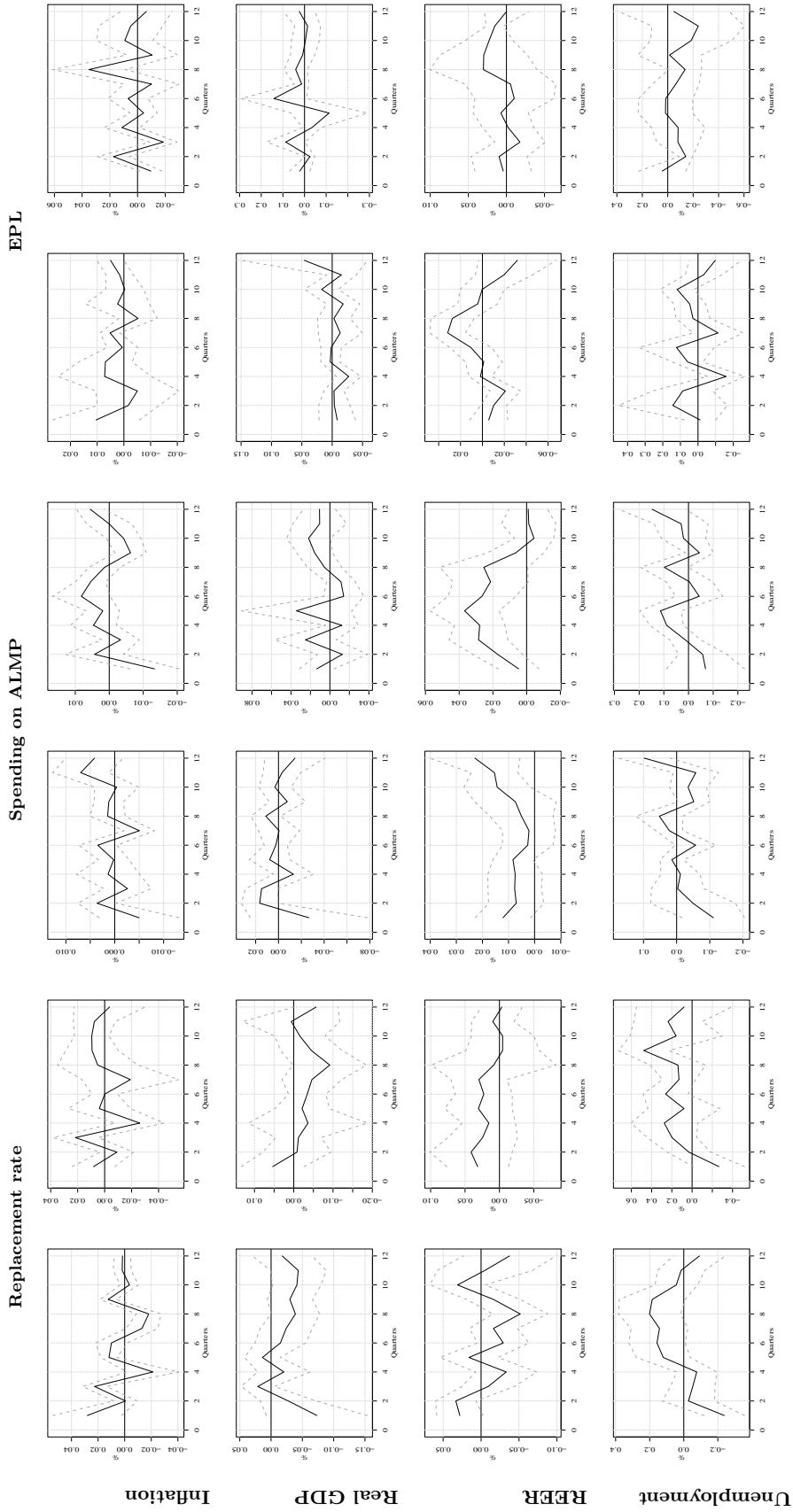

Figure E.2: Changes in macroeconomic variables due to the 1% increase in labor market policies (replacement rate, ALMP and EPL), falling interest rate (left) and rising interest rate (right), an expanded sample, model B

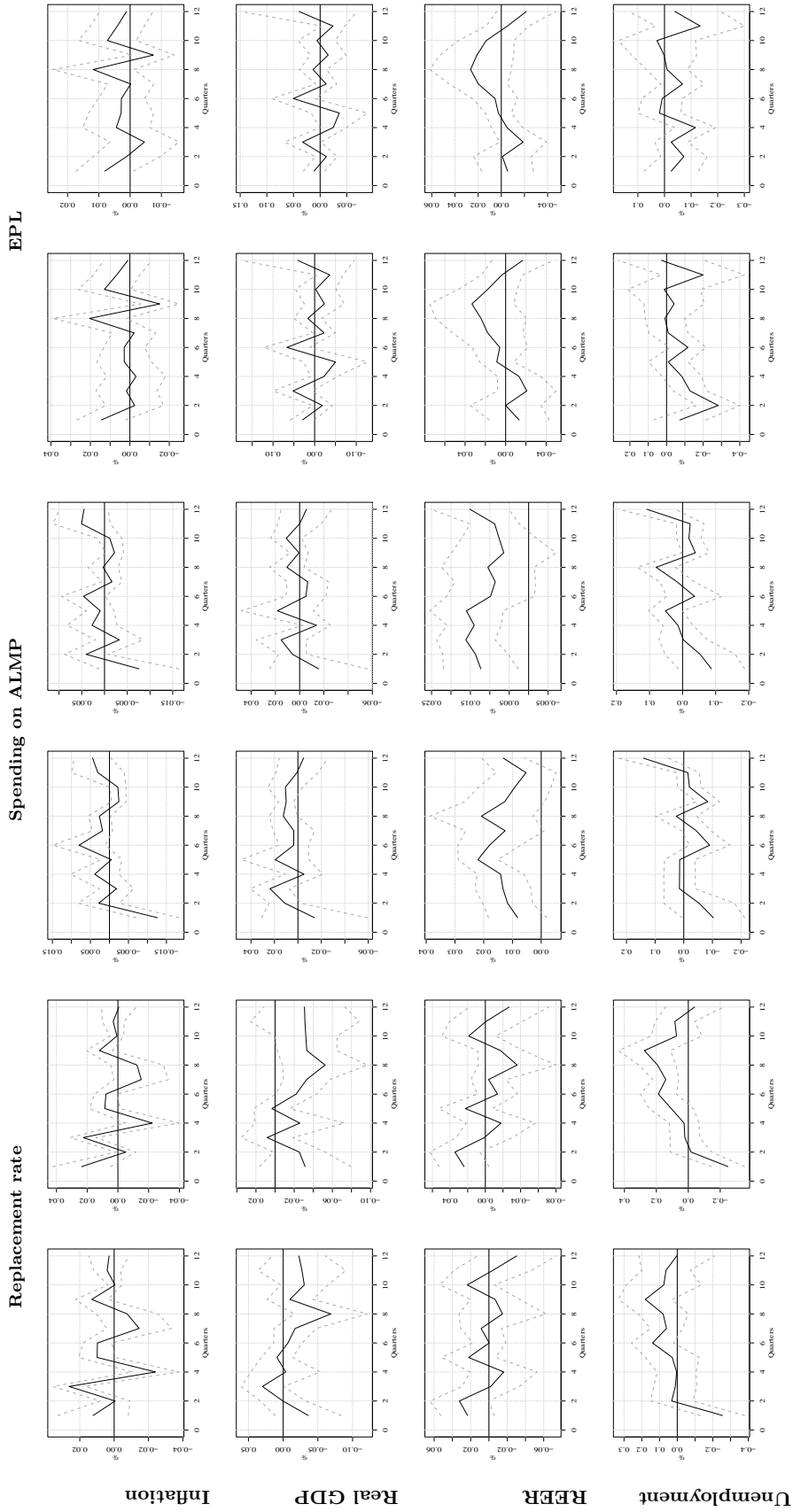

Figure E.3: Changes in macroeconomic variables due to the 1% increase in labor market policies (replacement rate, ALMP and EPL), first quarterly quartile (left) and third quarterly quartile (right), an expanded sample, model C

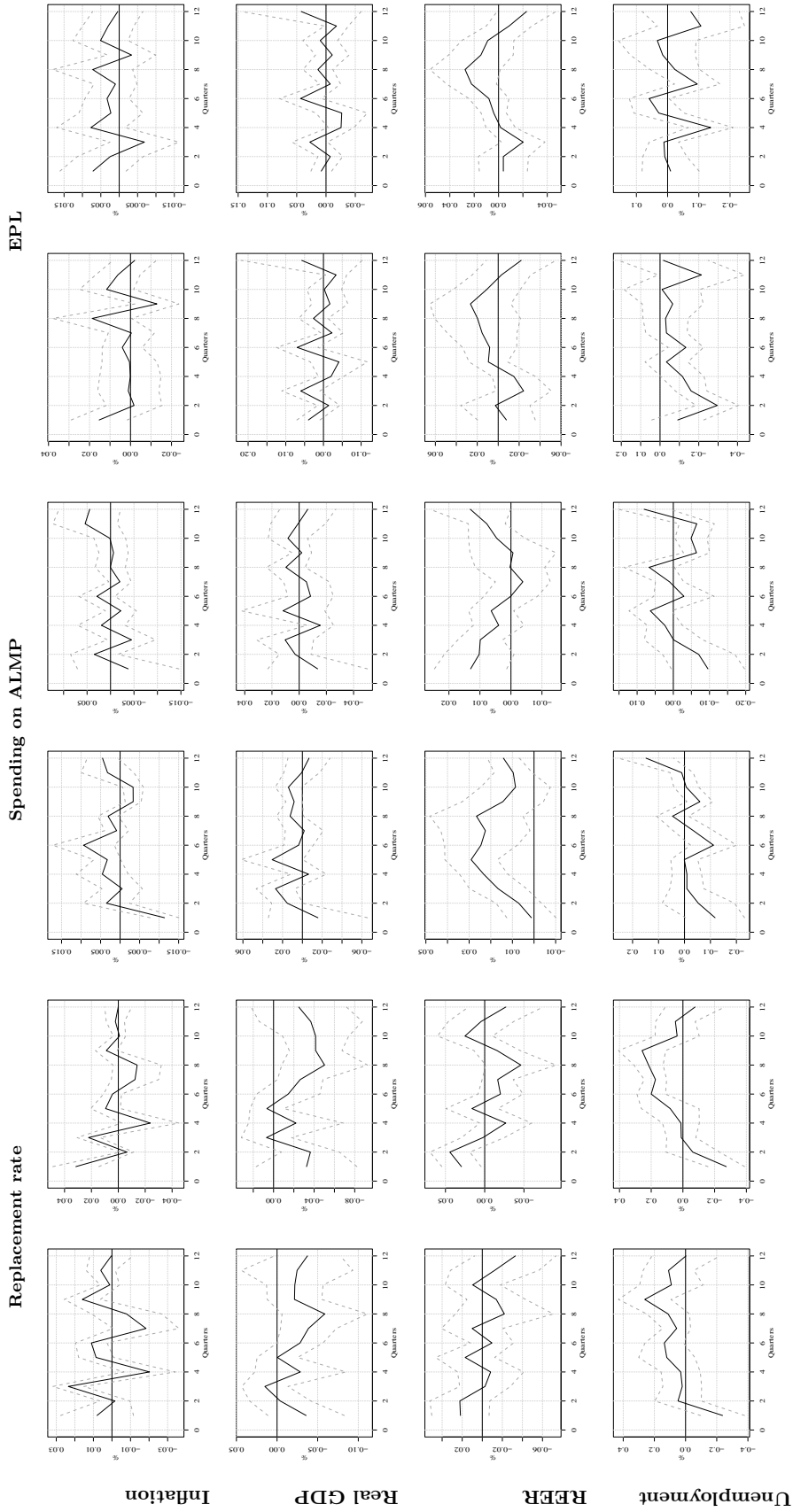

Figure E.4: Changes in macroeconomic variables due to the 1% increase in labor market policies (replacement rate, ALMP and EPL), first annual quartile (left) and third annual quartile (right), an expanded sample, model D

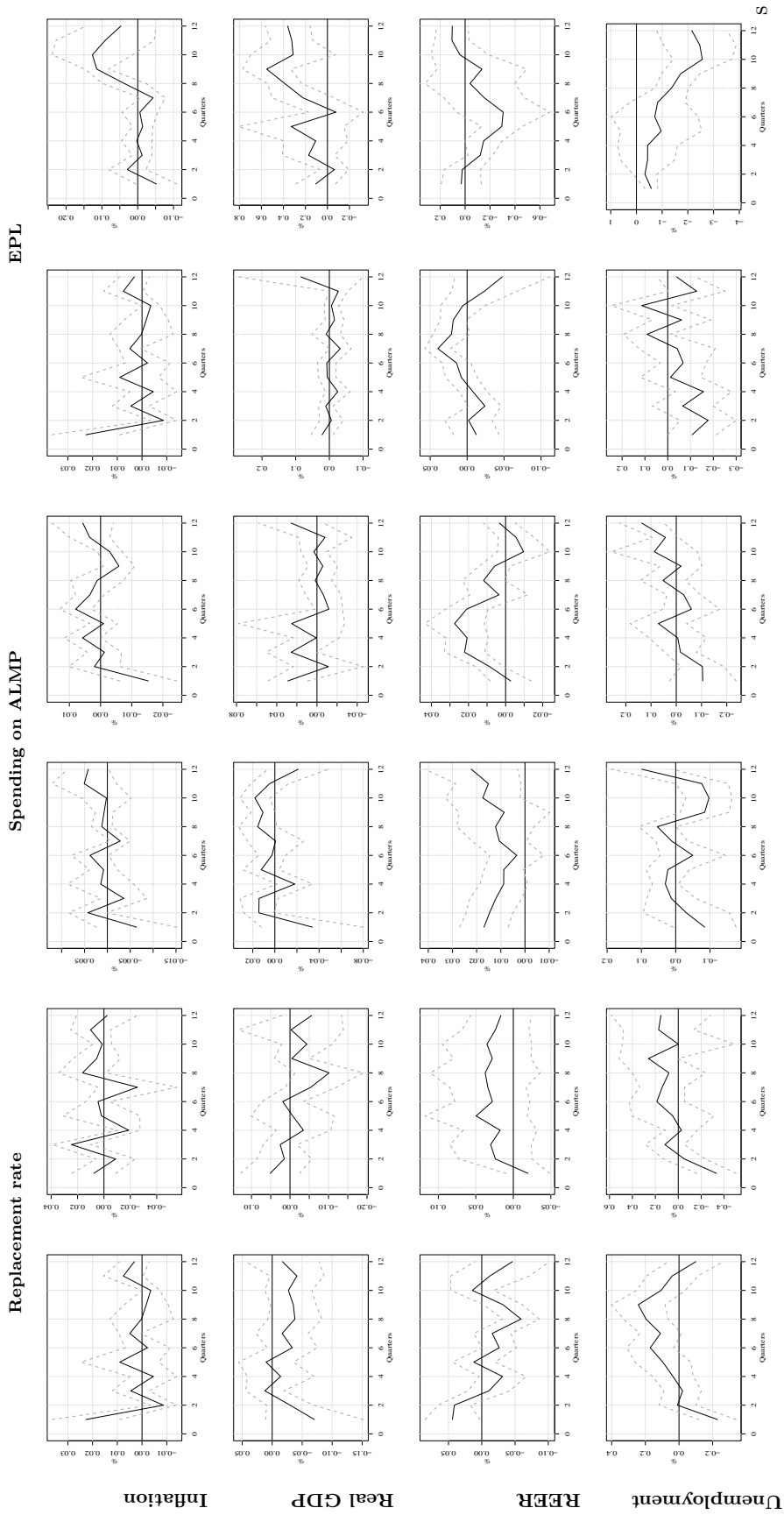

Figure E.5: Changes in macroeconomic variables due to the 1% increase in labor market policies (replacement rate, ALMP and EPL), falling interest rate (left) and rising interest rate (right), an expanded sample, model E

### E.1.2 Full Sample

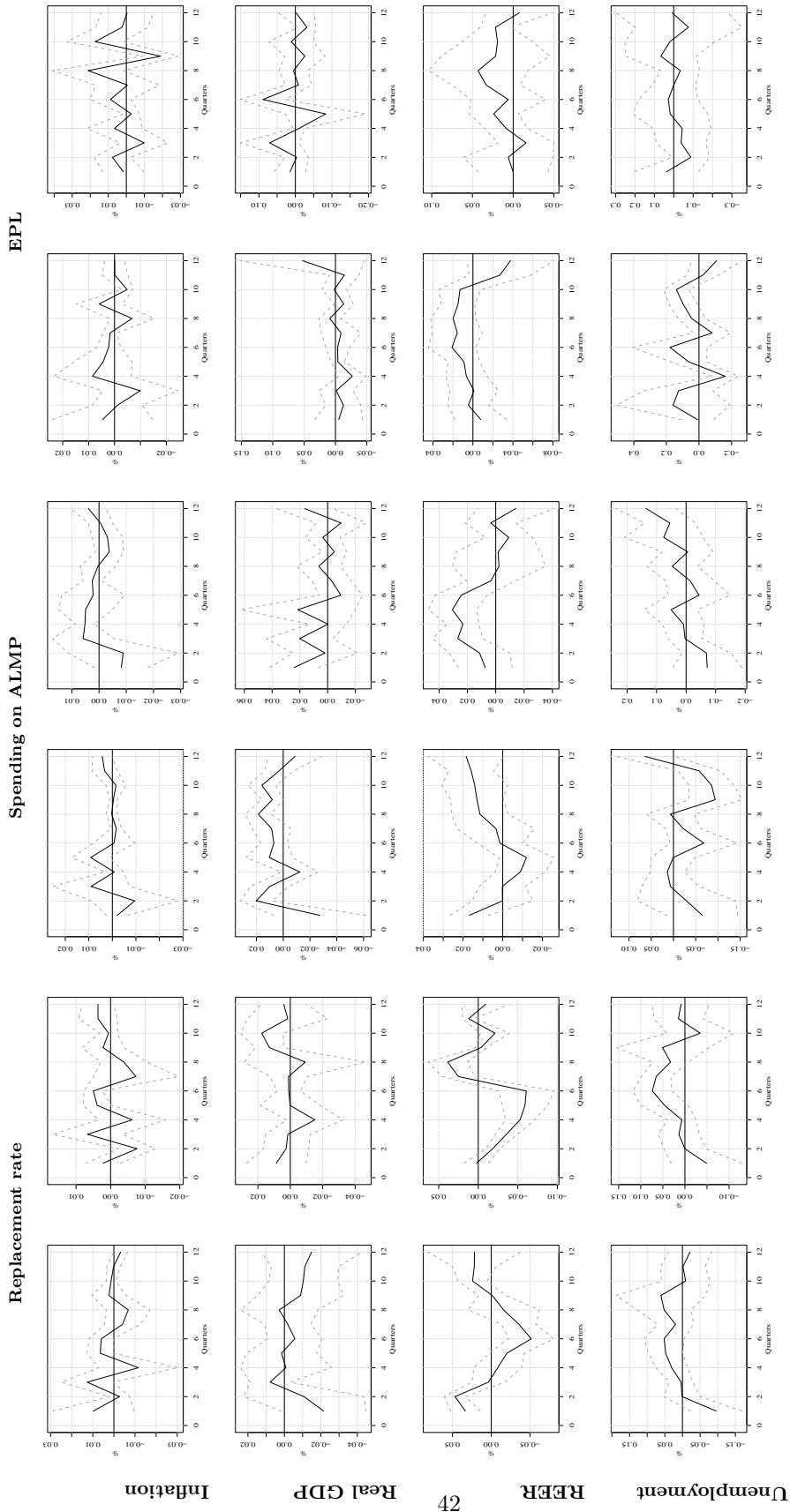

Figure E.6: Changes in macroeconomic variables due to the 1% increase in labor market policies (replacement rate, ALMP and EPL), falling interest rate (left) and rising interest rate (right), an expanded sample, model A

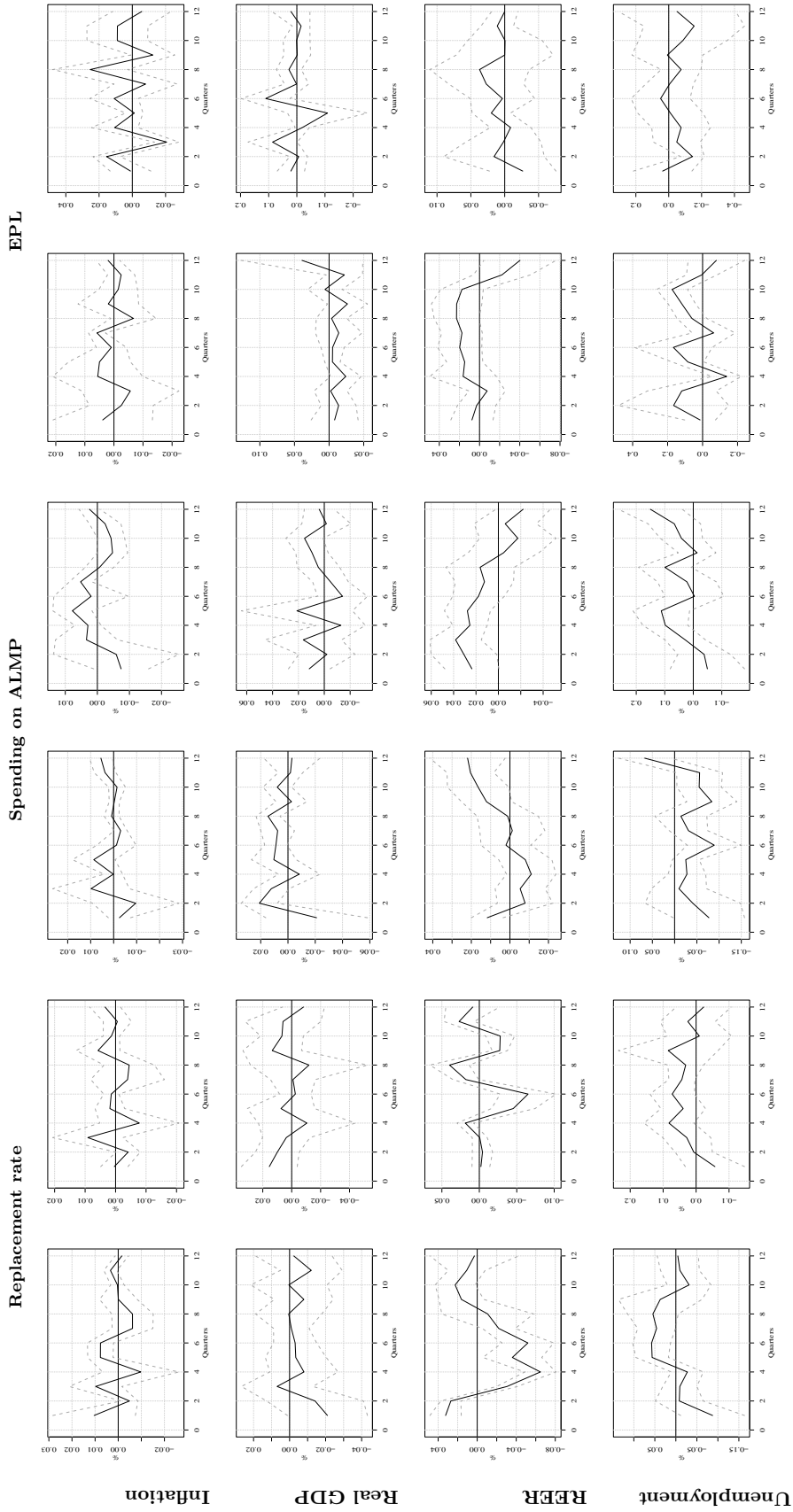

Figure E.7: Changes in macroeconomic variables due to the 1% increase in labor market policies (replacement rate, ALMP and EPL), falling interest rate (left) and rising interest rate (right), an expanded sample, model B

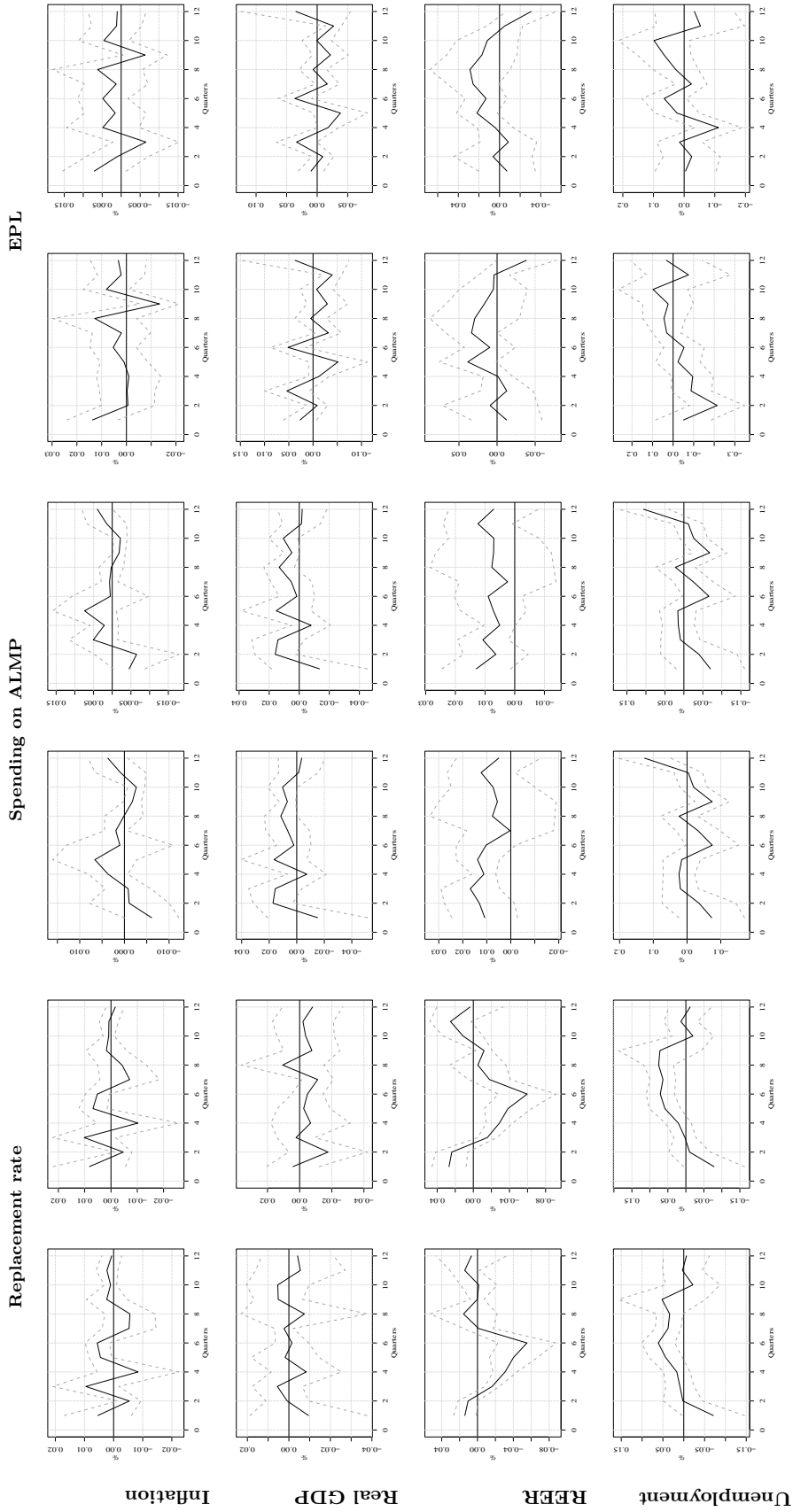

Figure E.8: Changes in macroeconomic variables due to the 1% increase in labor market policies (replacement rate, ALMP and EPL), first quarterly quartile (left) and third quarterly quartile (right), an expanded sample, model C

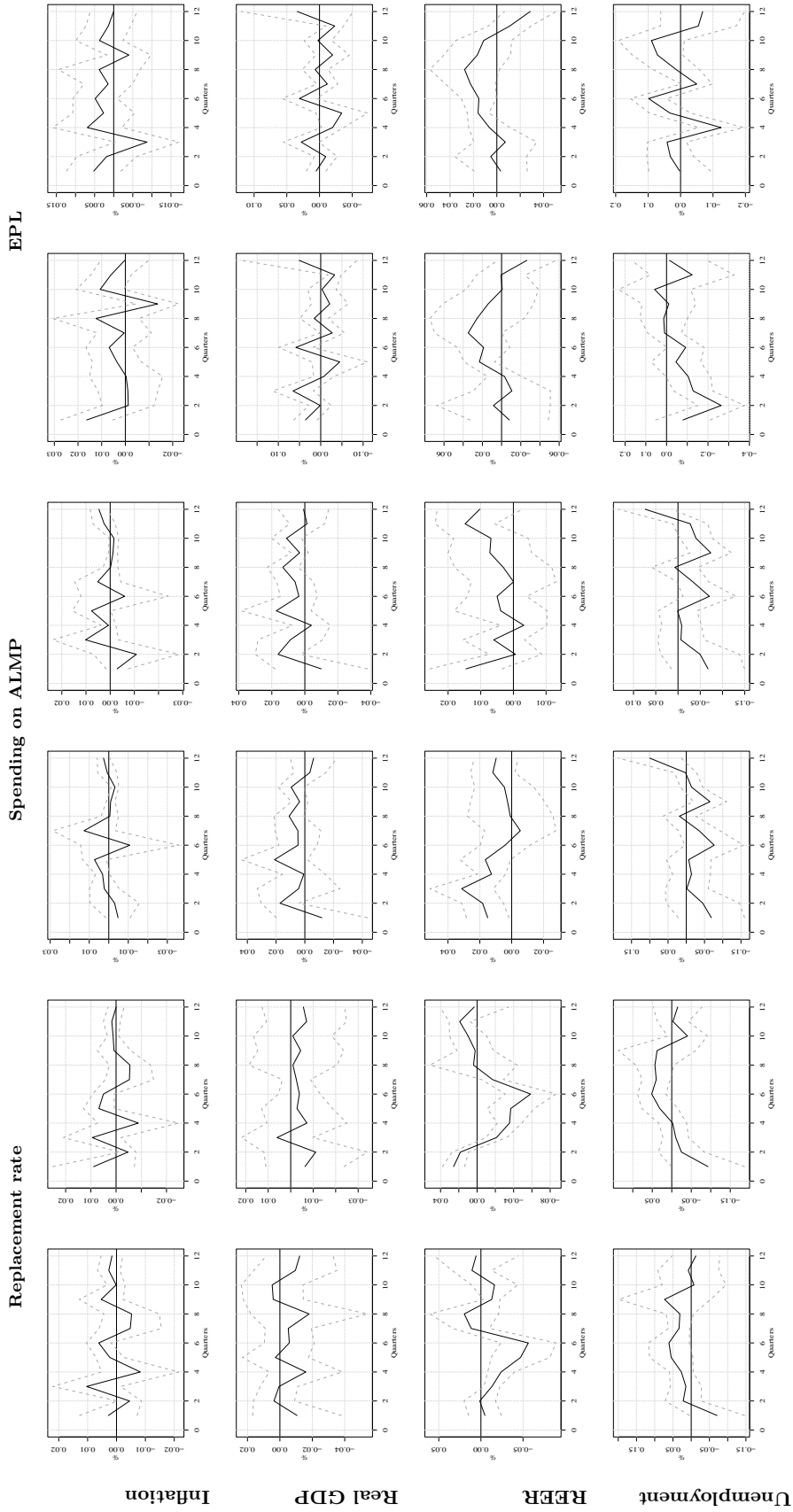

Figure E.9: Changes in macroeconomic variables due to the 1% increase in labor market policies (replacement rate, ALMP and EPL), first annual quartile (left) and third annual quartile (right), an expanded sample, model D

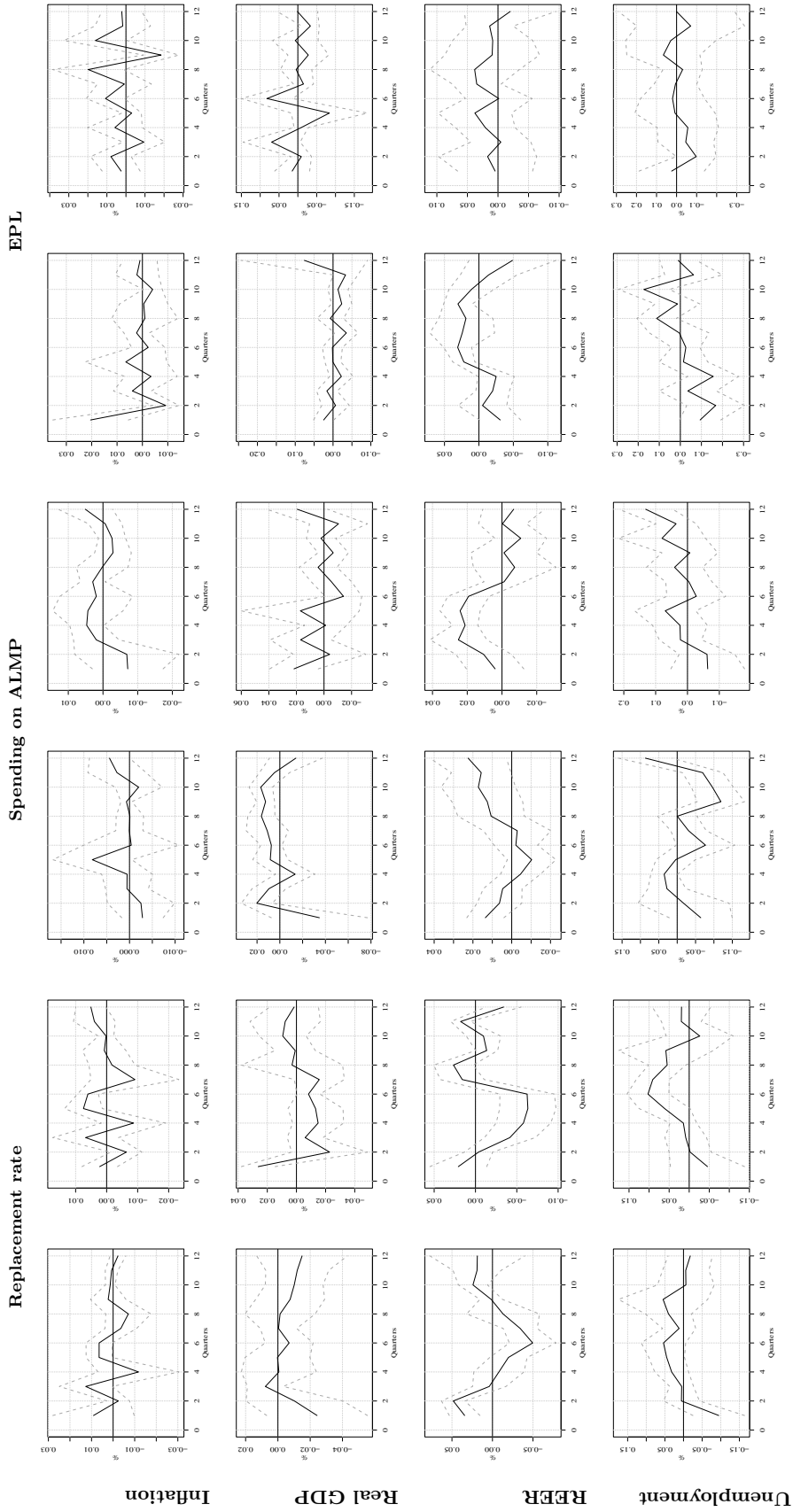

Figure E.10: Changes in macroeconomic variables due to the 1% increase in labor market policies (replacement rate, ALMP and EPL), falling interest rate (left) and rising interest rate (right), an expanded sample, model E

## E.2 Mallow's Weights: An Expanded Sample

## E.3 Average Local Projections: An Expanded Sample

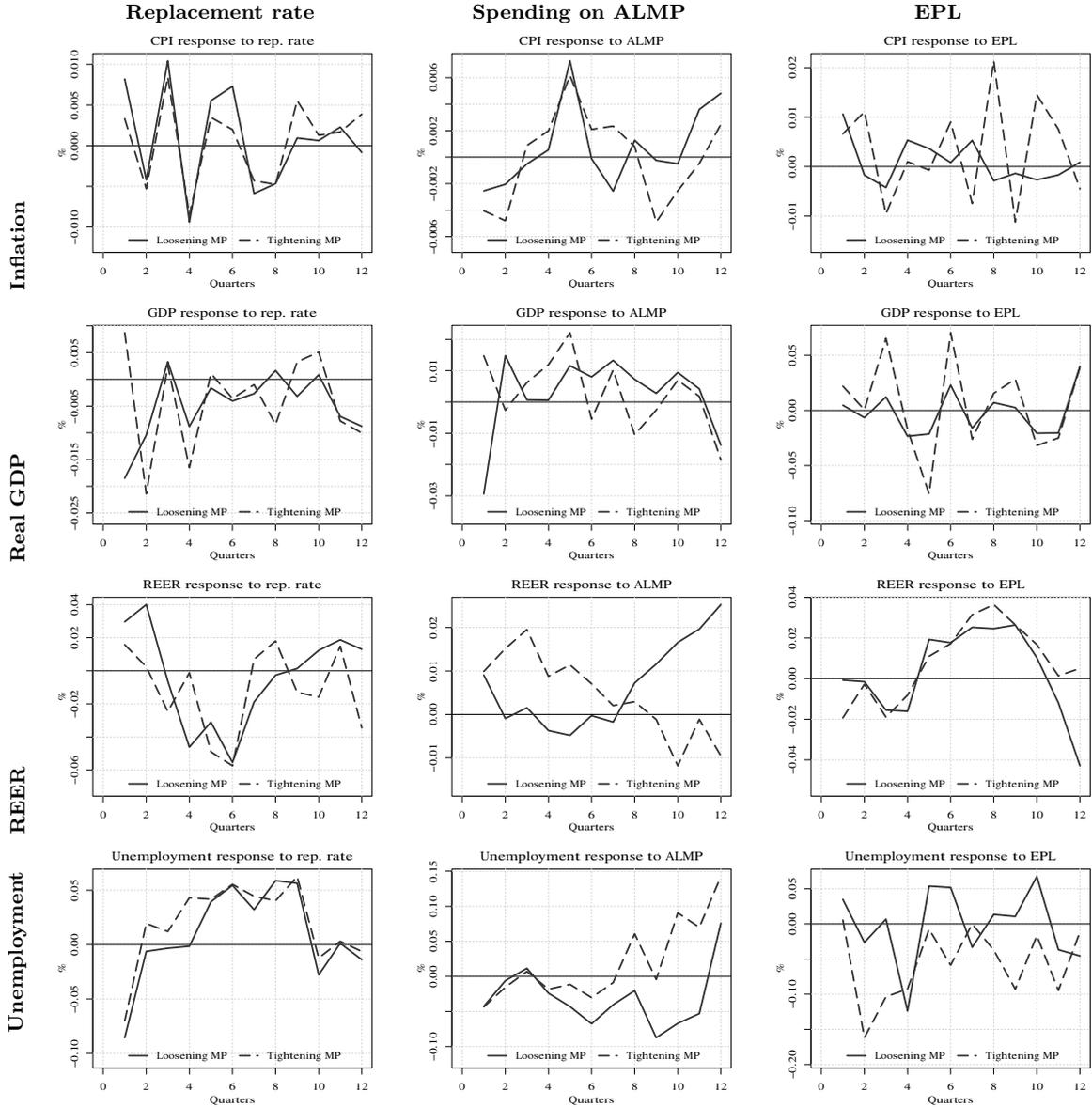

Figure E.13: Changes in macroeconomic variables due to the 1% increase in labor market policies (replacement rate, ALMP and EPL), weighted (Mallow's averaging) impulse responses (local projections) for an expanded sample

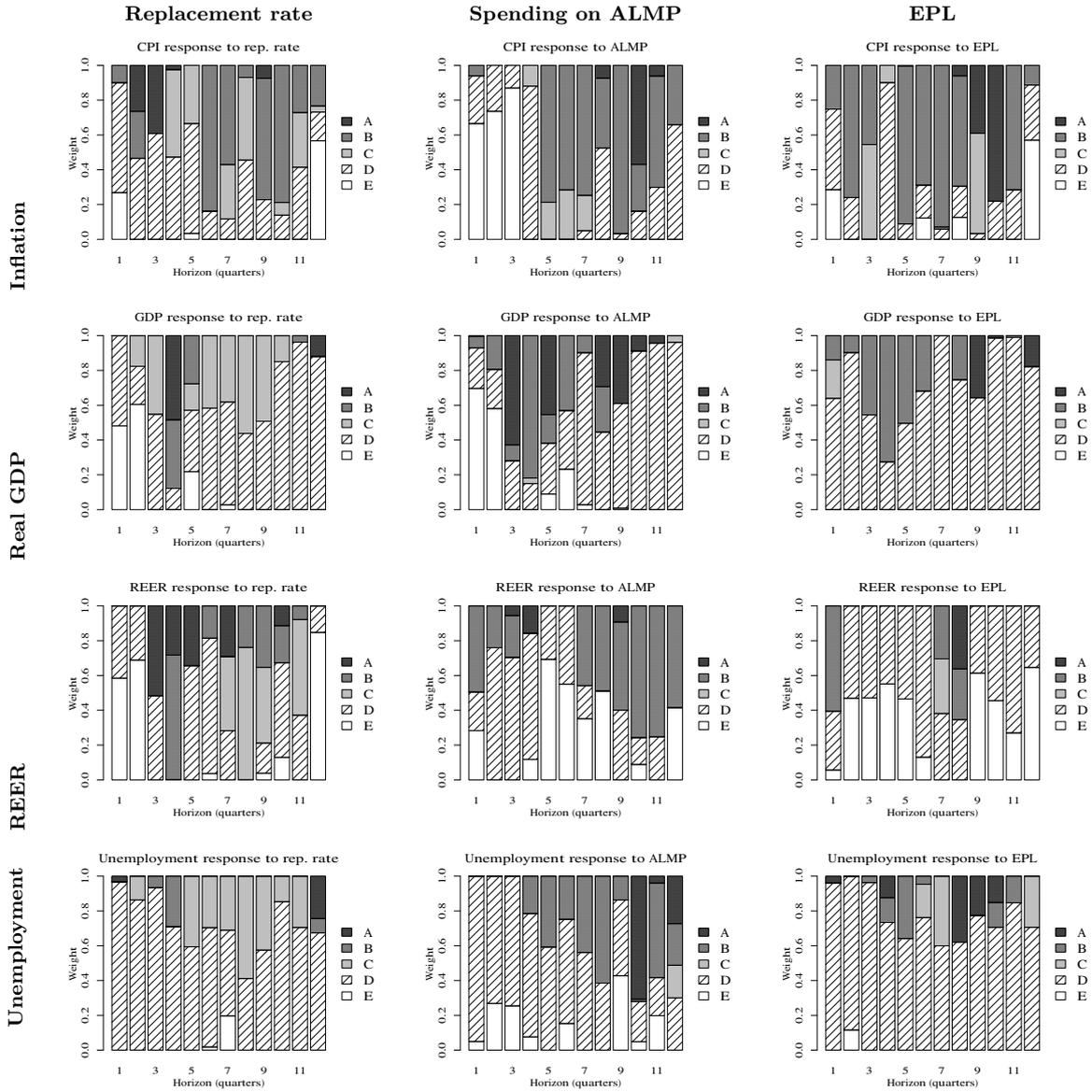

Figure E.11: Average model weights for each horizon, impulse and response variables (Mallow's criterion) for an expanded sample

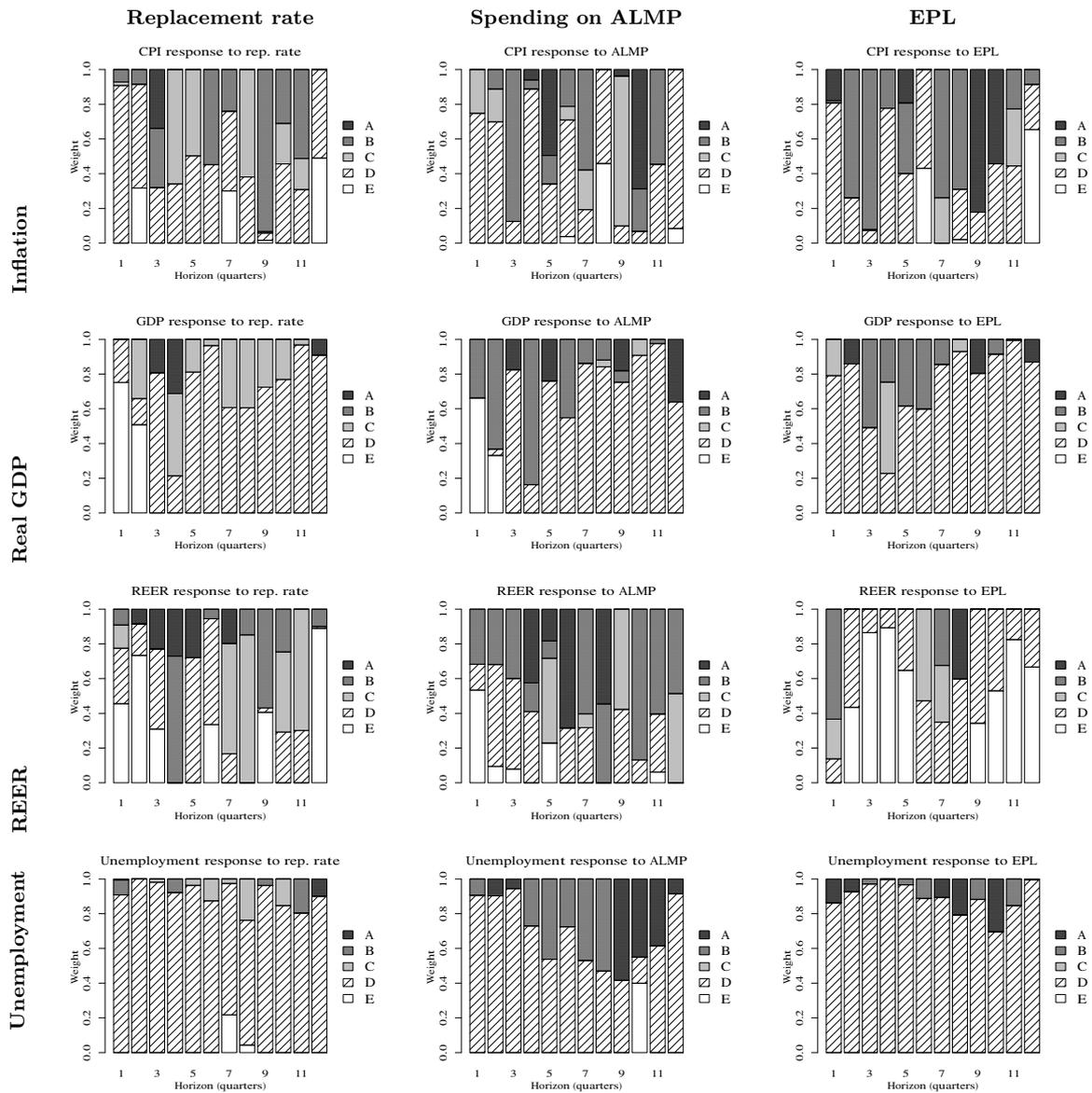

Figure E.12: Average model weights for each horizon, impulse and response variables (Mallow's criterion) with OECD GDP forecasts and HP-filtered output gaps for an expanded sample

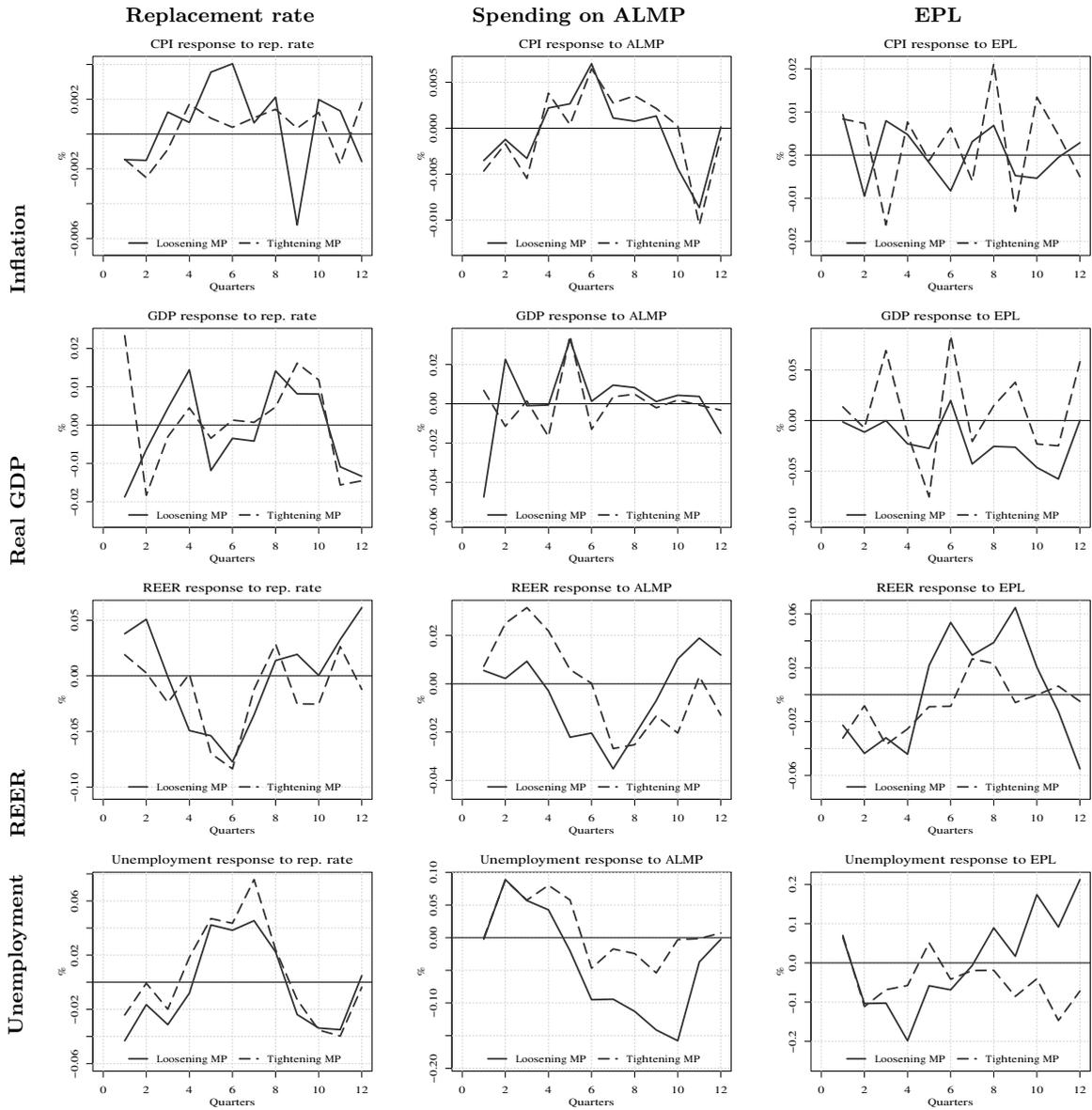

Figure E.14: Changes in macroeconomic variables due to the 1% increase in labor market policies (replacement rate, ALMP and EPL), weighted (Mallow's averaging) impulse responses (local projections) with OECD GDP forecasts and HP-filtered output gaps for an expanded sample